\documentclass[aps,prc,reprint,amsmath,amssymb,nofootinbib,floatfix,superscriptaddress]{revtex4-2}
\usepackage{graphicx}
\usepackage{dcolumn}
\usepackage{bm}
\usepackage{float}
\usepackage{dsfont} 
\usepackage{braket}
\usepackage{xcolor}
\usepackage{mathbbol}
\usepackage{multirow}
\usepackage{hhline}
\usepackage{array}
\usepackage{physics}
\usepackage{tabularx}
\usepackage{booktabs}
\usepackage{orcidlink}
\usepackage{xcolor}
\usepackage{hyperref}
\hypersetup{colorlinks=true,citecolor= blue}

\newcommand{\fett}[1]{\mbox{\boldmath $#1$}}

\usepackage{enumerate}
\usepackage{enumitem}

\newcommand{\beq}{\begin{equation}}
	\newcommand{\eeq}{\end{equation}}
\newcommand{\beqa}{\begin{eqnarray}}
                     \newcommand{\eeqa}{\end{eqnarray}}
\newcommand{\nn}{\nonumber \\ }                   

\newcolumntype{C}{>{\centering\arraybackslash}X}
\newcolumntype{R}{>{\raggedleft\arraybackslash}X}

\allowdisplaybreaks[4]
    
\begin{document}

\title{Spectroscopic basis for short-range three-nucleon forces}
\author{Josep~Sol\`{a} Cava\orcidlink{0009-0009-5970-5033}}
\affiliation
{Institut f\" ur Theoretische Physik II, Fakult\" at f\" ur Physik und Astronomie,
Ruhr-Universit\" at Bochum, D-44780 Bochum, Germany}
\author{Arseniy A.~Filin\orcidlink{0000-0002-7603-451X}}
\affiliation
{Institut f\" ur Theoretische Physik II, Fakult\" at f\" ur Physik und Astronomie,
Ruhr-Universit\" at Bochum, D-44780 Bochum, Germany}
\author{Sven Heihoff\orcidlink{0009-0000-3641-0640}}
\affiliation
{Institut f\" ur Theoretische Physik II, Fakult\" at f\" ur Physik und Astronomie,
Ruhr-Universit\" at Bochum, D-44780 Bochum, Germany}  
\author{Henri Paul Huesmann\orcidlink{0009-0006-2092-961X}}
\affiliation
{Institut f\" ur Theoretische Physik II, Fakult\" at f\" ur Physik und Astronomie,
Ruhr-Universit\" at Bochum, D-44780 Bochum, Germany}
\author{Evgeny Epelbaum\orcidlink{0000-0002-7613-0210}}
\affiliation
{Institut f\" ur Theoretische Physik II, Fakult\" at f\" ur Physik und Astronomie,
Ruhr-Universit\" at Bochum, D-44780 Bochum, Germany}

\begin{abstract}
We introduce a spectroscopic basis for the subleading contact
three-nucleon forces, which allows one to classify these interactions
according to the total angular momentum and parity quantum numbers in a 
transparent way. Using this new basis, we explore the sensitivity of
nucleon-deuteron observables to the three-nucleon short-range
interactions.
The low dimensionality of the variable-parameter space in 
the spectroscopic basis allows us to build a simple nucleon-deuteron scattering emulator using radial basis function interpolation. We perform exploratory fits of the
subleading contact three-nucleon interactions and demonstrate that $9$ of
$13$ low-energy constants can be reliably determined from elastic
nucleon-deuteron scattering data. 

\end{abstract}

\maketitle

\section{Introduction}

Three-nucleon forces (3NFs) are well known to play an important role
in nuclear physics \cite{Hammer:2012id,Endo:2024cbz}. While the bulk of nuclear properties can
be described using two-body potentials, tuned to reproduce 
low-energy nucleon-nucleon (NN) scattering data, the remaining
discrepancies between experimental data and {\it ab-initio}
calculations of nuclear structure and reactions are attributed to
poorly understood 3NFs \cite{Kalantar-Nayestanaki:2011rzs}. Three-nucleon interactions are
thus considered to be the main bottleneck in precision studies of
spectra and structure of light and medium-mass nuclei, as well as of the
equation of state of nuclear and neutron matter, relevant for
understanding the composition and mass-radius relations of neutron
stars \cite{Hebeler:2013nza}.  

The empirically established hierarchy of nuclear
interactions mentioned above is naturally explained in the framework
of chiral EFT
\cite{Weinberg:1990rz,vanKolck:1994yi,Epelbaum:2008ga}. In the last
decades, this model-independent and systematically improvable method
has established itself as the standard approach to describe low-energy
nuclear interactions. Combined with {\it ab-initio} few- and
many-body methods, it opens the way to describe nuclear systems
consistent with the chiral symmetry of QCD. In the two-body sector,
chiral EFT has been pushed to fifth expansion order (N$^4$LO) and successfully applied to perform
full-fledged partial-wave analysis of proton-proton and
neutron-proton scattering data up to pion production threshold
\cite{Reinert:2020mcu,Reinert:2022jpu}. The new generation of the N$^4$LO$^+$ NN potentials based on the semilocal
momentum-space regularization scheme (SMS) have been demonstrated to
provide a statistically perfect description of mutually consistent
NN scattering data \cite{Reinert:2017usi,Reinert:2020mcu,Epelbaum:2022cyo}. 

Three-nucleon forces first show up at third chiral order (N$^2$LO) in
the Weinberg power counting scheme. They originate from the tree-level
two-pion exchange, one-pion-exchange-contact and purely
contact diagrams \cite{vanKolck:1994yi,Epelbaum:2002vt}. The
longest-range two-pion exchange 3NF is parameter 
free, while the two shorter-range contributions are driven by the
low-energy constants (LECs) $D$ and $E$, which have to be determined
from experimental data beyond the two-nucleon sector. This dominant 3NF has
been extensively applied in calculations of nuclear structure and
reactions. The leading corrections to the
3NF at fourth expansion order (N$^3$LO) and most of the N$^4$LO contributions generated by
loop diagrams have been worked out using dimensional regularization
\cite{Ishikawa:2007zz,Bernard:2007sp,Bernard:2011zr,Krebs:2012yv,Krebs:2013kha},
see also Ref.~\cite{Krebs:2018jkc} for the long-range 3NF contributions from
intermediate $\Delta (1232)$ excitations. In
addition, at N$^4$LO, one needs to take into account 
tree-level contributions of the one-pion-exchange-contact type, which
involve new LECs $F_{1, \ldots , 16}$ \cite{Huesmann:2026khj}, and
purely short-range 3NFs that depend on the  
LECs $E_{1, \ldots , 13}$ \cite{Girlanda:2011fh}. 

To establish accurate and precise 3NF, it is necessary to take
into account the above-mentioned corrections beyond the dominant
N$^2$LO terms, which requires addressing two key challenges. On the
conceptual side, it was shown in Refs.~\cite{Epelbaum:2019kcf,Krebs:2019uvm} that using a naive
cutoff regularization of 3NFs and exchange current operators on top
of dimensional regularization violates chiral symmetry. Accordingly,
loop corrections to the 3NF starting from N$^3$LO need to be rederived
using a symmetry-preserving cutoff regulator. Only recently has the
required methodology for such calculations become available by merging
chiral EFT with the gradient flow method \cite{Krebs:2023ljo,Krebs:2023gge}. This novel scheme
has already been successfully applied to derive the N$^3$LO
contributions to the 3NFs, whose numerical implementation is in
progress by the LENPIC Collaboration.

The second challenge is a
computational one due to the need to determine a large
number of LECs entering the 3NF up to N$^4$LO from experimental
data. Besides $^3$H/$^3$He binding energies, three-nucleon
scattering data offer the most direct and accessible means for
constraining such LECs. The feasibility of extracting the LECs $E_i$,
which accompany the subleading contact 3NFs, was already explored in
Refs.~\cite{Girlanda:2018xrw,Witala:2022rzl}, see also
Ref.~\cite{Margaryan:2015rzg} for a related study
in the context of pionless EFT. 

Our work provides new insights into
this topic by introducing a spectroscopic basis for the purely
contact 3NFs. This basis offers a transparent interpretation of the
LECs entering contact 3NFs by disentangling their contributions to
partial waves in elastic nucleon-deuteron (Nd) scattering, see also
Ref.~\cite{Filandri:2026ori} for a related discussion.
We study the sensitivity of vector and tensor analyzing powers to the LECs in the spectroscopic
basis, denoted by $S_i$, and perform exploratory fits to selected
observables in elastic Nd scattering in order to assess the
feasibility of their determination from the available experimental
data. Our work thus provides an important step towards developing
precise 3NFs in the framework of chiral EFT. 

Our paper is organized as follows. In sec.~\ref{sec:SpecBasis}, we
define the spectroscopic basis for the subleading contact 3NFs and
provide triton-state expectation values for individual terms to serve as
benchmarks. Sec.~\ref{sec:Applications} presents a comprehensive investigation of the
impact of these short-range interactions on Nd scattering
observables. To go beyond the sensitivity analysis of specific
observables to the individual 3NF terms, we introduce a simple interpolation-based
emulator,
designed to accelerate the calculation of Nd observables for the
parametric updates of the Hamiltonian. This emulator is then used to
carry out exploratory fits of the 3NF LECs in elastic Nd
scattering. The main results of our paper are summarized in
sec.~\ref{sec:Summary}, while Appendices
\ref{Appendix:PWD}-\ref{Appendix:AccuracyEmulator} provide additional
technical details. In particular, in Appendix \ref{Appendix:PWD} we
give the explicit analytical expressions for partial-wave
decomposition of all $13$ terms in the contact part of the 3NF at
N$^4$LO.

\section{Spectroscopic basis}
\label{sec:SpecBasis}

The most general structure of the subleading contact 3NFs can be parametrized by 13 operators \cite{Girlanda:2011fh} in the form \cite{Witala:2022rzl}
\beqa
\label{DefEi}
V_{\rm 3N} &=& \sum_{i \neq j \neq
  k} \big( E_1 \fett q_i^2 + E_2 \fett q_i^2 \fett \tau_i \cdot \fett \tau_j \nn
&+&
E_3 \fett q_i^2 \fett \sigma_i \cdot \fett \sigma_j 
+ E_4 \fett q_i^2 \fett \tau_i \cdot \fett \tau_j\fett \sigma_i \cdot \fett \sigma_j\nn
&+& E_5 (3 \fett q_i \cdot \fett \sigma_i \fett q_i \cdot \fett \sigma_j -\fett q_i^2 \fett \sigma_i \cdot \fett \sigma_j)\nn
&+&
E_6  (3 \fett q_i \cdot \fett \sigma_i \fett q_i \cdot \fett \sigma_j -\fett q_i^2 \fett \sigma_i \cdot \fett \sigma_j)  \fett \tau_i \cdot \fett \tau_j \nn
&+&  i E_7 \fett q_i \times (\fett k_i - \fett k_j) \cdot (\fett \sigma_i + \fett \sigma_j) \nn
&+&  i E_8 \fett q_i \times (\fett k_i - \fett k_j) \cdot (\fett \sigma_i + \fett \sigma_j) \fett \tau_i \cdot \fett \tau_j \nn
&+& E_9 \fett q_i \cdot \fett \sigma_i \fett q_j \cdot \fett \sigma_j
+ E_{10} \fett q_i \cdot \fett \sigma_i \fett q_j \cdot \fett \sigma_j \fett \tau_i \cdot \fett \tau_j \nn
&+& E_{11} \fett q_i \cdot \fett \sigma_j \fett q_j \cdot \fett \sigma_i
+ E_{12} \fett q_i \cdot \fett \sigma_j \fett q_j \cdot \fett \sigma_i \fett \tau_i \cdot \fett \tau_j \nn
&+&E_{13} \fett q_i \cdot \fett \sigma_j \fett q_j \cdot \fett \sigma_i \fett \tau_i \cdot \fett \tau_k \big) ,
\eeqa
where $\fett q_i = \fett p_i' - \fett p_i $, $\fett k_i = (\fett p_i'
+ \fett p)/2$ with $\fett p_i$ and $\fett p_i'$  denoting the initial
and final momenta of nucleon $i$, respectively. Further, $\fett
\sigma_i$ ($\fett \tau_i$) refer to the Pauli spin (isospin) matrices
of nucleon $i$, while $E_i$ denote the corresponding
LECs. Partial-wave decomposition of the $E_1$ and $E_7$ terms is
discussed in Ref.~\cite{Epelbaum:2019zqc}, while partial-wave
decomposition of all $13$ terms in the $jJ$-coupling scheme is given
in Ref.~\cite{Witala:2022rzl}, see also Ref.~\cite{Filandri:2026ori}
for a related discussion. For the sake of completeness and since we
have been unable to reproduce some of the formulas from
Ref.~\cite{Witala:2022rzl}, we give in Appendix \ref{Appendix:PWD} the
expressions for the partial-wave decomposition of the first Faddeev
component of all terms in Eq.~(\ref{DefEi}) using the  $jJ$-coupling. 

To facilitate benchmarking, we provide in Table \ref{Table1} the expectation values of the individual terms in Eq.~(\ref{DefEi})\footnote{When calculating 3N observables, the 3NF in Eq.~(\ref{DefEi}) is multiplied with 
$\exp\phantom{}(-(\fett P_{ij}'^2 + \fett P_{ij}^2)/\Lambda^2) \exp(-3(\fett Q_k'^2 + \fett Q_k^2)/(4 \Lambda^2))$, where $\fett P_{ij}$, $\fett Q_k$ ($\fett P_{ij} '$, $\fett Q_k '$) are the initial (final) Jacobi momenta and the cutoff $\Lambda$ is set to $\Lambda = 450$~MeV.} in the triton, calculated using the SMS chiral N$^4$LO$^+$ potential of Ref.~\cite{Reinert:2017usi} with the cutoff $\Lambda = 450$~MeV. For this interaction, we find the triton binding energy of $E_{\rm ^3 H} = 8.143$~MeV.
\begin{table}[!tp]
  \vspace{0.2cm}
  \begin{ruledtabular}
	\begin{tabular*}{0.49\textwidth}{@{\extracolsep{\fill}}rrr}
          $i$ & $\langle V_{E_i} \rangle_{^3\rm H}$ (MeV)& $\langle V_{S_i} \rangle_{^3\rm H}$  (MeV)\\\hline
          $1$ & $0.348$ & $0.098$ \\
          $2$ & $-0.419$ & $0.187$ \\
          $3$ & $-0.277$ & $-0.615$ \\
          $4$ & $-1.043$ & $-0.020$ \\
          $5$ & $-0.457$ & $0$ \\           
          $6$ & $1.371$ & $0$ \\
          $7$ & $0.000$ & $0$ \\
          $8$ & $0.009$ & $0$ \\
          $9$ & $0.324$ & $0$ \\
          $10$ & $-0.749$ & $0$ \\
          $11$ & $0.306$ & $0$ \\
          $12$ & $-0.732$ & $0$ \\
          $13$ & $-0.094$ & $0$
        \end{tabular*}
        \caption{Expectation values of the subleading contact 3N interactions in the original operator basis defined in Eq.~(\ref{DefEi}) and in the spectroscopic basis in the triton. The triton wave function is calculated using the SMS chiral N$^4$LO$^+$ potential of Ref.~\cite{Reinert:2017usi} with the cutoff $\Lambda = 450$~MeV. The expectation values of the individual $E_i$-terms ($S_i$-terms) in the original (spectroscopic) basis for the corresponding dimensionless coefficients set to $e_i = 1$ ($s_i = 1$) are shown in the second (third) column.}
    \label{Table1}                                    
   \end{ruledtabular}
 \end{table}
Throughout this paper, we adopt the usual convention by expressing
the LECs $D$, $E$ and $E_i$ in terms of the corresponding
dimensionless parameters $c_D$, $c_E$ and $e_i$, respectively,  via
 \beq
D = \frac{c_D}{F_\pi^2 \Lambda_\chi},  \qquad  E = \frac{c_E}{F_\pi^4 \Lambda_\chi},  \qquad E_i = \frac{e_i}{F_\pi^4 \Lambda_\chi^3}, 
 \eeq
 where $F_\pi = 92.4$~MeV is the pion decay constant and the hard scale $\Lambda_\chi$  is set to $\Lambda_\chi = 700$~MeV. 
We expect the values quoted in Table \ref{Table1} to be accurate at the level of a few keV. Notice that the expectation values for $E_5$, $E_6$ and $E_{9, \ldots , 13}$ differ from those of Ref.~\cite{Witala:2022rzl}. 

Each of the operators in Eq.~(\ref{DefEi}) contributes to multiple
$J^P(T)$-states and to multiple partial waves in Nd scattering. More
precisely, the subleading short-range 3NF has nonvanishing matrix
elements in $30$ partial waves in total: $10$ in the $J^P(T) = 1/2^+ (1/2)$,
$6$ in the $J^P(T) = 1/2^- (1/2)$, $3$ in the $J^P(T) = 1/2^- (3/2)$,
$6$ in the $J^P(T) = 3/2^- (1/2)$,  $3$ in the $J^P(T) = 3/2^- (3/2)$,
and $2$ in the $J^P(T) = 5/2^- (1/2)$ channels. To define the
spectroscopic basis for the terms in Eq.~(\ref{DefEi}), we performed analytically
partial-wave decomposition of the {\it full} antisymmetrized 3NF (in contrast to its
first Faddeev component as discussed in Appendix \ref{Appendix:PWD}). 
We then found that in each  $J^P(T)$-channel, only a small number of linear combinations of $E_i$ contribute. Accordingly, we define a new set of LECs $S_i$ via the relations 
\beqa
\label{SpectroLECs}
S_1 &=& \frac{1}{2}   (2 E_1 - E_2 - 3 E_3 - 6 E_4 + E_9 + E_{10} + E_{11} \nn
&& {} + E_{12} - 
2 E_{13}),\nn
S_2 &=& -\frac{2}{9}   (-6 E_1 + 9 E_2 + 3 E_3 + 18 E_4 + E_9 -3 E_{10} \nn
&& {} + E_{11} - 3 E_{12}),\nn
S_3 &=& \frac{\sqrt{2}}{6} (3 E_5 - 9E_6 -2 E_9 + 6 E_{10} -2 E_{11} +6 E_{12}),\nn
S_4 &=& -\frac{2}{9} (E_2 - E_3 -5 E_5 + 15 E_{6} +2 E_{8} +6 E_{9} \nn
&& {} - 12 E_{10} + 2 E_{11} - 8 E_{12} + E_{13}), \nn
S_5 &=& -\frac{1}{18} (6 E_1- 6E_2  - 6 E_3 +18 E_4 -6 E_{7} +6 E_{8} \nn
&& {} +7 E_{9}  - 15 E_{10} -5 E_{11} - 3 E_{12} + 9 E_{13}), \nn
S_6 &=& -\frac{1}{9} (6 E_1 + 6 E_3 -30 E_5 - 15 E_{7} +45 E_{8} +4 E_{9} \nn
&& {} + 12 E_{10} + 4 E_{11} +12 E_{12} -12  E_{13}), \nn
S_7 &=& -\frac{\sqrt{2}}{18} (15 E_5 + 45 E_6 +3 E_7 -6  E_{8} -10 E_{9} +6 E_{10} \nn
&& {} - 10 E_{11} -6  E_{12} +18 E_{13}), \nn
S_8 &=& \frac{1}{18} (-6 E_1 +6 E_2 +6 E_3 -18 E_{4} -3 E_{7} +3 E_{8} \nn
&& {} + 2 E_{9} + 6 E_{10} - 4 E_{11} +12 E_{12}), \nn
S_9 &=& -\frac{1}{9} (6E_1 + 6E_3 +24 E_5 -6 E_{7} +18 E_{8} -5 E_{9} \nn
&& {} - 15 E_{10} -5 E_{11} - 15 E_{12} + 15 E_{13}), \nn
S_{10} &=&\frac{\sqrt{5}}{18} (3 E_5 + 9 E_6 -3 E_7 + 6 E_{8} -2 E_{9} -6 E_{10} \nn
&& {} - 2 E_{11} + 6 E_{12}), \nn
S_{11} &=& -\frac{1}{3} (2 E_1 + 2 E_3 -2 E_5 + 3 E_{7} -9 E_{8}), \nn
S_{12} &=& -\frac{1}{3} (2 E_1 + 2 E_2 -4 E_7 -4 E_{8} -3 E_{9} -3 E_{10} \nn
&& {} + E_{11} +  E_{12} + E_{13}), \nn
S_{13} &=& -\frac{2}{3} (E_1 + E_2 + E_7 + E_{8} - E_{11} - E_{12}
-E_{13}) .
\eeqa
In a close analogy to the LECs $E_i$, we parametrize the new
 LECs in terms of the corresponding dimensionless
coefficients $s_i$ via
 \beq
S_i = \frac{s_i}{F_\pi^4 \Lambda_\chi^3}. 
 \eeq
In contrast to $E_i$'s, the new LECs $S_i$ contribute to single $J^P(T)$-channels as shown in Table \ref{tab:Spectro_summary}.

\begin{table}[!h]
  \centering
  \vspace{0.2cm}
  \renewcommand{\arraystretch}{1.2}
  \setlength{\tabcolsep}{13.5pt}
  \begin{tabular}{c|c|c|l}
    \hline\hline
    LEC & Isospin & $J^P$ & \phantom{X}Nd transition \\
    \hline
    $S_1$    & $T=1/2$ & $1/2^+$ & \phantom{X}$^2S \leftrightarrow {}^2S$ \\
    $S_2$    &         &         & \phantom{X}$^2S \leftrightarrow {}^2S$ \\
    $S_3$    &         &         & \phantom{X}$^2S \leftrightarrow {}^4D$ \\
    $S_4$    &         &         & \phantom{X}$^2S \leftrightarrow \text{breakup}$ \\
    \cline{3-4}
    $S_5$    &         & $1/2^-$ & \phantom{X}$^2P \leftrightarrow {}^2P$ \\
    $S_6$    &         &         & \phantom{X}$^4P \leftrightarrow {}^4P$ \\
    $S_7$    &         &         & \phantom{X}$^2P \leftrightarrow {}^4P$ \\
    \cline{3-4}
    $S_8$    &         & $3/2^-$ & \phantom{X}$^2P \leftrightarrow {}^2P$ \\
    $S_9$    &         &         & \phantom{X}$^4P \leftrightarrow {}^4P$ \\
    $S_{10}$ &         &         & \phantom{X}$^2P \leftrightarrow {}^4P$ \\
    \cline{3-4}
    $S_{11}$ &         & $5/2^-$ & \phantom{X}$^4P \leftrightarrow {}^4P$ \\
    \cline{2-4}
    $S_{12}$ & $T=3/2$ & $1/2^-$ & \phantom{X}--- \\
    $S_{13}$ &         & $3/2^-$ & \phantom{X}--- \\
    \hline\hline
  \end{tabular}
  \caption{Contributions of the subleading contact 3NFs in the spectroscopic basis to specific partial waves in Nd scattering.}
  \label{tab:Spectro_summary}
\end{table}

In line with Refs.~\cite{Witala:2022rzl,Filandri:2026ori}, we find two
linear combinations of LECs contributing solely to the isospin $T=3/2$ channels, which are denoted as $S_{12}$ and $S_{13}$. The remaining LECs $S_{1, \ldots , 11}$ possess a simple and transparent interpretation in terms of partial
waves and mixing angles contributing to Nd scattering. Specifically,
using the 
channel spin operator $\fett \Sigma = \fett J_d + \fett
s_N$, the S-matrix $S^J_{\lambda' \Sigma ' , \lambda \Sigma}$ for
elastic Nd scattering in the channel spin representation \cite{Seyler:1969sii} becomes a $2 \times 2$ ($3 \times 3$) matrix in
the $J=1/2$ states (all other states).
The expressions connecting the S-matrix in the $jJ$ and channel spin representations can be found in Ref.~\cite{Gloeckle:1995jg}.
For $J > 1/2$, the S-matrix in a parity $P = (-1)^{J \pm 1/2}$ state
can
be parametrized in
terms of three eigen phases $\delta^J_{\lambda \Sigma}$, namely a
single doublet ($\Sigma = 1/2$) partial wave with $\lambda = J \pm
1/2$ and two quartet  ($\Sigma = 3/2$)  partial waves with $\lambda =
J \pm 1/2$ and $\lambda = J \mp 3/2$, and by the corresponding (three)
mixing angles, see Appendix \ref{Appendix:Smatrix}. The subleading contact 3NFs contribute to $S$-
($\lambda = 0$) and $P$- ($\lambda = 1$) waves in elastic Nd
scattering and to the mixing angles between different $P$-waves and between $S$-
and $D$-waves. 
Using the spectroscopic notation $^{2 \Sigma + 1}\lambda_{J}$ with $S, P, D, \ldots$ referring to $\lambda = 0, 1, 2, \ldots$,  the matrix elements of the short-range 3NFs contributing to elastic Nd scattering take the form
\beqa
\label{SpectroscopicTransitions}
\langle ^2S_{1/2}  | V| ^2S_{1/2}  \rangle &=& 24 \pi^2 \bigg[ {}- E + S_1 (Q_1^2 + Q_1^{\prime 2})  \nn
&+& \Big( S_2 + \frac{3}{2} S_3 \Big) (P_{23}^2 + P_{23}^{\prime \, 2}) \bigg], \nn
\langle ^4D_{1/2}  | V| ^2S_{1/2}  \rangle  &=& 24 \pi^2 S_3 \, Q_1^{\prime 2}, \nn
\langle ^2S_{1/2}  | V| ^4D_{1/2}  \rangle  &=& 24 \pi^2 S_3 \, Q_1^{2}, \nn
\langle ^2P_{1/2}  | V| ^2P_{1/2}  \rangle &=& 24 \pi^2 S_5 \, Q_1 Q_1^\prime, \nn
\langle ^4P_{1/2}  | V| ^4P_{1/2}  \rangle &=& 24 \pi^2 S_6 \, Q_1 Q_1^\prime, \nn
\langle ^4P_{1/2}  | V| ^2P_{1/2}  \rangle  &=& \langle ^2P_{1/2}  | V| ^4P_{1/2} \rangle  =  24 \pi^2 S_7 \, Q_1 Q_1^\prime, \nn
\langle ^2P_{3/2}  | V| ^2P_{3/2}  \rangle &=& 24 \pi^2 S_8 \, Q_1 Q_1^\prime, \nn
\langle ^4P_{3/2}  | V| ^4P_{3/2}  \rangle &=& 24 \pi^2 S_9 \, Q_1 Q_1^\prime, \nn
\langle ^4P_{3/2}  | V| ^2P_{3/2}  \rangle  &=& \langle ^2P_{3/2}  | V| ^4P_{3/2} \rangle  =  24 \pi^2 S_{10} \, Q_1 Q_1^\prime, \nn
\langle ^4P_{5/2}  | V| ^4P_{5/2}  \rangle &=& 24 \pi^2 S_{11} \, Q_1 Q_1^\prime, 
\eeqa
where
\begin{displaymath}
  \langle ^{2 \Sigma '+ 1}\lambda'_{J} | V | ^{2 \Sigma + 1}\lambda_{J} \rangle
  \equiv
  \langle ^{2 \Sigma '+ 1}\lambda'_{J} P_{23}' Q_1' | V| ^{2 \Sigma + 1}\lambda_{J} P_{23} Q_1 \rangle
\end{displaymath}
with $P_{23}$, $Q_1$,  $P_{23}'$ and $Q_1'$ denoting the corresponding
Jacobi momenta. 
In the above expressions, we have chosen the quantum numbers of the $(23)$-subsystem to be those of the deuteron, i.e., $l_{23}$,  $l_{23}'$ equal to $0$ or $2$, $s_{23} = s_{23}' = 1$ and $j_{23} = j_{23}' = 1$. 
Notice that while the $S_4$-term does not possess non-vanishing Nd
$\to$ Nd matrix elements, it still contributes to elastic Nd
scattering through non-vanishing Nd $\to$ breakup
transitions. Specifically, it contributes to the transitions to the
$|(l_{23}s_{23})j_{23} (\lambda_1 s_1) I_1\rangle $ three-nucleon
states $\big |(1 1)0 (1 \frac{1}{2}) \frac{1}{2} \big\rangle$,
$\big|(1 0)1 (1 \frac{1}{2}) \frac{1}{2} \big\rangle$, $\big|(1 0)1 (1
\frac{1}{2}) \frac{3}{2} \big\rangle$, $\big|(1 1)1 (1 \frac{1}{2})
\frac{3}{2} \big\rangle$ and $\big|(1 1)2 (1 \frac{1}{2}) \frac{3}{2}
\big\rangle$ with $J^P = \frac{1}{2}^+$ and isospin
$T=\frac{1}{2}$. We, therefore, expect Nd elastic scattering
observables to be less sensitive to the LEC $S_4$ as compared with the
other LECs in the isospin $T=1/2$ channel. This is also
consistent with the significantly smaller-in-magnitude contribution of
$s_4$ to the triton binding energy as compared with those of the LECs $s_1$, $s_2$ and $s_3$, see Table \ref{Table1}. 
Furthermore, from the expressions in
Eq.~(\ref{SpectroscopicTransitions}), we expect the LEC $S_2$ to be
largely redundant when calculating elastic Nd scattering
observables. Indeed, in the Born approximation, its contribution to
the scattering amplitude can be obtained by performing the integration
over the momenta $P_{23}$ and $P_{23}'$ in the deuteron wave functions
and has the same form as the one of the $E$-term at N$^2$LO. Our results
for the LECs $S_{1, \ldots, 3}$ and $S_{5, \ldots , 11}$ in
Eq.~(\ref{SpectroLECs}), as well as
for the two linear combinations of purely isospin-$T=3/2$ operators, are
consistent with the ones reported in
Ref.~\cite{Filandri:2026ori}.

The spectroscopic basis for the subleading short-range 3NFs offers
obvious advantages compared to the original basis. It provides
a clear and transparent relationship between 3N observables and
contact 3N operators
contributing to specific partial waves, and the LECs $S_i$ are
expected to be less correlated than $E_i$'s. In addition, it
simplifies the estimation of a natural range for the LECs. Indeed, in
the two-nucleon sector of chiral EFT, all spectroscopic LECs entering
the SMS potentials of Ref.~\cite{Reinert:2017usi} turn out to be of
natural size (except for the largest considered cutoff value of
$\Lambda = 550$~MeV), see also Ref.~\cite{Epelbaum:2019kcf} for a
discussion.
Based on the experience in the NN sector \cite{Reinert:2017usi,Epelbaum:2019kcf} and taking into account results
shown in Table \ref{Table1}, we expect the natural range of the N$^4$LO
contact terms to roughly correspond to $| s_i | \lesssim 2$. 
Notice that a spherically symmetric prior probability
distribution in the space of the LECs $s_i$ would result in a non-spherically
symmetric distribution for $e_i$'s, which makes it more difficult to
estimate their natural range. Last but not least, we emphasize that the linear relationship
between the sets $\{E_i\}$ and $\{S_i\}$ is a one-to-one
correspondence, see Appendix \ref{Appendix:InverseSpectroBasis} for
the inverse basis transformation.

\section{Applications to nucleon-deuteron scattering}
\label{sec:Applications}

We are now in the position to explore the impact of the N$^4$LO
short-range 3NFs on selected three-nucleon scattering observables. To
this aim, we solve the discretized form of the momentum-space Faddeev
integral equation for the Nd scattering operator in the partial-wave
basis as explained in detail in Ref.~\cite{Gloeckle:1995jg}. Following the standard
calculation setup as used, e.g., in Refs.~\cite{LENPIC:2018ewt,Epelbaum:2019zqc,Maris:2020qne,LENPIC:2022cyu}, we take into
account the NN interaction with a total angular momentum $j \leq 5$,
whereas in the three-nucleon system we include all channels with $J
\leq 25/2$. 3NF contributions are taken into account in the partial
waves with $J \leq 7/2$.

\begin{table}[!tp]
  \vspace{0.2cm}
  \begin{ruledtabular}
	\begin{tabular*}{0.49\textwidth}{@{\extracolsep{\fill}}c|cccc}
        $E_N$ (MeV) & $d\sigma/d\Omega$ & $A_y$ & $A_{ij}$ & $C_{ij}$ \\[2pt]
          \hline
          &&&& \\[-7pt]
        10 & \cite{Sperisen:1984bjw}\phantom{$^{*}$} & \cite{Sperisen:1984bjw}$^{*}$ & \cite{Sperisen:1984bjw} & $-$ \\ 
        70 & \cite{Sekiguchi:2002sf}$^{*}$ & \cite{Sekiguchi:2002sf,SUDA2007745}$^{*}$ & \cite{Sekiguchi:2002sf,SUDA2007745}$^{*}$ & $-$ \\
        135 & \cite{Sekiguchi:2002sf,Sekiguchi:2005vq}$^{*}$, \cite{Ermisch:2005kf} & \cite{Sekiguchi:2002sf,SUDA2007745,vonPrzewoski:2003ig}$^{*}$ & \cite{Sekiguchi:2002sf,SUDA2007745,vonPrzewoski:2003ig}$^{*}$ & \cite{vonPrzewoski:2003ig}  \\
        200 & $-$ & \cite{vonPrzewoski:2003ig} & \cite{vonPrzewoski:2003ig} & \cite{vonPrzewoski:2003ig}  \\          
        \end{tabular*}
        \vskip 0.1 true cm
        {$^*$ Data included in the fits described in Sec. \ref{sec:Fits}. \hfill}
        \caption{References to the experimental data on elastic proton-deuteron scattering used in this study. Experimental data at $10$ MeV have been corrected for Coulomb effects as described in Ref.~\cite{Epelbaum:2002vt}. In the plots below, the data are color coded as follows: solid dark blue triangles for \cite{Sperisen:1984bjw}, dark orange crosses for \cite{Sekiguchi:2002sf}, open purple circles for \cite{SUDA2007745}, open light blue squares for \cite{Sekiguchi:2005vq}, solid yellow circles for \cite{Ermisch:2005kf} and open green inverted triangles for \cite{vonPrzewoski:2003ig}.
     }
    \label{tab:exp_database_sources}  
   \end{ruledtabular}
 \end{table}
 
Throughout this study, we use the experimental data for Nd elastic
scattering observables collected  in Table \ref{tab:exp_database_sources}. The
total experimental uncertainty  is obtained by adding the
statistical and systematic errors in squares, and we also take into
account the angular resolution.

\subsection{Sensitivity to the individual LECs}
\label{sec:Sensitivity}

To assess the impact of the subleading short-range 3NFs, 
it is instructive to first quantify the sensitivity of Nd scattering observables
to individual $s_i$'s.  In Ref.~\cite{Witala:2022rzl}, such a sensitivity analysis
was carried out using the LECs $E_i$ in the original operator basis. As
explained in the previous section, employing the spectroscopic
basis allows one to separate effects of individual partial
waves on the considered scattering observables. In addition, one 
benefits from a more reliable estimation of the expected natural size
of the 3NF contributions. 

\begin{figure}[t]
	\centering
	\includegraphics[width=0.975\linewidth]{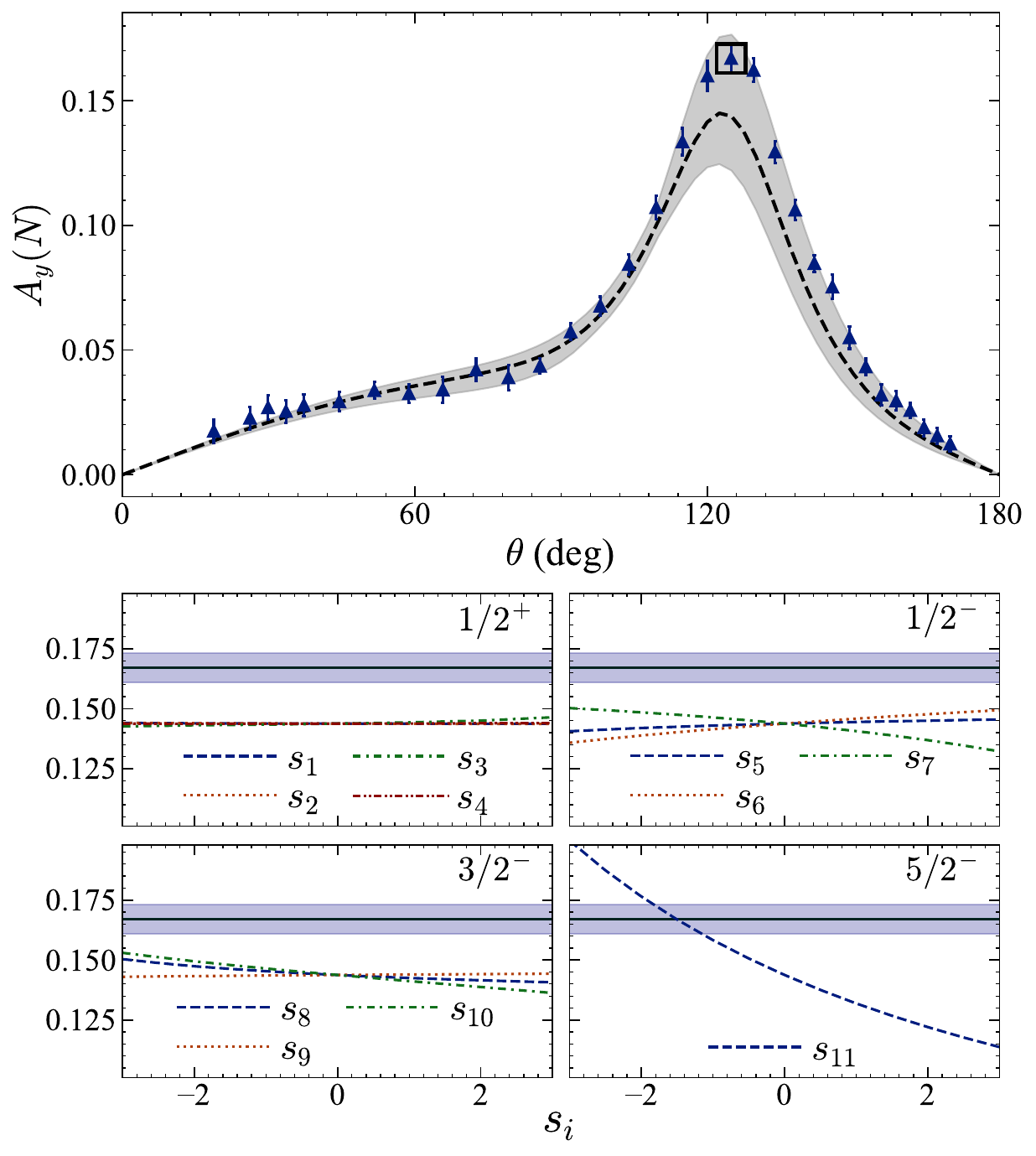}
    \vspace{-0.3cm}
    \caption{Sensitivity of the analyzing power $A_y(N)$ at $10$~MeV to the LECs $s_i$
      of the subleading contact 3NF. In the upper panel, dashed
      line shows $A_y (N)$ as a function of the center-of-mass
      scattering angle $\theta$, calculated using the SMS N$^4$LO$^+$
      NN potential of Ref.~\cite{Reinert:2017usi} in combination with
      the parameter-free two-pion exchange 3NF at N$^2$LO
      \cite{Maris:2020qne}, while the light-shaded gray band visualizes the
      maximal effect caused by a variation of a single $s_i$ in the
      range of $|s_i | \leq 2$. References to the experimental data and their respective color coding 
      are given in Table \ref{tab:exp_database_sources}. The empty
      square in the upper plot marks the reference datum used to visualize the impact of
      the LECs $s_i$ in the $J^P = 1/2^+$,  $1/2^-$,  $3/2^-$, and
      $5/2^-$ channels
      in the four subplots of the lower panel.
      The experimental value of $A_y (N)$ is shown by black solid lines, with light-shaded violet bands corresponding to its uncertainty.
      }
	\label{fig:sensitivity_Ay(N)_10}
     \vspace{-0.2cm}
\end{figure}

\begin{figure}[t]
	\centering
	\includegraphics[width=0.975\linewidth]{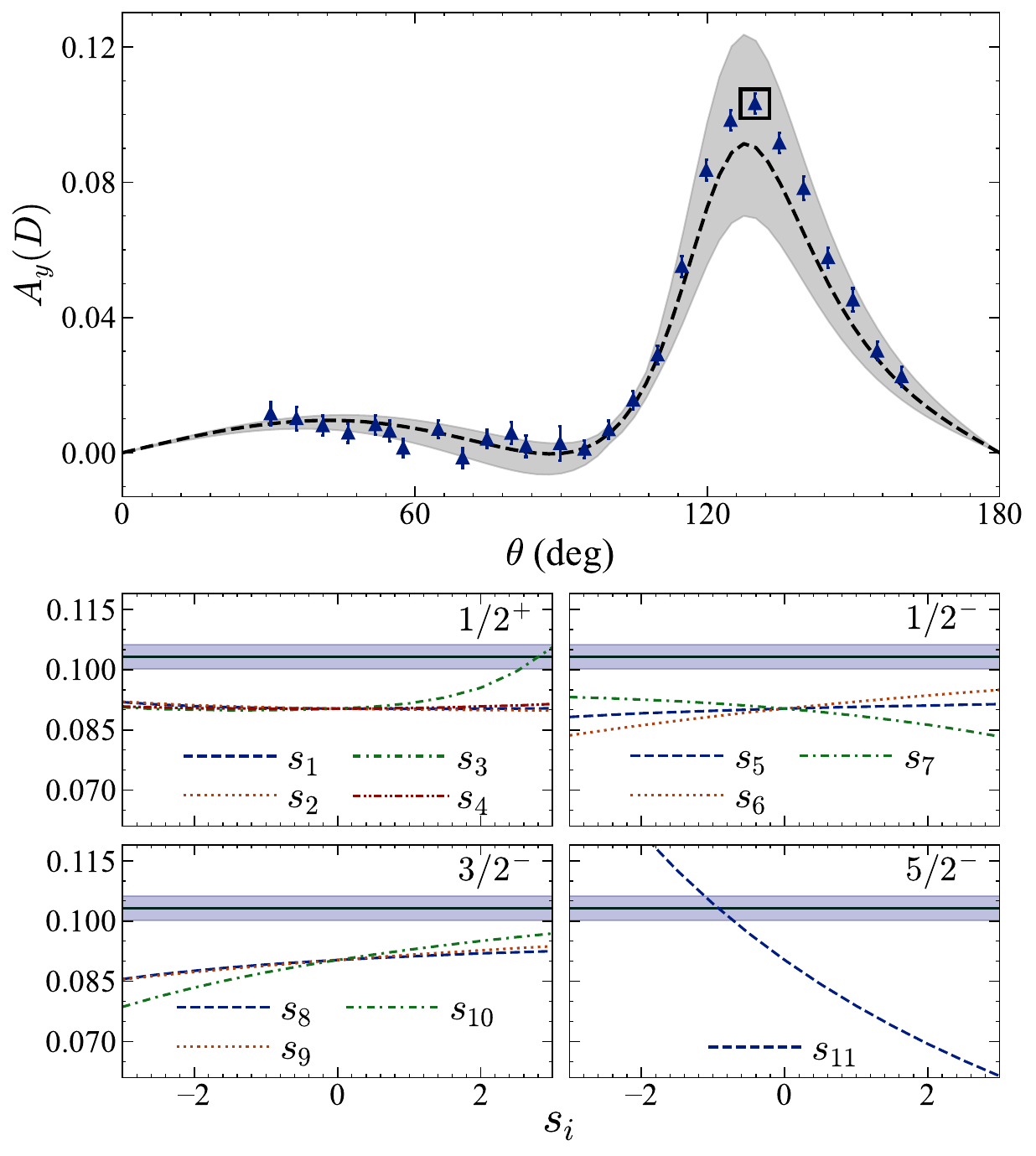}
    \vspace{-0.3cm}
	\caption{Same as Fig.~\ref{fig:sensitivity_Ay(N)_10}, but for $A_y (D)$ at 10 MeV.
    }
	\label{fig:sensitivity_Ay(D)_10}
     \vspace{0.1cm}
   \end{figure}

\begin{figure}[t]
	\centering
	\includegraphics[width=0.98\linewidth]{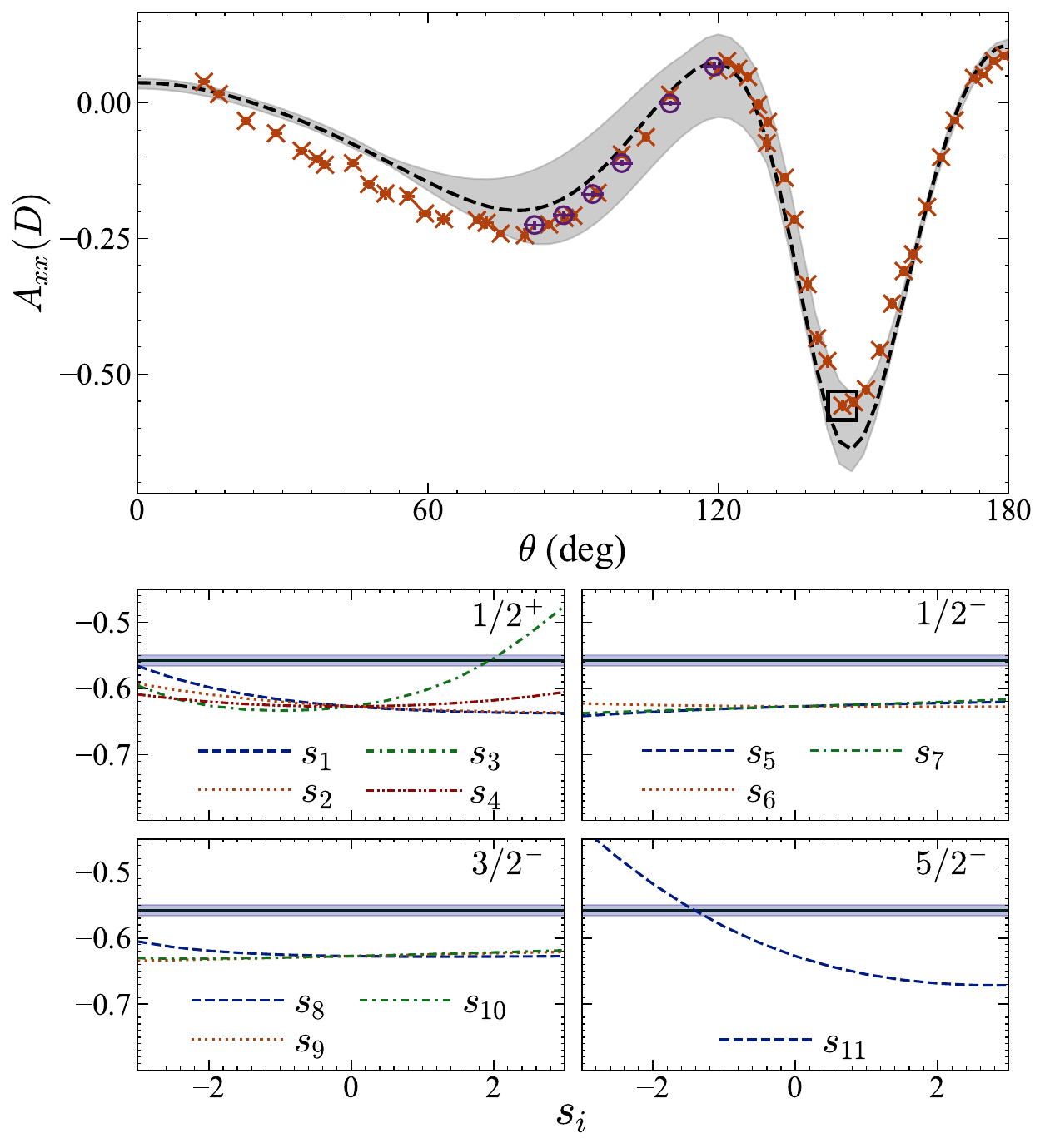}
    \vspace{-0.3cm}
	\caption{Same as Fig.~\ref{fig:sensitivity_Ay(N)_10}, but for $A_{xx}$ at 70 MeV.
    }
    \label{fig:sensitivity_Axx_70}
        \vspace{0.1cm}
   \end{figure}

\begin{figure}[t]
	\centering
	\includegraphics[width=0.96\linewidth]{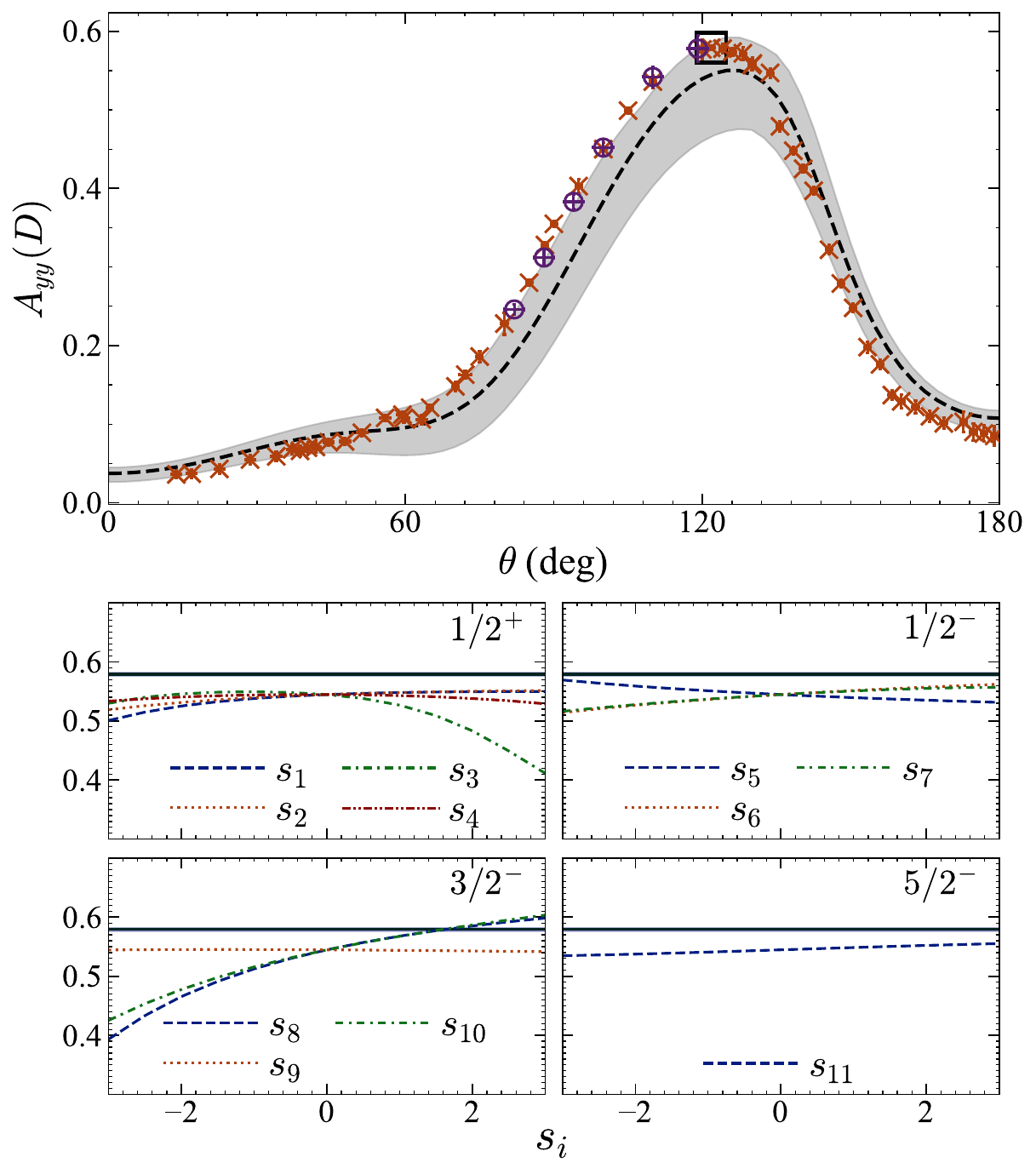}
    \vspace{-0.3cm}
	\caption{Same as Fig.~\ref{fig:sensitivity_Ay(N)_10}, but for $A_{yy}$ at 70 MeV.
    }
	\label{fig:sensitivity_Ayy_70}
     \vspace{0.1cm}
   \end{figure}

\begin{figure}[t]
	\centering
	\includegraphics[width=0.96\linewidth]{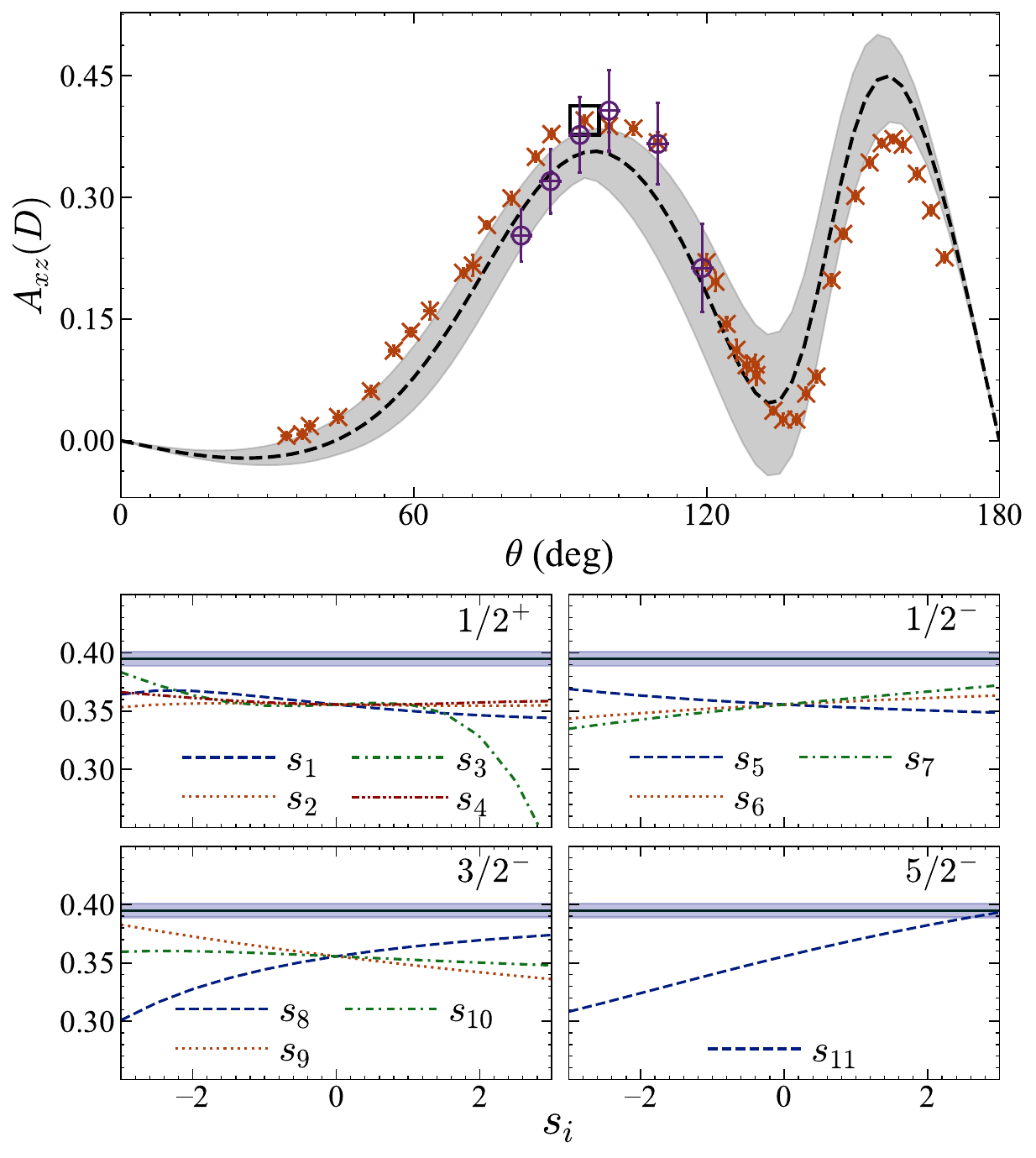}
    \vspace{-0.3cm}
	\caption{Same as Fig.~\ref{fig:sensitivity_Ay(N)_10}, but for $A_{xz}$ at 70 MeV.
    }
    \label{fig:sensitivity_Axz_70}
        \vspace{0.1cm}
   \end{figure}

\begin{figure}[t]
	\centering
	\includegraphics[width=0.978\linewidth]{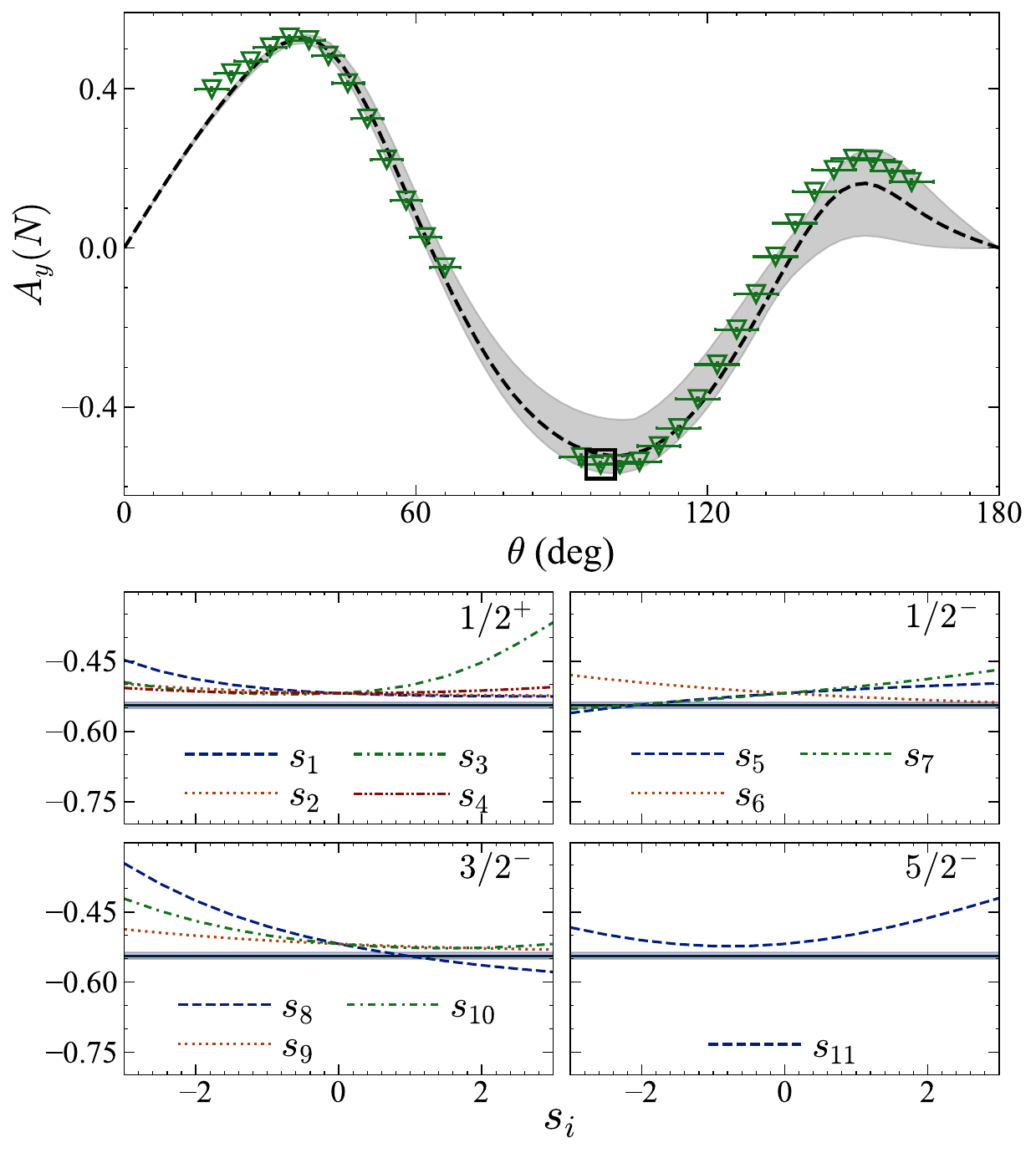}
    \vspace{-0.3cm}
	\caption{Same as Fig.~\ref{fig:sensitivity_Ay(N)_10}, but for $A_{y} (N)$ at 135 MeV.
    }
	\label{fig:sensitivity_Ay(N)_135}
     \vspace{0.1cm}
   \end{figure}

\begin{figure}[t]
	\centering
	\includegraphics[width=0.978\linewidth]{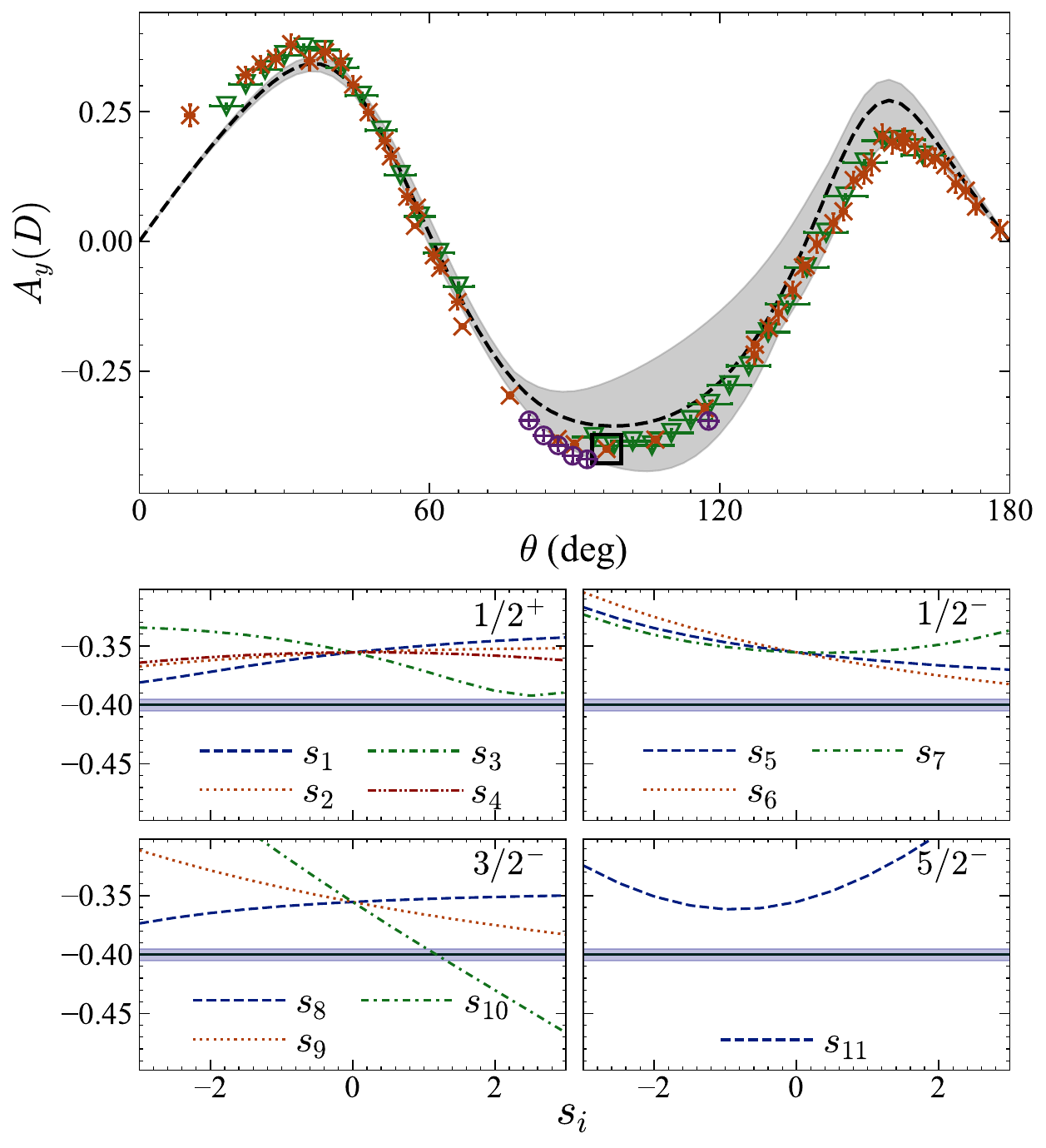}
    \vspace{-0.3cm}
	\caption{Same as Fig.~\ref{fig:sensitivity_Ay(N)_10}, but for $A_{y} (D)$ at 135 MeV.
    }
    \label{fig:sensitivity_Ay(D)_135}
        \vspace{0.1cm}
   \end{figure}

\begin{figure}[t]
	\centering
	\includegraphics[width=0.98\linewidth]{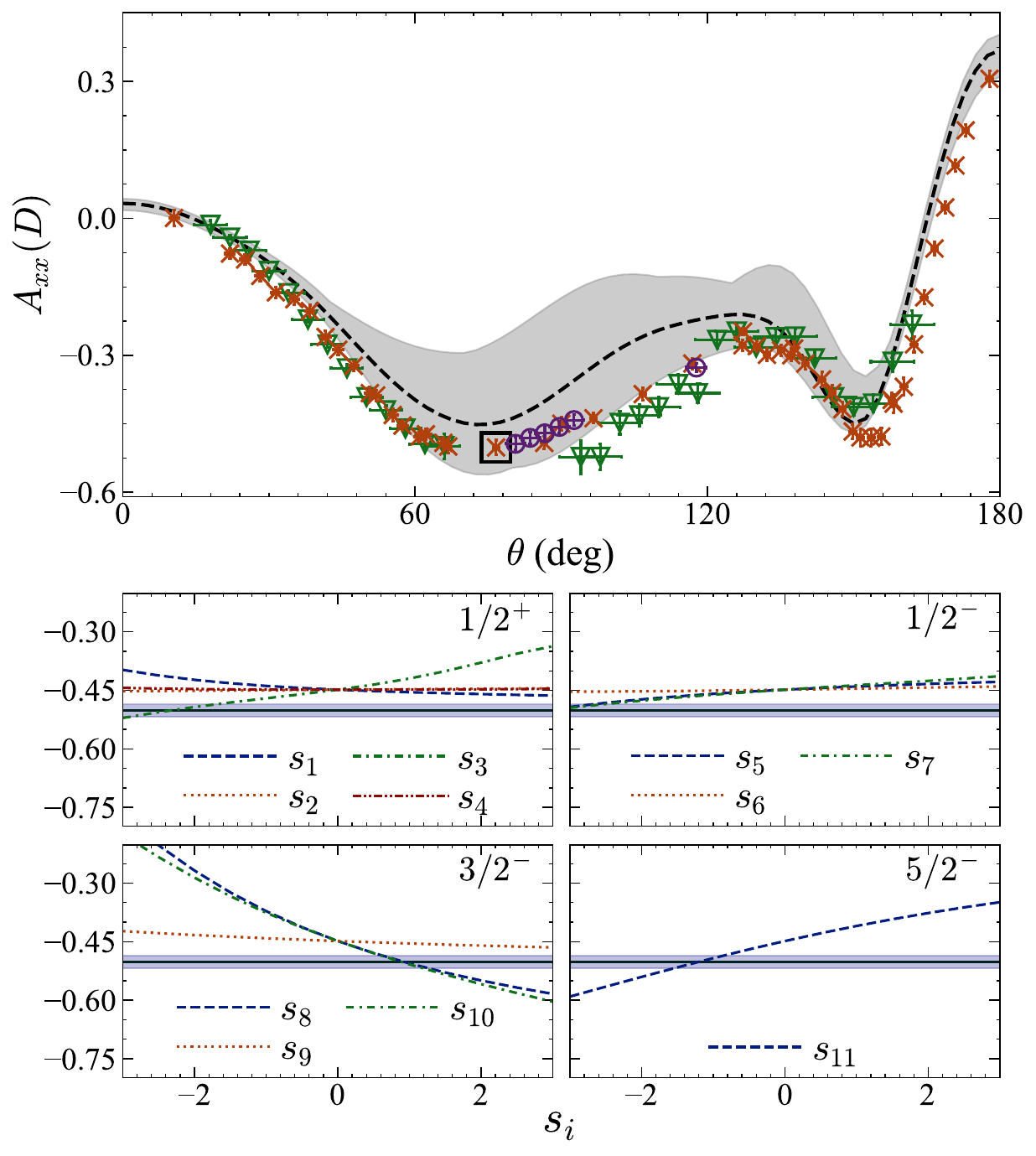}
    \vspace{-0.3cm}
	\caption{Same as Fig.~\ref{fig:sensitivity_Ay(N)_10}, but for $A_{xx}$ at 135 MeV.
    }
	\label{fig:sensitivity_Axx(D)_135}
     \vspace{0.1cm}
   \end{figure}

Given that Nd scattering is restricted to
isospin $T=1/2$, we discard in the following the LECs $s_{12}$ and $s_{13}$, which
contribute solely to the $T=3/2$ channel. 
In Figs.~\ref{fig:sensitivity_Ay(N)_10}--\ref{fig:sensitivity_Axx(D)_135}, we show the sensitivity of the
selected scattering observables to the individual variations of $s_{1,
  \ldots , 11}$.  

At very low energies, Nd scattering
observables are known to be largely insensitive to 3NFs except
for the nucleon and deuteron vector analyzing powers $A_y(N)$ and
$A_y(D)$, respectively \cite{Gloeckle:1995jg}. The experimental data for $A_y$ at
low-energies are significantly underpredicted by calculations using
high-precision two-nucleon potentials and phenomenological 3NF models, which is often
referred to as the $A_y$-puzzle
\cite{Gloeckle:1995jg,Kievsky:2013yva}.  Thus,  at the lowest
considered energy of $10$~MeV, we only show the
results for $A_y(N)$ and $A_y(D)$. At
higher energies of $70$ and $135$~MeV, we study the impact of the
subleading short-range 3NFs also for tensor analyzing powers,
for which experimental data are available. 

In each figure, the plot in the upper panel shows with the dashed line
the predictions based on the N$^4$LO$^+$ SMS NN potential \cite{Reinert:2017usi},
in combination with the consistently regularized (parameter-free)
long-range part of the 3NF at N$^2$LO, i.e.~we use the expression for
the 3NF from Ref.~\cite{Maris:2020qne} with $c_D = c_E =
0$. Light-shaded bands in the upper plots 
depict the maximal achievable effect due to a variation of individual N$^4$LO 3NF
terms in the range of $s_i \in [-2, 2]$. The four plots in the
lower panels of Figs.~\ref{fig:sensitivity_Ay(N)_10}--\ref{fig:sensitivity_Axx(D)_135} show the individual
impact of the LECs contributing to the $J^P=1/2^+$, $1/2^-$,   $3/2^-$, 
and $5/2^-$ channels.
Since the 3NF contributions to the considered observables at different
values of the scattering angle $\theta $ are highly correlated, we restrict
ourselves in these plots to a fixed scattering angle (the
corresponding experimental datum is marked with an open square in the
upper plots).

Multiple interesting conclusions can be drawn from the sensitivity
plots. 
First of all, one observes that certain $s_i$'s
have a potential to resolve the $A_y$ puzzle at $10$~MeV. It
is known that the vector analyzing powers at low energies are sensitive
to quartet
$P$-wave phase shifts \cite{Ishikawa:1999rg,Ishikawa:2002ti}. Our
results demonstrate that $A_y (N)$ and $A_y (D)$ are largely
insensitive to the $^2P_{1/2}$ partial wave governed by $s_5$, while  
showing some sensitivity to
the LECs $s_6$ and  $s_8$, which contribute to the  $^4P_{1/2}$ and 
$^2P_{3/2}$ partial waves, as well as to the mixing
of the doublet and quartet $P$-waves in the $J^P=1/2^-$ and $3/2^-$
channels driven by the LECs $s_7$ and $s_{10}$, respectively. 
However, none of the individual variations of these LECs alone within the
natural range is capable of resolving the $A_y$-puzzle. On the other hand, both
analyzing powers are highly sensitive to the $^4P_{5/2}$
phase shift governed by the LEC $s_{11}$, and the discrepancies between
theory and data can be removed by choosing $s_{11} \sim - 1.5\ldots
{-1}$. This strong sensitivity of $A_y$ to the  $^4P_{5/2}$ phase shift
is consistent with the findings of Ref.~\cite{Ishikawa:1999rg}:
Indeed, the approximate
relationship between the analyzing power and the quartet $P$-wave
phase shifts, derived in that paper, shows a large numerical
coefficient in front of the phase shift $\delta_{^4P_{3/2}}$. Interestingly and remarkably, the $P$-wave LECs in the $3/2^-$
channel have an opposite impact 
on $A_y (N)$ and $A_y (D)$, thus providing a mechanism
for a simultaneous reproduction of these observables. Notice
further that the large effect of $s_3 \gtrsim 2$ on $A_y (D)$ should be
taken with care given $|\langle V_{s_3 = 1}\rangle_{^3\rm H} |\sim
600$~keV, see Table \ref{Table1}, which suggests that the values $s_3
\gtrsim 2$ likely lead to unnaturally large 3NFs beyond the validity
range of the employed chiral EFT formulation
\cite{Epelbaum:2019kcf}. This conclusion is in
line with the strongly nonlinear behavior observed for most of the
considered observables for large positive values of $s_3$.

At the intermediate energy of $70$~MeV, 3NF effects become more
prominent, and the sensitivity plots show a more complex
pattern. Negative values of the LEC $s_{11}$ that were shown to
be capable of resolving
the $A_y$-puzzle at low energies also lead to an improved description
of $A_{xx}$, while worsening the reproduction of $A_{xz}$ and having
almost no impact on  $A_{yy}$. On the other hand, we observe significant
effects on $A_{yy}$ and $A_{xz}$ from the LECs in the $J^P =
3/2^-$ channel, while $A_{xx}$ appears to be sensitive to the LECs
$s_1$ and $s_3$ in the $J^P = 1/2^+$ channel. At  the
highest considered energy of $135$~MeV, the effects of the 3NFs
further increase in importance, but  the situation remains  
qualitatively similar to that at $70$~MeV. 

To conclude, our results show that the Nd analyzing powers are
strongly affected by the phase shifts and mixing angles in the $J^P=5/2^-$ and
$J^P=3/2^-$ channels, while we observe less sensitivity to the LECs contributing to the
$J^P=1/2^+$ (apart from large positive values of $s_3$) and
$J^P=1/2^-$ channels. In line with the arguments put forward in sec.~\ref{sec:SpecBasis}, we
see almost no effects from  the LEC $s_4$. 

\begin{figure}[t]
   \vspace{-0.1cm}
	\centering
	\includegraphics[width=0.94\linewidth]{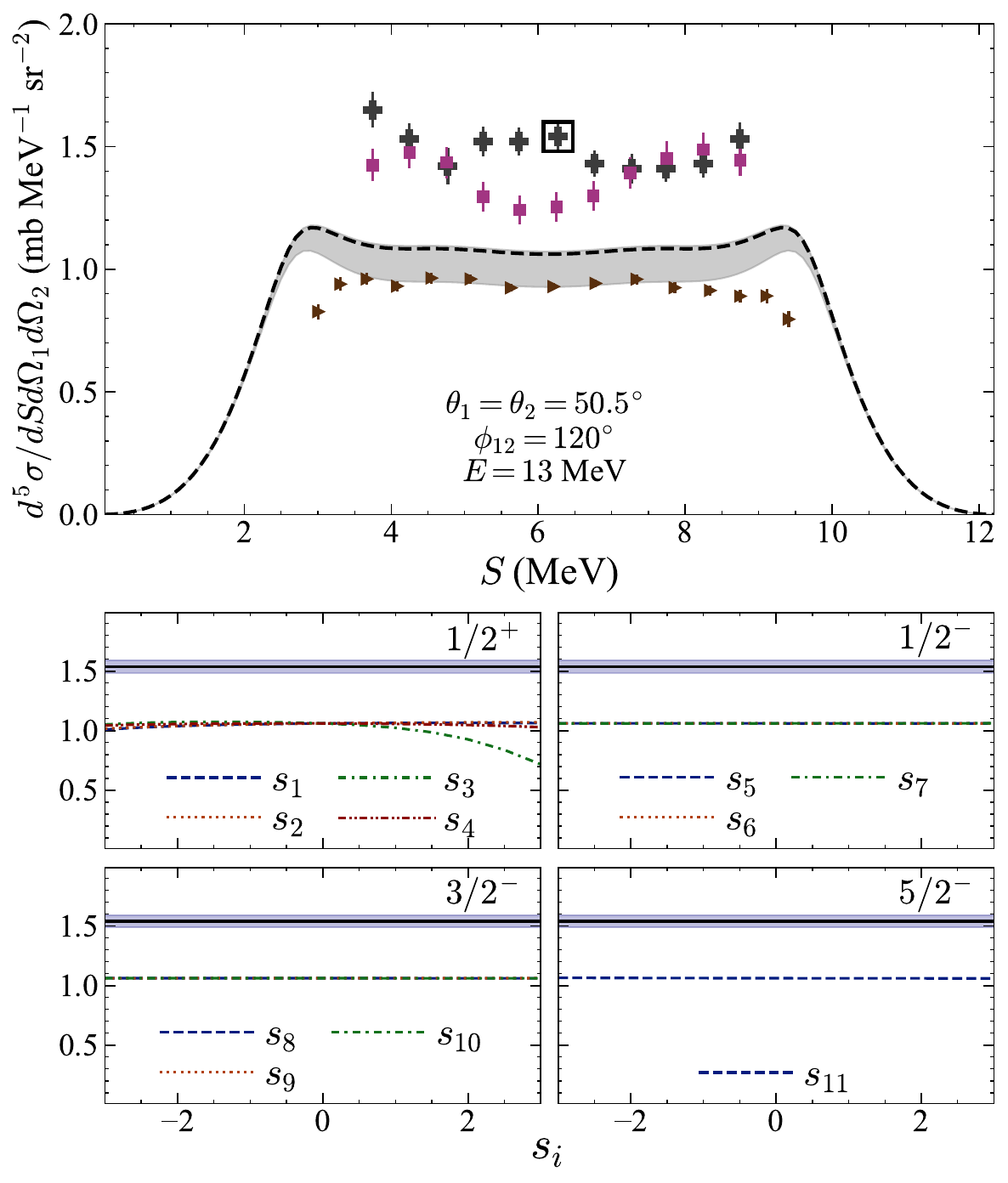}
    \vspace{-0.3cm}
	\caption{Same as Fig.~\ref{fig:sensitivity_Ay(N)_10}, but for
          the differential cross section of the symmetric space-star
          configuration in the Nd breakup reaction at 13 MeV, plotted
          against the kinematical locus variable $S$ (see Ref.~\cite{Gloeckle:1995jg}
          for the definition). The nd breakup reaction data from \cite{Strate:1989lle} and \cite{Setze:2005jk} are denoted by pink squares and black crosses, respectively, while the pd data from \cite{Rauprich:1991cmg} are indicated by brown triangles.
          }
    \label{fig:sensitivity_SST}
        \vspace{0.1cm}
   \end{figure}

While the main focus of this study is on elastic Nd scattering, we
also 
briefly comment on another well-known puzzle in the three-nucleon
continuum, which is related to the symmetric space-star (SST) breakup
configuration \cite{Gloeckle:1995jg,Kalantar-Nayestanaki:2011rzs}.  This configuration is characterized by the
outgoing nucleons in the
center-of-mass system  having the same magnitudes of momenta and forming a ``mercedes-star'' in the plane perpendicular to
the beam direction. Several sets of cross-section data for both
neutron- (nd) and proton-deuteron (pd) scattering in the SST configuration are
available at low energies between $10$ and $25$~MeV. Here, we consider
the SST breakup configuration at $E_N = 13$~MeV, for which the nd \cite{Strate:1989lle,Setze:2005jk} and
pd \cite{Rauprich:1991cmg} data differ from each other by a factor of $\sim 1.5$, with
theoretical predictions lying between the nd and pd data sets \cite{Gloeckle:1995jg,Kalantar-Nayestanaki:2011rzs,Witala:2021zmb}, see Fig.~\ref{fig:sensitivity_SST}.  
Theoretical predictions for the SST cross section are known to be very
robust, and the 3NF models considered so far were found to have almost no impact on
this observable at low energies, see Ref.~\cite{Witala:2021zmb} and
references therein. The significant discrepancies between theory and
data and also between the pd 
and nd data sets, given the smallness of the Coulomb
effect \cite{Deltuva:2005cc}, constitute an unsolved puzzle. Our
results in Fig.~\ref{fig:sensitivity_SST} show that the 
puzzles related to the SST breakup configuration can unlikely be resolved by the N$^4$LO contact 3NF in Eq.~(\ref{DefEi}). We emphasize, however, that the cross section may still be affected by the N$^3$LO and N$^4$LO contributions to the 3NF not included in this study. Notice further that
contrary to a naive expectation, the LECs $s_4$ also leaves this
breakup observable unaffected.

Before closing this section, it is important to address the
limitations of the presented analysis. While the sensitivity of the
specific observables to $s_i$'s are unlikely to be affected by the
inclusion of long-range 3NF components beyond N$^2$LO, their
contributions needed to describe specific scattering observables will certainly change upon performing a more complete
analysis. In addition, our considerations so far are limited to effects
caused by single LECs $s_i$ and do not take into account quadratic
and higher-order contributions. Below, we present a more complete and
conclusive analysis by performing exploratory fits of the short-range
3NF components to elastic Nd scattering observables.

\subsection{Building an emulator for elastic Nd scattering}
\label{sec:Emulator}

To determine the LECs $c_D$, $c_E$ and $s_{1, \ldots , 11}$ from Nd observables, it
is necessary to speed up the process of repeatedly solving the Faddeev equation using
an appropriate emulator. Different approaches for emulating
three-nucleon scattering observables have been considered in the
literature including a perturbation-theory-based method of
Refs.~\cite{Witala:2021ufh,Witala:2021xqm},
eigenvector continuation emulators of
Refs.~\cite{Zhang:2021jmi,Gnech:2025lbg} and the
Woodbury-identity-based reduced-basis method of Ref.~\cite{Heihoff:2026ycq}.  
In this study, we follow a different path and employ a simple
emulator by approximating the complex-valued Nd transfer matrix $U$ in Eq.~(\ref{UMatrix}) at a fixed energy, viewed as
a function of the variable LECs, by utilizing 
a radial basis function (RBF)
interpolation method
\cite{hardy1971multiquadric,kansa1990multiquadrics2}. 
The key feature
that makes the RBF-based emulator feasible in our case is the usage of
the spectroscopic basis for the subleading short-range 3NFs, which
strongly reduces the dimensionality of the parameter space as
shown in Table \ref{tab:lecs_per_pw}. 
\begin{table}[!tp]
  \vspace{0.2cm}
  \begin{ruledtabular}
	\begin{tabular*}{0.49\textwidth}{@{\extracolsep{\fill}}cc}
 $J^P$ channel & contributing 3NF LECs \\[2pt]
          \hline
          & \\[-7pt]
       $1/2^+$ & $c_E,\,c_D,\,s_1,\,s_2,\,s_3,\,s_4$ \\
        $1/2^-$ & $c_D,\,s_5,\,s_6,\,s_7$ \\
        $3/2^-$ & $c_D,\,s_8,\,s_9,\,s_{10}$ \\
        $5/2^-$ & $c_D,\,s_{11}$ \\
        $3/2^+$, $5/2^+$, $7/2^\pm$ & $c_D$ \\
        \end{tabular*}
        \caption{Dependence of the S-matrix elements for elastic Nd scattering in different
          $J^P$ channels on the LECs $c_D$, $c_E$ and $s_{1, \ldots ,
            11}$. 3NF matrix elements are taken into account in all
          channels with $J \leq 7/2$.} \label{tab:lecs_per_pw}              
   \end{ruledtabular}
 \end{table}

To fit the LECs entering the 3NF, we restrict ourselves to the
energies of $10$, $70$ and $135$~MeV and build a simple
interpolation-based emulator, 
where each channel is treated independently for each
energy. For one-dimensional interpolations in the $J^P = 3/2^+$, $5/2^+$
and $7/2^\pm$ channels, we use an evenly spaced grid of $30$ points in
the interval $c_D \in [-8.0, 8.0]$, for which the matrix $U^J_{\lambda '
  \Sigma', \, \lambda \Sigma}$ is calculated by solving the Faddeev
equation at each of the considered energies.

In the 
$J^P= 1/2^-$, $3/2^-$ and $5/2^-$ channels, one needs a larger number
of grid points to account for the increased dimensionality of the
parameter space. We found that $\sim 10^3$ exact calculations per
energy are sufficient at the accuracy level of our analysis.
The sample points in
the space of $s_i$'s used to build the interpolation database, i.e.~the set of exact solutions, 
are
determined via Latin hypercube sampling (LHS) \cite{mckay1979comparison}, ensuring a complete
coverage of the desired range of values for these LECs. Specifically,
we expect the actual values of the LECs $s_i$ to lie within the range
$s_i \in [-3.0, 3.0]$. However, we found that the preferred values for
$s_6$ and $s_9$ lie outside of this range and have, therefore, extended their
sampled values to $s_6 \in [-8.0, 8.0]$ and $s_9 \in [-6.0, 6.0]$.  
For $c_D$, we keep the same interval $c_D \in [-8.0, 8.0]$.

\begin{table*}[t!]
  \vspace{0.2cm}
  \setlength{\tabcolsep}{7pt}
 \begin{ruledtabular}  
   \begin{tabular*}{\textwidth}{@{\extracolsep{\fill}}cc|rrrr}
     \multicolumn{2}{c|}{} & \multicolumn{4}{c}{Channels of the contact N$^4$LO 3NF included in the fit} \\[2pt]
     $J^P$ & LEC & $5/2^{-}$ & $5/2^{-},\,\,3/2^{-}$ & $5/2^{-},\,\,3/2^{-},\,\,1/2^{-}$ & $5/2^{-},\,\,3/2^{-},\,\,1/2^{-},\,\,1/2^{+}$ \\[2pt]
     \hline
     &&&&&\\[-7pt]
    & $c_D$     & $-4.098 \pm 0.314$ & $-2.533 \pm 0.331$ & $-3.384 \pm 0.328$ & $-2.427\pm 0.386$ \\[2pt]
     \hline
     &&&&&\\[-7pt]
                                                & $c_E$    & $0.297 \pm 0.039$ & $0.095 \pm 0.044$ & $0.206 \pm 0.042$ & $0.319 \pm 0.054$ \\[2pt]
     $1/2^{+}$  &       $s_1$    & {$\,\,\,\;-$}   & {$\,\,\,\;-$}   & {$\,\,\,\;-$}   & $-0.033\pm 0.377$ \\[2pt]
                                                &                            $s_3$    & {$\,\,\,\;-$}   & {$\,\,\,\;-$}   & {$\,\,\,\;-$}   & $-1.684\pm 0.239$ \\[2pt]
     \hline
     &&&&&\\[-7pt]    
 & $s_5$    & {$\,\,\,\;-$}   & {$\,\,\,\;-$}   & $0.647 \pm 0.406$  & $1.712\pm 0.614$ \\   [2pt]
     $1/2^{-}$       &                            $s_6$    & {$\,\,\,\;-$}   & {$\,\,\,\;-$}   & $6.222 \pm 0.466$  & $5.978\pm 0.445$ \\  [2pt]
        &                            $s_7$    & {$\,\,\,\;-$}   & {$\,\,\,\;-$}   & $-0.273 \pm 0.325$ & $-0.658\pm 0.475$ \\  [2pt]
    \hline
                           &&&&&\\[-7pt]
 & $s_8$    & {$\,\,\,\;-$}   & $-0.447 \pm 0.128$ & $0.029 \pm 0.140$ & $-0.220\pm 0.132$ \\ [2pt]
           $3/2^{-}$                    &                            $s_9$    & {$\,\,\,\;-$}   & $-3.295 \pm 0.170$ & $-3.672 \pm 0.140$ & $-3.818\pm 0.135$ \\ [2pt]
                           &                            $s_{10}$ & {$\,\,\,\;-$}   & $1.311 \pm 0.104$  & $0.825 \pm 0.098$  & $1.125\pm 0.112$ \\ [2pt]
     \hline
    &&&&&\\[-7pt]        
     $5/2^{-}$                  & $s_{11}$ & $-0.768 \pm 0.058$ & $-0.331 \pm 0.068$ & $-0.145 \pm 0.065$ & $-0.079\pm 0.069$ \\  [2pt]
   \hline
                           &&&&&\\[-7pt]
     \multicolumn{2}{c|}{$\chi^2$/datum} &   $12.30$ & $9.74$ & $5.37$ & $4.91$ \\  [2pt]
   \end{tabular*}
       \caption{Values of LECs and statistical errors obtained in fits
         involving different numbers of LECs $s_i$.
         The statistical
         errors were calculated by taking the square root of the
         diagonal elements of the covariance matrix, while the
         systematic errors have not been estimated. The values of the
         $\chi^2$ are to be compared with the quality of the
         description of the considered observables with all adjustable
         parameters set to
         zero, leading to  $\chi^2/{\rm datum}=20.09$.
         } 
       \label{tab:fit_results}
   \end{ruledtabular}      
 \end{table*}

Finally, in the $J^P=1/2^+$ channel, we follow the usual approach
by using the $^3$H binding energy of $8.482$~MeV as a constraint to fix the value of
the LEC $c_E$. To this aim, we employ a grid of $10^4$ points using
LHS for $s_{1, \ldots ,4} \in [-3.0, 3.0]$ and equally-spaced points for
$c_D \in [-8.0, 8.0]$.   For each of these points in the ${s_{1}, s_{2}, s_{3}, s_{4},
  c_D}$-space, the LECs $c_E$ is adjusted to reproduce the triton
binding energy. Finally, the interpolation database for the transfer matrix $U$ is
obtained by solving the Faddeev equation using the above-mentioned
grid of the LECs $s_i$ and $c_D$, along with the constrained values of the
LEC $c_E$.

To verify that the number of grid points is sufficient for our
purposes, we evaluate the accuracy of the emulated solutions at all
considered scattering energies in appendix~\ref{Appendix:AccuracyEmulator}.

\subsection{Exploratory fits}
\label{sec:Fits}

Having a fully functional tool to calculate Nd elastic scattering
observables for arbitrary values of $c_D$ and $s_i$'s within the interpolation domain
at low computational cost, it is straightforward to extract these
parameters from experimental data. We fix the LECs via a least squares $\chi^2$
minimization based on the Trust Region Reflective method \cite{2020SciPy-NMeth},  which allows one to search for
local minima in a bounded region of the parameter space.
Regarding the treatment of experimental uncertainties for an
observable ${O}$, we include
statistical, systematic and angular errors $\delta{O}_\text{stat}$, $\delta{O}_\text{sys}$
and $\delta\theta_i$, respectively. To account for the angular resolution, we employ the
effective variance method \cite{orear1982least} to propagate the
uncertainty $\delta\theta_i$ into an additional uncertainty $\delta
O$. The relationship between the angular resolution in the laboratory
and center-of-mass frames is discussed in Appendix
\ref{Appendix:angular_error}. All three types of uncertainties are combined together, assuming
their uncorrelated nature, via 
\begin{equation}
    \left(\delta{O}_\text{eff}\right)^2 =
    \left(\delta{O}_\text{stat}\right)^2 +
    \left(\delta{O}_\text{sys}\right)^2 +
    \left(\frac{\partial{O}_\text{th}}{\partial\theta}\right)^2
    \delta\theta^2,
    \label{eq:eff_variance}
  \end{equation}
where ${O}_\text{th}$  denotes the theoretical prediction for the
observable under consideration.
The $\chi^2/\text{datum}$ for the $N_\text{dat}$ experimental points is then defined by comparing each experimental value ${O}^i_\text{exp}$ with the corresponding interpolated theoretical value ${O}^i_\text{th}$ at the same angle:
\begin{equation}
    \chi^2/\text{datum} = \frac{1}{N_\text{dat}}\sum_{i=1}^{N_\text{dat}}
    \frac{\left({O}^i_\text{th} - {O}^i_\text{exp}\right)^2}
    {\left(\delta{O}^i_\text{eff}\right)^2}.
    \label{eq:chi2}
\end{equation}
Because the interpolated predictions ${O}^i_\text{th}$ depend on the
fit parameters, the effective uncertainty changes at every $\chi^2$
evaluation: As the parameters vary, the slope of  ${O}_\text{th}$
and thus also the effective uncertainty $\delta O_\text{eff}^i$ change
for every angle.  Therefore, the effective variances were
recomputed iteratively.  

Not all of the experimental data listed in Table
\ref{tab:exp_database_sources} are used in the fitting procedure. As
already emphasized above, we do not include the data at the highest
energy of $200$~MeV in the fits due to the expected large EFT
truncation error. These experimental data will be compared with the
theoretical predictions in the next subsection.  
Furthermore, because of the generally small impact of the 3NF at low
energies and the missing Coulomb effects in our calculations, we only include
the experimental data for the analyzing powers $A_y(N)$
and $A_y(D)$ at $E_N = 10$~MeV, for which clear
discrepancy between theory and data are observed. At $70$ and $135$~MeV,
we include experimental data for the differential cross
section and analyzing powers quoted in Table \ref{tab:exp_database_sources}, while the data for
spin-correlation coefficients $C_{ij}$ will be used to assess the predictive
power of the resulting Hamiltonian.  Finally, since we do not take
into account the Coulomb interaction, we exclude from the fits
experimental data at forward and backward angles, keeping
only data for center-of-mass scattering angle in the range $\theta \in
[60^\circ, 160^\circ]$. This leaves
us with the total of $N_\text{dat} = 501$ data points
used in the fits. 

\begin{figure}[t!]
	\centering
	\includegraphics[width=0.98\linewidth]{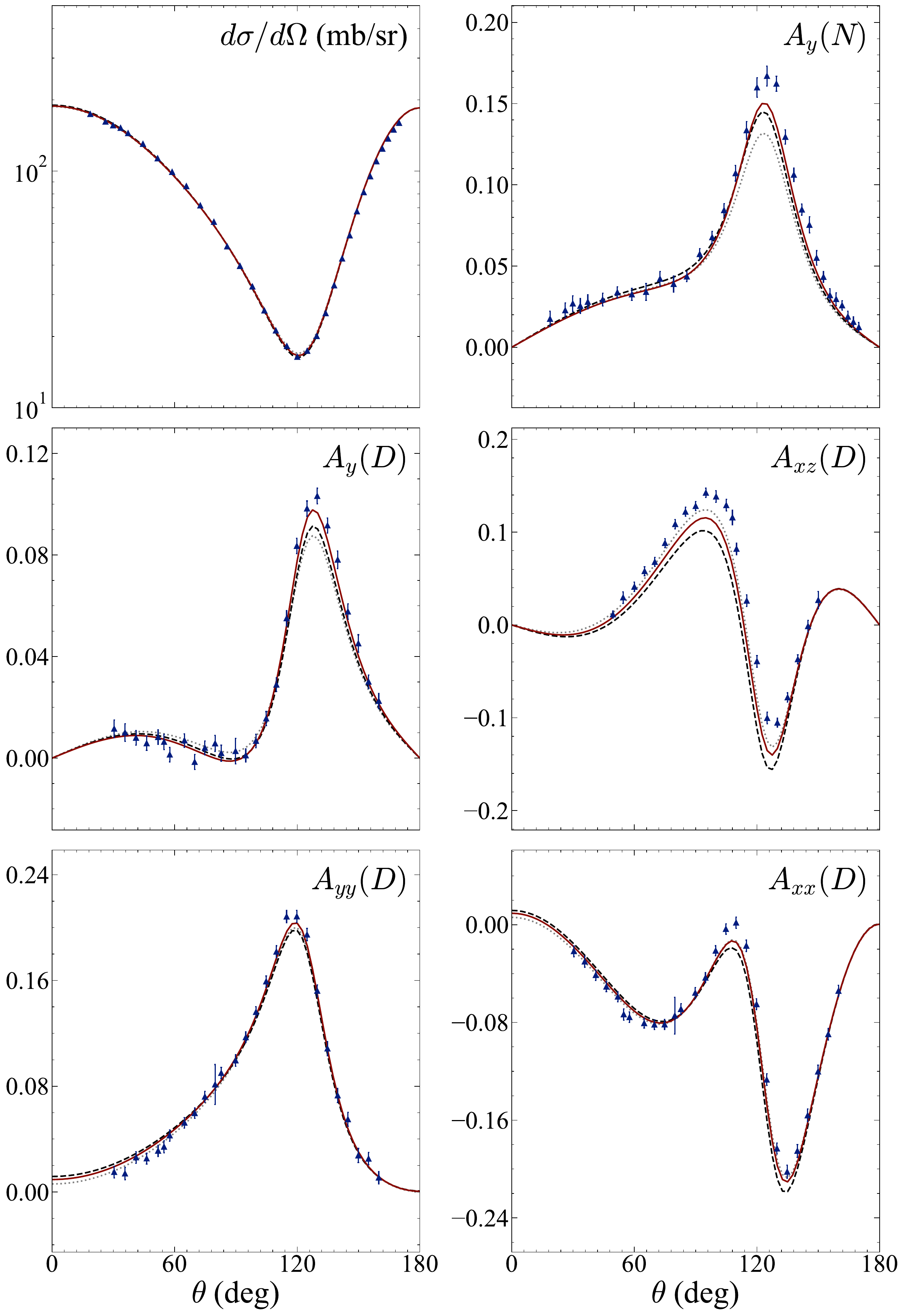}
    \caption{Differential cross section $d\sigma/d\Omega$, vector
      analyzing powers $A_y (N)$ and $A_y (D)$ as well as tensor
      analyzing powers $A_{xz}$, $A_{yy}$ and $A_{xx}$ at the
      scattering energy of $E_N = 10$~MeV. Gray dotted lines
      correspond to the predictions based on the SMS N$^4$LO$^+$ NN
      potential, while black dashed lines show the effects of taking
      into account the (parameter-free) two-pion
      exchange N$^2$LO 3NF. Red solid lines are the results of the
      full fit specified in the last column of
      Table.~\ref{tab:fit_results}. All results correspond to the
      cutoff value of $\Lambda = 450$~MeV. For references to the
      experimental data and marker legend see Table \ref{tab:exp_database_sources}. 
    }
	\label{fig:Fit_10MeV}
     \vspace{0.1cm}
   \end{figure}

\begin{figure}[t!]
	\centering
	\includegraphics[width=0.995\linewidth]{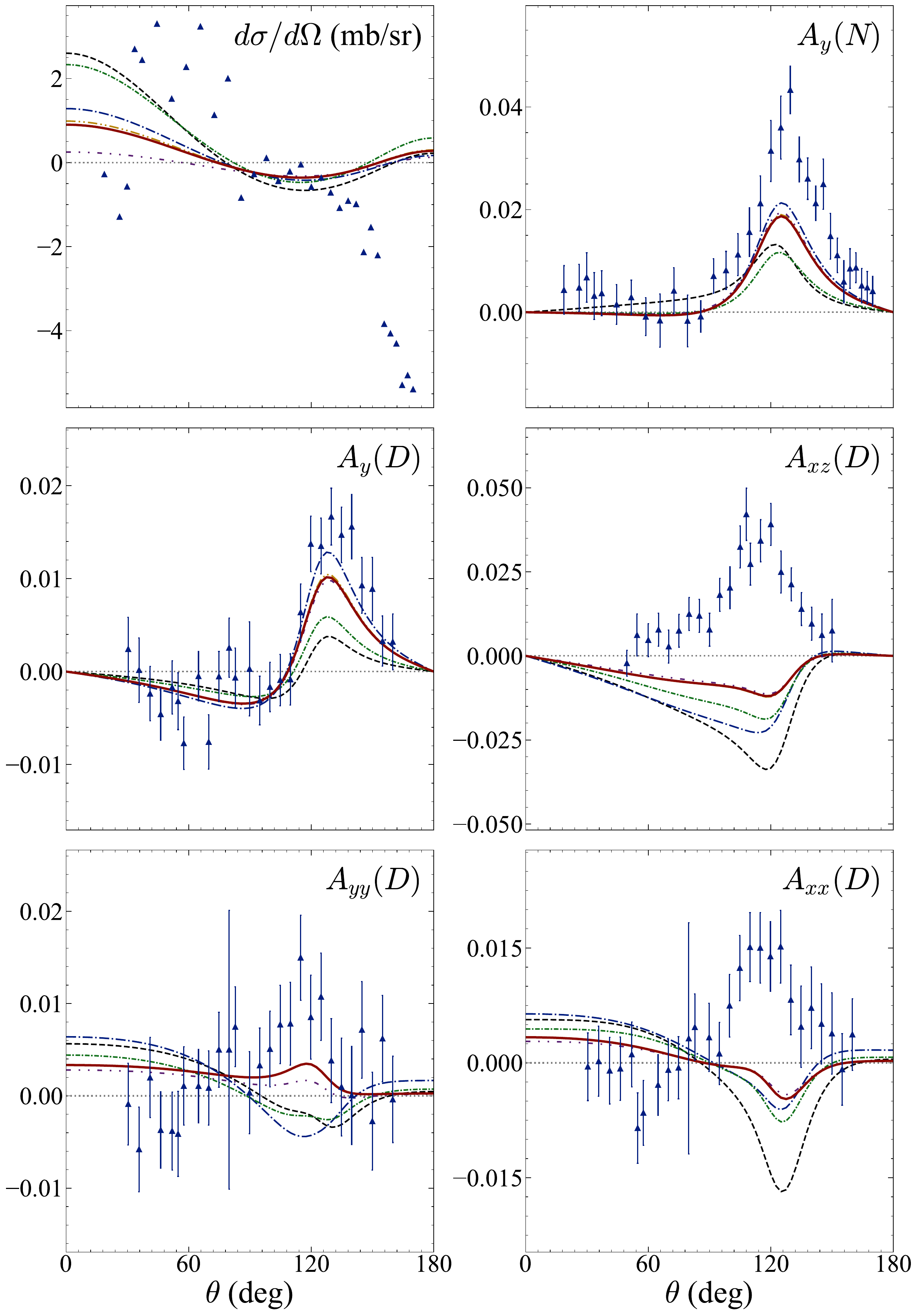}
    \vspace{-0.3cm}
	\caption{Same as Fig.~\ref{fig:Fit_10MeV}, but with the
         contributions of the NN force being subtracted.  Blue
         long-dashed-dotted, green short-dashed dotted, purple
         short-dashed-double-dotted and red solid lines show the
         results of the fits including the subleading contact 3NFs
         in the $J^P = 5/2^-$,  $5/2^- + 3/2^-$,  $5/2^- + 3/2^- +
         1/2^-$ and  $5/2^- + 3/2^- + 1/2^- + 1/2^+$ channels using the
         LECs specified in the third, fourth, fifth and sixth columns
         of Table   \ref{tab:exp_database_sources},
         respectively. Orange long-dashed-double-dotted lines
         show the impact of  including the cross section data
         at $E_N = 135$~MeV from Ref.~\cite{Ermisch:2005kf} in the full fit. 
    }
	\label{fig:Fit_10MeV_subtracted}
     \vspace{0.1cm}
   \end{figure}

\begin{figure}[t!]
	\centering
	\includegraphics[width=0.99\linewidth]{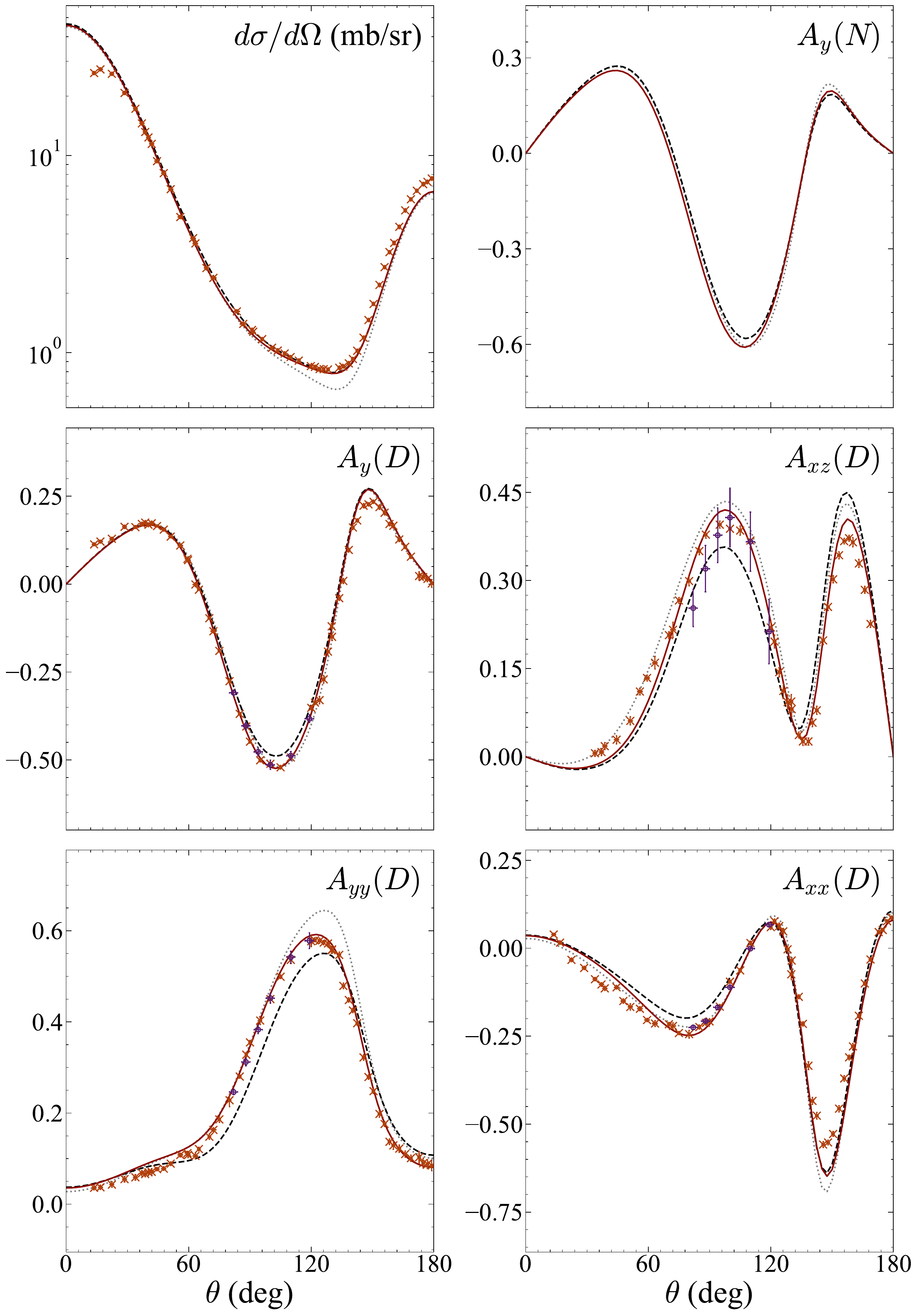}
    \vspace{-0.3cm}
	\caption{Same as Fig.~\ref{fig:Fit_10MeV}, but for $E_N=70$~MeV.
    }
	\label{fig:Fit_70MeV}
     \vspace{0.1cm}
   \end{figure}

\begin{figure}[t!]
	\centering
	\includegraphics[width=0.99\linewidth]{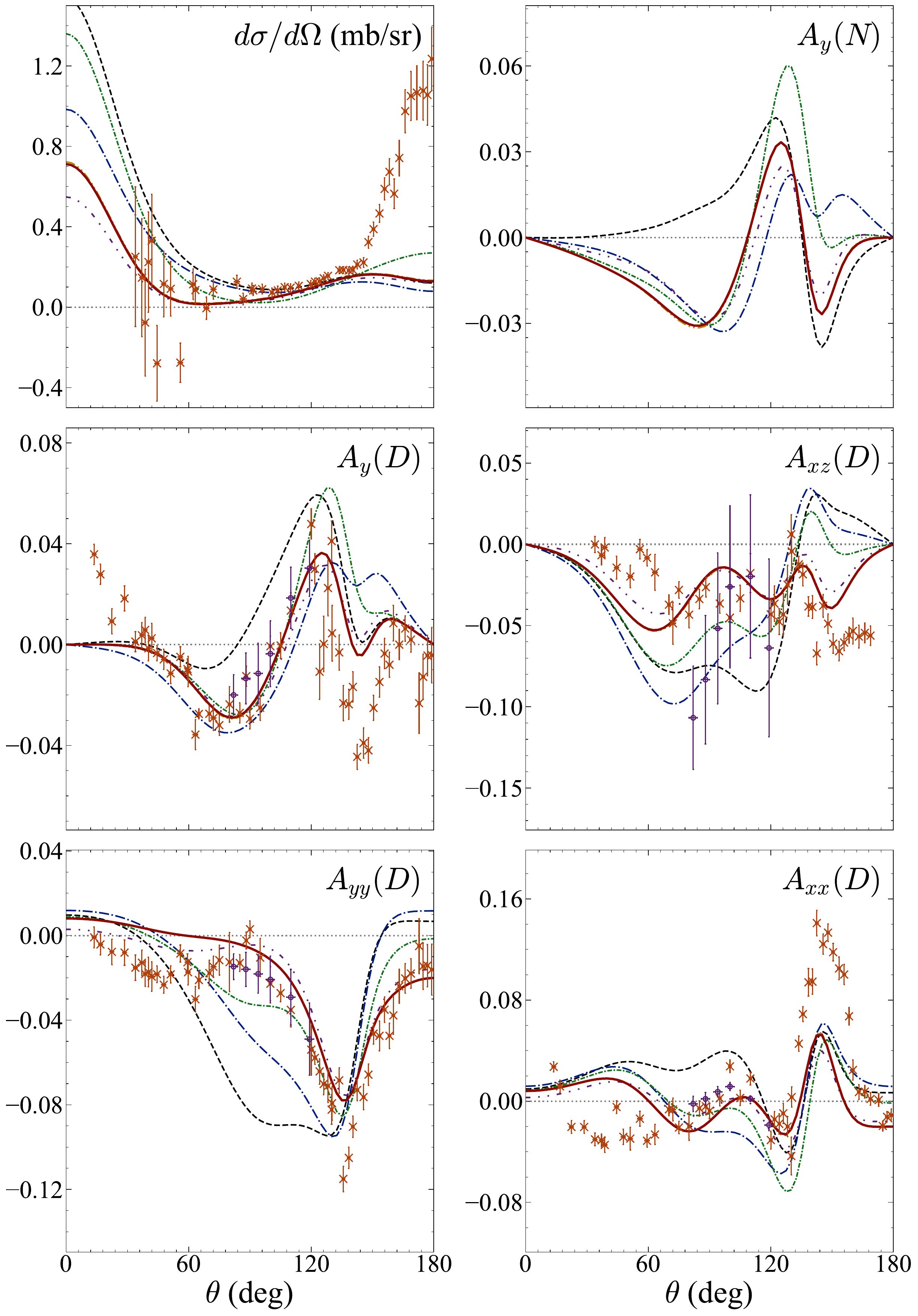}
    \vspace{-0.3cm}
	\caption{Same as Fig.~\ref{fig:Fit_10MeV_subtracted}, but for $E_N=70$~MeV.
    }
	\label{fig:Fit_70MeV_subtracted}
     \vspace{0.1cm}
   \end{figure}

\begin{figure}[t!]
	\centering
	\includegraphics[width=0.99\linewidth]{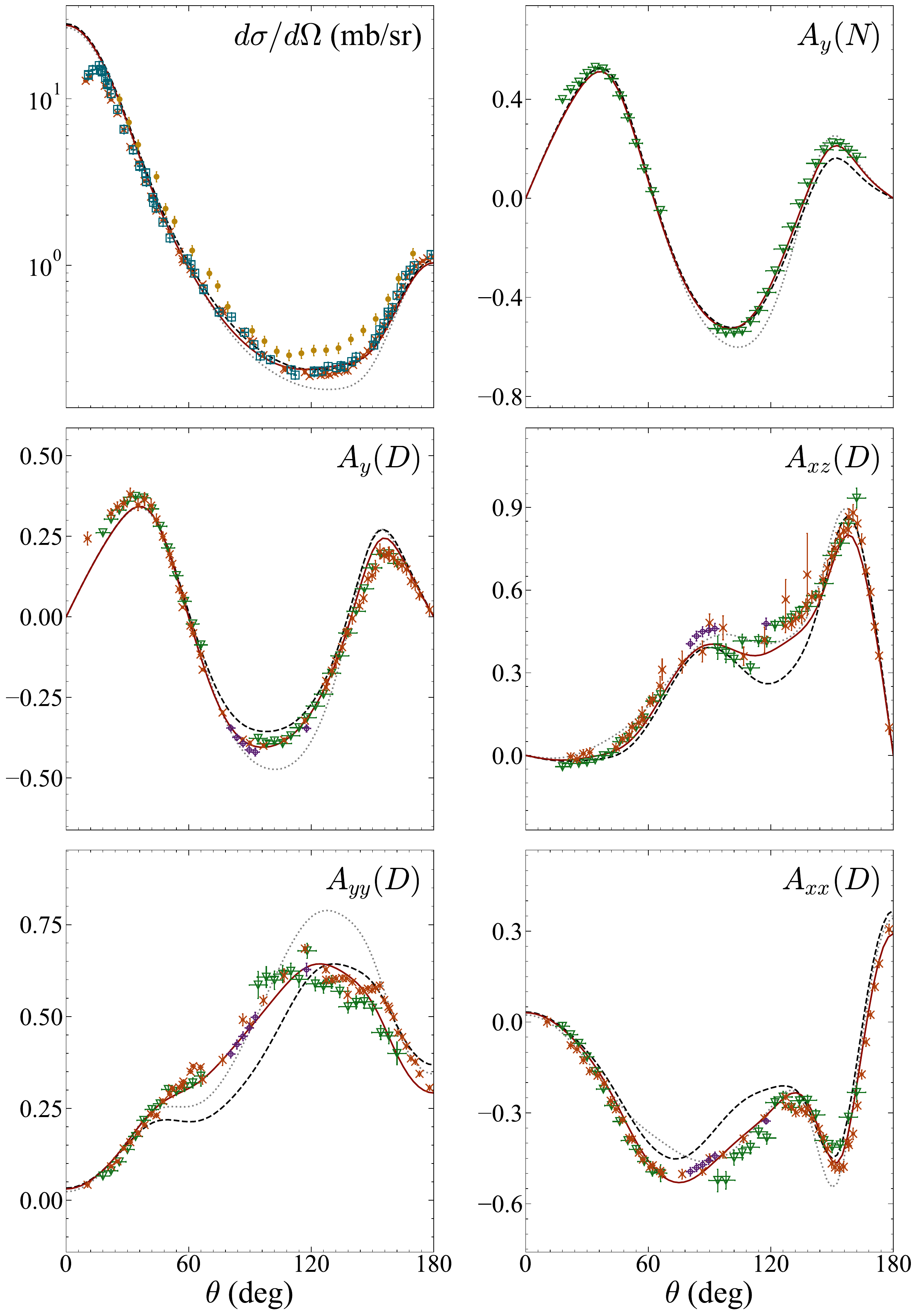}
    \vspace{-0.3cm}
	\caption{Same as Fig.~\ref{fig:Fit_10MeV}, but for $E_N=135$~MeV.
    }
	\label{fig:Fit_135MeV}
     \vspace{0.1cm}
   \end{figure}

\begin{figure}[t!]
	\centering
	\includegraphics[width=0.99\linewidth]{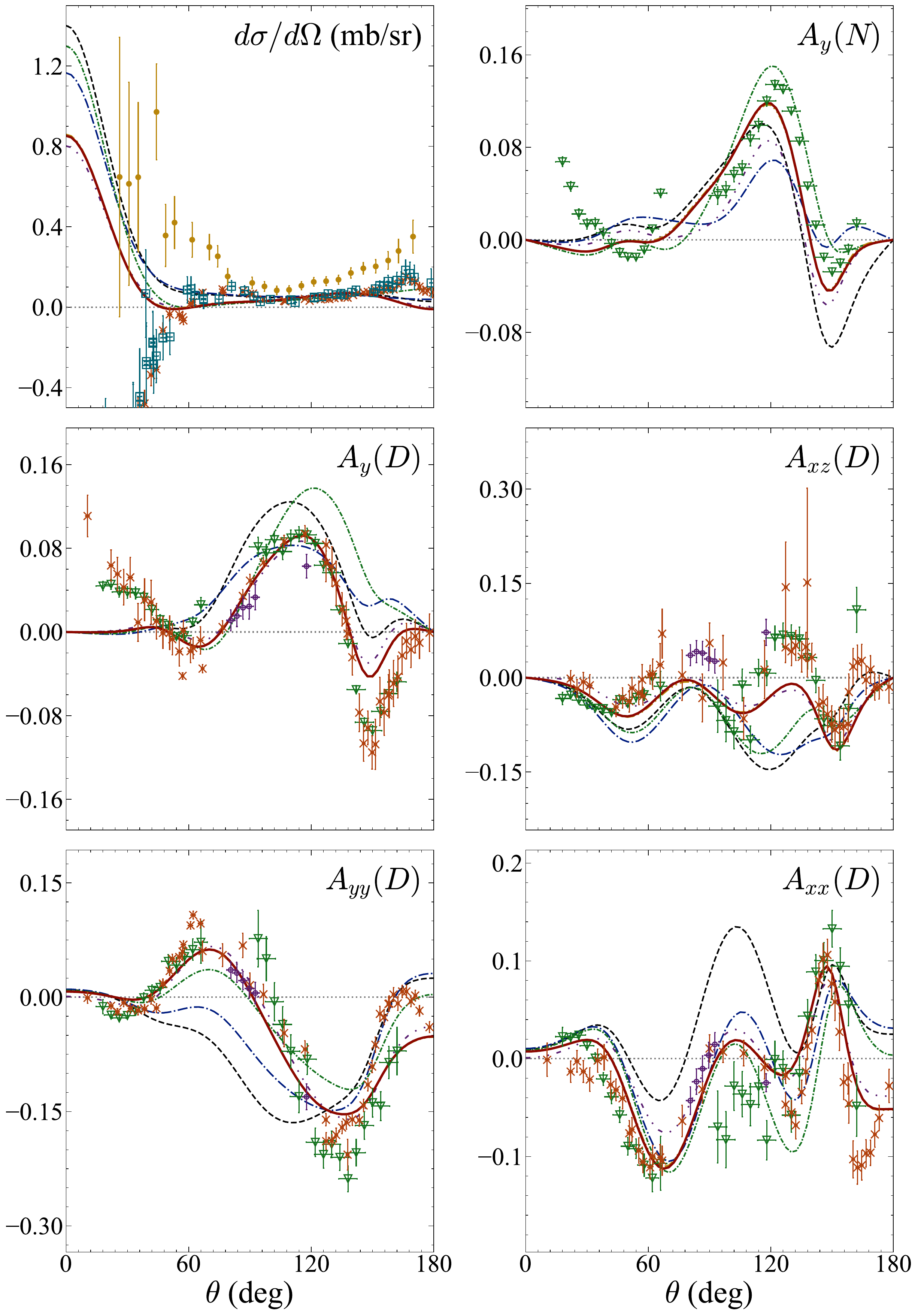}
    \vspace{-0.3cm}
	\caption{Same as Fig.~\ref{fig:Fit_10MeV_subtracted}, but for $E_N=135$~MeV.
    }
	\label{fig:Fit_135MeV_subtracted}
     \vspace{0.1cm}
   \end{figure}

In Table \ref{tab:fit_results}, we collect the values of various LECs,
along with the corresponding $\chi^2/{\rm datum}$,
resulting from fits of increasing complexity, in which we subsequently include the
N$^4$LO contact 3NFs in the $J^P = 5/2^-$,  $J^P = 3/2^-$, $J^P =
1/2^-$ and $J^P = 1/2^+$ channels. The resulting description of
the scattering observables is shown in
Figs.~\ref{fig:Fit_10MeV}-\ref{fig:Fit_135MeV_subtracted}.


In agreement with the sensitivity analysis of sec.~\ref{sec:Sensitivity},
the inclusion of the $^4P_{5/2}$ LEC $s_{11}$, along with the N$^2$LO
short-range 3NFs driven by the LECs $c_D$ and $c_E$, already leads to a
very significant improvement in the description of experimental data,
resulting in the decrease of $\chi^2/{\rm datum}$ from $\chi^2/{\rm
  datum} = 20.09$ in
calculations using the parameter-free part of the N$^2$LO 3NF to
$\chi^2/{\rm datum} = 12.30$. As expected from the results of
sec.~\ref{sec:Sensitivity}, the LEC $s_{11}$ takes a negative value,
which leads to a strongly (moderately) improved description of $A_y
(D)$ ($A_y (N)$) at $E_N = 10$~MeV. On the other hand, there is
still room for improvement in the
description of tensor analyzing powers at intermediate energies, with
the exception of $A_{xx} (D)$ at $E_N = 135$~MeV.  Interestingly, one
also observes a strong change in the values of the N$^2$LO 3NF LECs.
In particular, in the absence of the N$^4$LO 3NF contributions,
the LECs $c_D$ and $c_E$ take the values of $c_D = 0.892$ and $c_E =
-0.386$ \cite{Jimenez:2026xgh} using the standard LENPIC fitting
protocol \cite{LENPIC:2018ewt,Maris:2020qne,LENPIC:2022cyu}. The main
reason for a large shift of $c_D$, which is strongly constrained by
the cross section data at $E_{N}=70$~MeV \cite{LENPIC:2018ewt,Maris:2020qne,Jimenez:2026xgh}, is a considerable increase
of the differential cross section around the minimum region for
negative values of the LEC $s_{11}$. This effect then has to be compensated
by decreasing the value of $c_D$. Interestingly, the resulting
large and negative value of the LEC $c_D$ appears to be consistent
with the constraint from the tritium $\beta$-decay, see Fig.~1 of
Ref.~\cite{Jimenez:2026xgh}. In fact, negative values of this LEC
seem to be a robust result of including the N$^4$LO short-range 3NFs
with all  fits yielding $c_D \sim -4.1 \ldots -2.4$.
We emphasize, however, that these results can only serve as
an indication that the discrepancy for the $^3$H
Gamow-Teller matrix element at N$^2$LO \cite{Jimenez:2026xgh} may get
resolved (at least  partially) by the change in the LEC $c_D$ caused by the inclusion of
higher-order contact 3NFs. More conclusive
analysis will require taking into account the longer-range contributions
to the 3NF and the axial current operator beyond N$^2$LO. Our
preliminary results indicate that the parameter-free two-pion exchange and the
two-pion-one-pion exchange 3NF contributions at N$^3$LO tend to
shift the value of $c_D$ back in the positive direction. 

The inclusion of the LECs $s_{8,9,10}$ in the $J^P = 3/2^-$ channel
leads to a further improvement in the description of the experimental
data, which results in the $\chi^2/{\rm datum}$ value of $9.74$. The
most visible improvement is observed for the tensor analyzing power
$A_{yy}$ at $E_N = 70$ and $135$~MeV. On the other hand, this fit
results in a smaller-in-magnitude negative value of the LEC $s_{11}$,
which worsens the description of $A_y$ at the lowest energy.    

Somewhat unexpectedly (in view of the sensitivity studies described in
sec.~\ref{sec:Sensitivity}), the inclusion of the LECs $s_{5,6,7}$
contributing to the $J^P = 1/2^-$ channel is found to provide a 
substantial improvement in the description of the experimental data,
leading to  $\chi^2/{\rm datum} = 5.37$. In particular, one observes a
visible improvement in the description of the tensor analyzing powers
$A_{xz}$ and  $A_{yy}$ at both intermediate energies of $E_N = 70$ and
$135$~MeV, and a slightly better reproduction of $A_{xx}$ at $70$~MeV.  
It is also comforting to see that the values of the $J^P = 3/2^-$ LECs
remain fairly stable upon inclusion of the 3NF contributions in the
$J^P = 3/2^-$ channel (apart from $s_8$). In spite the fact that the
LEC $s_{11}$ further decreases in magnitude, the description of the
vector analyzing powers at low energy significantly improves compared
to the previous fit. This can probably be traced back to the
large and positive value of the LEC $s_6$ (cf.~Figs.~\ref{fig:sensitivity_Ay(N)_10} and \ref{fig:sensitivity_Ay(D)_10}), which governs
the quartet $P$-wave in the  $J^P = 1/2^-$ channel. This can also
explain a visible improvement in the description of $A_{yy}$ at $E_N
= 10$~MeV. 

Finally, we turn to the most complete fit performed in this study by
also including the $s_i$'s contributing to the $J^P = 1/2^+$
channel, see the last column of Table \ref{tab:fit_results}. When
carrying out this fit, we have observed, in agreement with the
considerations in sec.~\ref{sec:SpecBasis}, that the LECs
$s_2$ and $s_4$ are largely redundant when looking at elastic
Nd scattering observables. For example, using the values of $s_2 =
\pm 2.0$ ($s_4 =
\pm 2.0$) only changes the value of $\chi^2/{\rm datum}$ by $0.8\%$ ($0.6\%$) 
compared to the result for $s_2 = s_4 = 0$, which is at least an
order of magnitude smaller than the $\chi^2/{\rm datum}$
changes for a similar variation of other LECs.\footnote{While the description of elastic scattering
  observables is almost unaffected by the values of $s_2$ and
  $s_4$, a variation of the LEC $s_2$ leads to significant shifts in the values
  of $s_1$ and $s_3$, as one may expect from Eq.~(\ref{SpectroscopicTransitions}).} Keeping the nearly redundant $s_2$ and $s_4$ as free
parameters in the fit, therefore, destabilizes the determination of
other LECs. For this reason, we excluded $s_2$ and $s_4$ from
the fit by setting their values to $s_2 = s_4 = 0$. The most
significant effect from the inclusion of the $J^P=1/2^+$ LECs $s_1$
and $s_3$ is observed for the nucleon vector analyzing power at
intermediate energies, leading to a visibly improved description of
the experimental data at $E_N = 135$~MeV.  We also observe a better
description of the tensor analyzing power $A_{xx}$ at forward
scattering angles at the same energy. 

\begin{figure}[t]
	\centering
	\includegraphics[width=0.99\linewidth]{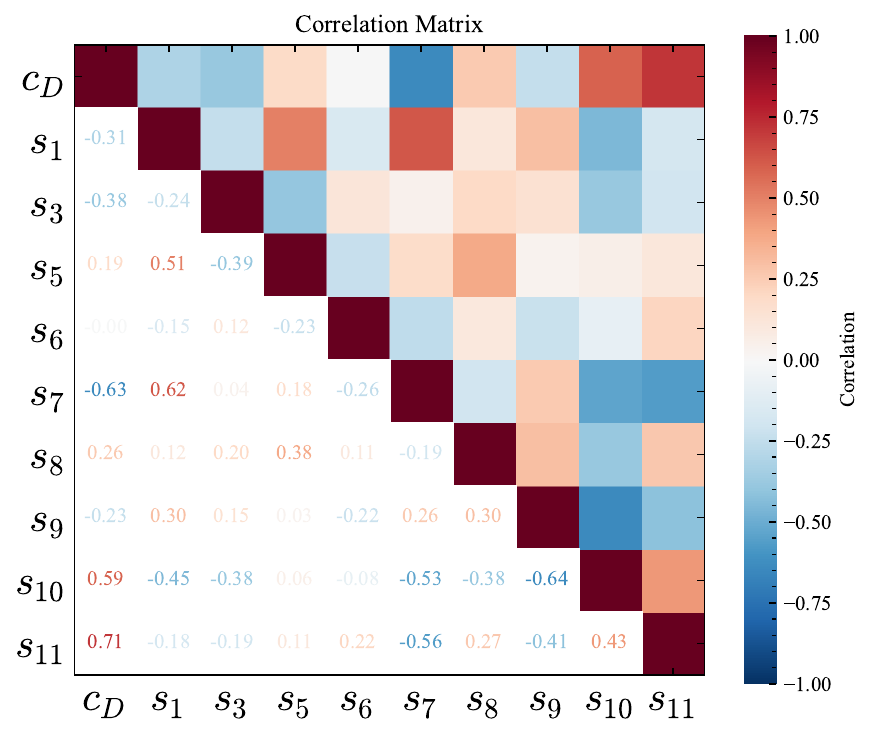}
    \caption{Correlation matrix for the full fit specified in the last column of Table~\ref{tab:fit_results}. The lower triangle gives the correlation coefficients.
    }
	\label{fig_correlations}
      \end{figure}

It is important to emphasize that the results of the full fit remain 
stable in spite of a large number of the adjustable parameters. The
stability was verified by performing fits using randomly chosen
starting points in the parameter space within the considered range,
which always resulted in the same minimum. We further emphasize
that the LECs $s_i$ show, as expected, a fairly small amount of correlations (after
excluding the LECs $s_2$ and $s_4$ from the fit), which was one of the
motivations for introducing the spectroscopic basis. In
Fig.~\ref{fig_correlations}, we provide the resulting correlation
matrix for the full fit.  The strongest correlation of $0.71$
appears between LECs $c_D$ and $s_{11}$. Among the $s_i$'s,
the strongest (anti-) correlations are observed between $s_9$ and
$s_{10}$ in the $J^P = 3/2^-$ channel, and between the LECs $s_1$  and
$s_7$. 

It is also instructive to look at the statistical uncertainties of the
LECs $s_i$. First of all, one observes that the uncertainties of the
LECs contributing to the $J^P=5/2^-$, $3/2^-$ and $1/2^-$ channels do
not significantly increase when performing more complete fits with a
larger number of LECs. This is expected given that the LECs contribute
to different partial waves and are largely uncorrelated with each
other. Nevertheless, this observation provides an important
consistency check of our results and confirms their
robustness. Furthermore, looking at the statistical uncertainties of
the LECs $s_i$ from the full fit, we observe a clear hierarchy with
$\delta s_i^{5/2^-} < \delta s_i^{3/2^-} < \delta s_i^{1/2^+} < \delta
s_i^{1/2^-}$.  
This pattern is consistent with the expectations based on the
sensitivity analysis in sec.~\ref{sec:Sensitivity}, which revealed
that the most significant impact on elastic Nd scattering is generated
by the LEC $s_{11}$ in the $J^P = 5/2^-$ channel, followed by the LECs
$s_{8,9,10}$ contributing to the  $J^P = 3/2^-$ channel. The low
sensitivity to the LECs contributing to the $J^P = 1/2^-$ partial
waves observed in sec.~\ref{sec:Sensitivity} is consistent with
the largest statistical uncertainties of the corresponding LECs.
We also observe that the changes in $s_i$'s from
the simplest to the most complete fit typically exceed their
statistical errors. This points towards sizable systematic
uncertainties, which are not considered in this study.

While our results from the full fit show a clear improvement in
the description of the considered observables, some discrepancies 
remain at intermediate energies, 
in particular for the differential cross section at
backwards scattering angles (not included in the fits) and the tensor analyzing
power $A_{xx}$ at $E_N = 70$~MeV. This is also reflected in the value of
$\chi^2/{\rm datum}= 4.91$, which is significantly larger than 
$\chi^2/{\rm datum} \approx 1$ expected for a statistically consistent
fit. These remaining discrepancies are not unexpected given the exploratory nature of
this study and the still missing contributions of the long- and
intermediate range 3NFs beyond N$^2$LO. On the other hand, judging the
quality of the fit based on the resulting value of $\chi^2/{\rm
  datum}$ alone might be misleading given that the employed database
has not been analyzed to exclude mutually incompatible experimental
data, as it is usually done in the two-nucleon sector
\cite{Stoks:1993tb,Gross:2008ps,NavarroPerez:2013mvd,NavarroPerez:2016eli,Reinert:2020mcu,Reinert:2022jpu}.
In particular, about $22\ldots 24\%$ of neutron-proton scattering data and  $
5\ldots 7\%$ of proton-proton data below $E_{\rm lab} = 300$~MeV are
rejected from the two-nucleon database using the $3\sigma$ criterion \cite{Reinert:2022jpu}. 
We,
therefore, expect that a significant fraction of the resulting
$\chi^2/{\rm datum}$ value might originate from outliers and
conflicting experimental data (sets), see
Fig.~\ref{fig:Fit_135MeV_subtracted} for some examples. One of the
well-known examples of such conflicting data sets concerns the
experimental data for the differential cross section at $E_{N} =
135$~MeV taken at KVI \cite{Ermisch:2005kf}(shown by the solid yellow
circles in Figs.~\ref{fig:Fit_135MeV}, \ref{fig:Fit_135MeV_subtracted})  and at 
RIKEN \cite{Sekiguchi:2002sf,Sekiguchi:2005vq} (shown by dark orange crosses
and open light blue squares in these figures).  We have
only included the cross section data from
Refs.~\cite{Sekiguchi:2002sf,Sekiguchi:2005vq} in our fits,
from which we also use the data for polarization observables and
the differential cross section at $E_{N} = 70$~MeV. To quantify the
impact of the KVI cross section data at $135$~MeV on our results, we
have redone the most complete fit including this data set.  
The results of this fit are visualized in Figs.~\ref{fig:Fit_10MeV_subtracted},
\ref{fig:Fit_70MeV_subtracted} and \ref{fig:Fit_135MeV_subtracted}
with orange dashed-double-dotted lines. 
The effect of including the KVI data is barely visible since the solid red line, corresponding to our complete fit, almost completely overlaps with it. This is due to the resulting LECs being very close to the values of Table \ref{tab:fit_results}, and the only real difference being a slightly larger $\chi^2/{\rm datum}$. We thus conclude
that the inclusion of these cross section data has a minor effect on the fit
results, probably due to the larger experimental uncertainties as
compared with the RIKEN data.

The impact of the subleading contact
3NF on Nd scattering observables has also been considered in
Refs.~\cite{Girlanda:2018xrw,Witala:2022rzl}. However, comparing our findings with theirs is not
straightforward. In particular, the authors of Ref.~\cite{Girlanda:2018xrw} focused on
the description of proton-deuteron scattering observables at a single
energy of $E_N = 3$~MeV using the AV 18 NN potential \cite{Wiringa:1994wb}, accompanied by
the Urbana IX 3NF model \cite{Pudliner:1997ck}, along with the N$^2$LO
and contact N$^4$LO 3NF contributions.
Their results show that one can achieve a nearly perfect
description of $\sim 300$ available experimental data at this energy,
including those for the nucleon and deuteron analyzing powers. On the
other hand, limiting oneself to very low energies makes a reliable
determination of the subleading contact 3NF challenging
\cite{Girlanda:2018xrw}. As shown in Figs.~\ref{fig:sensitivity_Ay(N)_10} and
\ref{fig:sensitivity_Ay(D)_10}, we also find that it is, in
principle, possible to simultaneously resolve the nucleon and
deuteron $A_y$ puzzles at low energies by appropriately tuning the
LECs $s_{10}$ and $s_{11}$. However,  the values of these LECs
have to comply with 
experimental data at higher energies. Using the
presently available (incomplete) 3NF model, tuned to Nd scattering
observables in a broad energy range, we find an improved
but still not perfect description of $A_y (N)$ and $A_y(D)$ at low
energies, see Figs.~\ref{fig:Fit_10MeV} and \ref{fig:Fit_10MeV_subtracted}.   
This conclusion is in line with the findings of Ref.~\cite{Witala:2022rzl} based on
the same NN potential as used in our analysis. Our description of
the differential cross section and $A_y (N)$ at the three considered
energies is similar to that presented in this work, while the results
for the deuteron vector and tensor analyzing powers show significant 
differences. It is difficult to make definite statements regarding the
origin of these disagreements, given the discrepancies observed already
at the level of the $^3$H expectation values in Table
\ref{Table1}. Additional possible source of differences could be the reliance
on a perturbative emulator for Nd scattering observables in
Ref.~\cite{Witala:2022rzl}.

\subsection{Selected predictions}
      
\begin{figure}[t]
	\centering
	\includegraphics[width=0.96\linewidth]{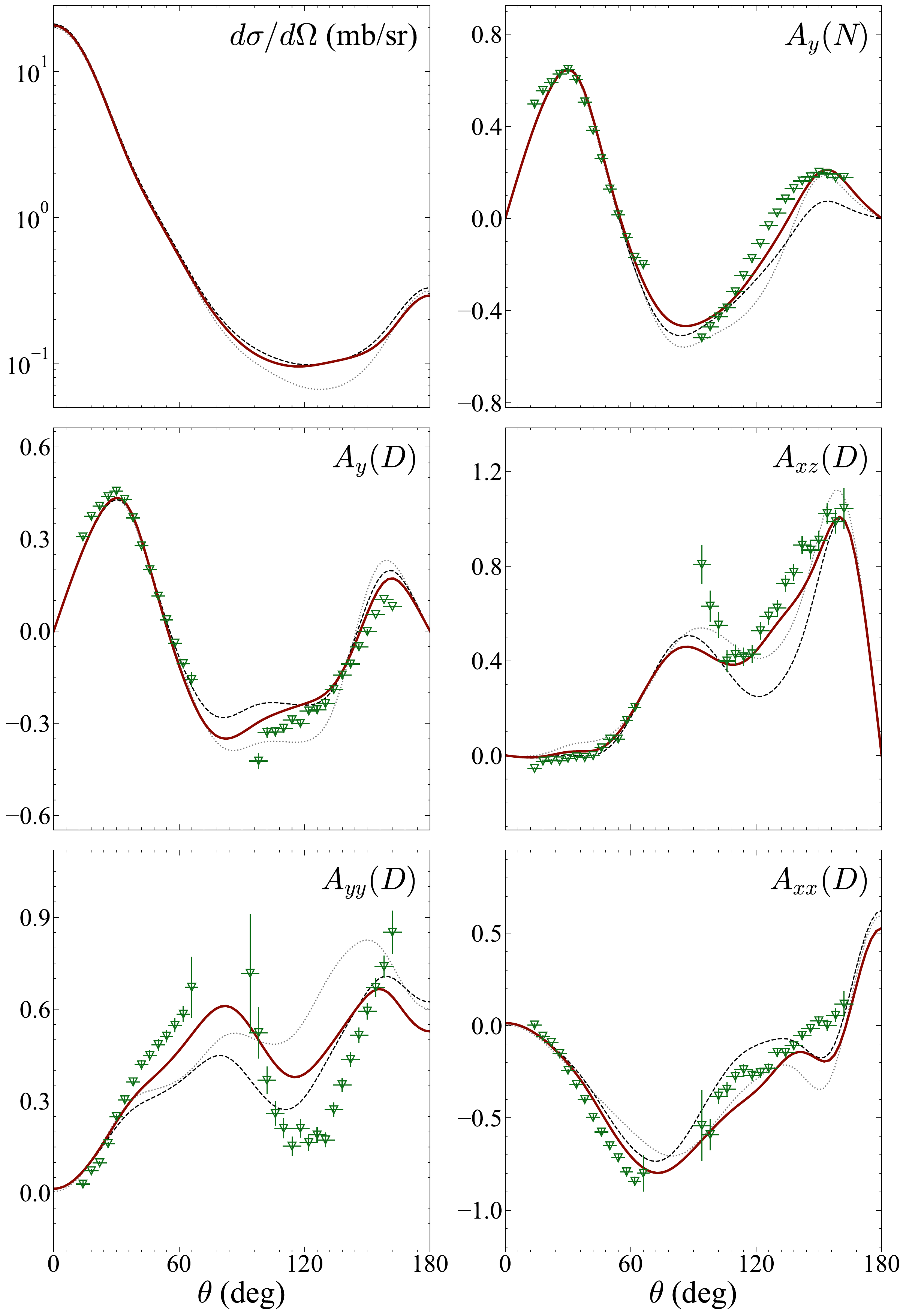}
	\caption{Predictions for the differential cross section as
          well as 
          vector and tensor analyzing powers in Nd elastic scattering
          at $E_N = 200$~MeV. Gray dotted lines show the results
          obtained using the SMS N$^4$LO$^+$ NN potential of
          Ref.~\cite{Reinert:2017usi} with the cutoff $\Lambda = 450$~MeV, while
          black dashed lines take into account the two-pion
          exchange 3NF at N$^2$LO. Red solid lines also include the
          contributions from the $c_D$- and $c_E$-terms at N$^2$LO and
          from the $s_i$-terms at N$^4$LO using the values from the
          last column of Table \ref{tab:fit_results} (full
          fit). References to the experimental data are given in Table
          \ref{tab:exp_database_sources}. 
    }
	\label{fig:predictions200}
     \vspace{0.1cm}
   \end{figure}

Having determined the short-range part of the 3NF from some
experimental data for elastic Nd scattering, it is now interesting to
confront the resulting model with observables not used in the fitting
procedure. In Fig.~\ref{fig:predictions200}, 
\begin{figure}[t]
	\centering
	\includegraphics[width=0.96\linewidth]{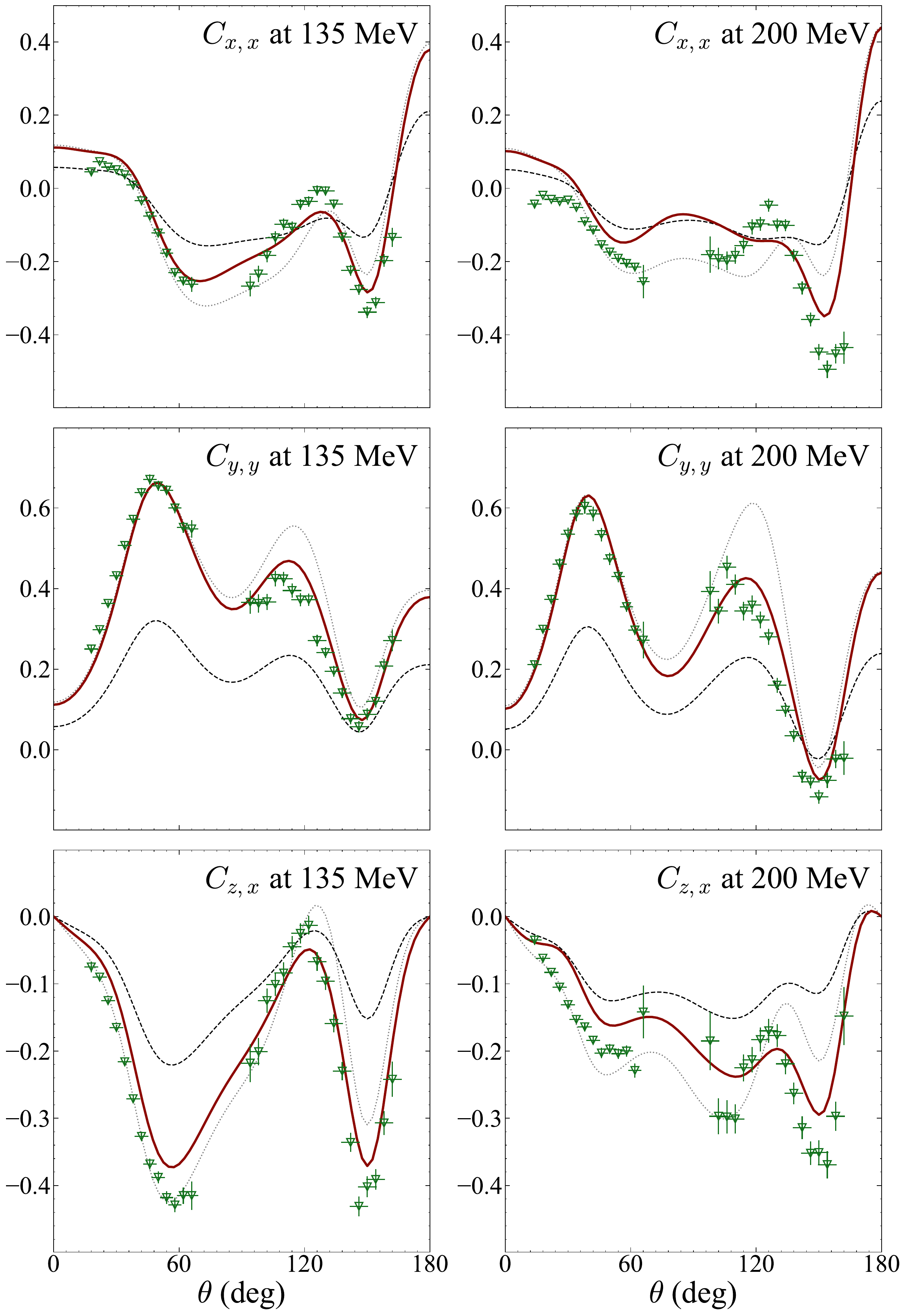}
	\caption{Predictions for the selected spin correlation
          coefficients at $E_N = 135$~MeV (left plots) and $E_N =
          200$~MeV (right plots). For remaining notation see
          Fig.~\ref{fig:predictions200}. 
    }
	\label{fig:predictionsCij}
     \vspace{0.1cm}
   \end{figure}
we show the predictions
for the differential cross section and analyzing powers at the highest
considered 
energy of $E_{N} = 200$~MeV, while Fig.~\ref{fig:predictionsCij}
presents the results for selected spin correlation coefficients
at  $E_{N} = 135$~MeV and  $E_{N} = 200$~MeV.
The description of the analyzing powers at $E_{N} = 200$~MeV using the
3NF from the full fit is visibly improved compared with the
results based on the NN interaction only, shown
by gray dotted lines, and those including the parameter-free two-pion
exchange 3NF, shown by the black-dashed lines.
This also applies to
the considered spin correlation coefficients with the only exception
of $C_{z,x}$ in
the forward direction, where the experimental data are better
described based on the NN interaction only. Interestingly, we observe
that the leading long-range contribution to the 3NF due to two-pion
exchange has a very large impact on the spin correlation coefficients
$C_{y, y}$ and $C_{z,x}$, especially at the lower of the two
considered energies. The effects from these parameter-free 3NF
contributions move theoretical predictions away from experimental
data, but they are largely compensated by the short-range 3NF
components. For $C_{y, y}$ and $C_{z, x}$ at backward angles,
the resulting description of the data is substantially improved
compared to that obtained using the NN interaction only.
Nevertheless, our results indicate that chiral EFT predictions
for these double-polarization observables may potentially suffer from
sizable truncation uncertainties, even at the N$^4$LO order. 
Taking the differences between the gray dotted and black dashed
lines as a measure of the size of the long-range N$^2$LO 3NF
components, the remaining discrepancies between
theory and data are compatible with the expected size of
higher-order long- and intermediate-range corrections to the 3NF.
It will be interesting to confront our results with the future data
on spin correlation coefficients at $E_N = 100$~MeV at RIKEN \cite{Saito:2026dzr}.  

Another interesting observation concerns the differential cross
section. The experimental data for the Nd cross section at high
energies and backward angles are well known to be considerably
underpredicted by theoretical calculations
\cite{Kalantar-Nayestanaki:2011rzs,Endo:2024cbz}.
As shown in Fig.~\ref{fig:predictions200}, the currently
employed 3NF model is not capable of resolving these discrepancies, see
Ref.~\cite{Witala:2022rzl} for a similar conclusion. We also find
a similar (slight) underestimation of the total cross section at $E_N
= 200$~MeV as reported in that paper. The robustness and significance of these
discrepancies can only be quantified upon performing more complete
calculations by taking into account the long- and intermediate-range
components of the 3NF and estimating the truncation uncertainties,
which is beyond the scope of our study.

\section{Summary}
\label{sec:Summary}

The main results of our paper can be summarized as follows.
\begin{itemize}
\item  
We provide explicit analytical expressions for partial-wave
decomposition of the subleading contact 3NF in
Eq.~(\ref{DefEi}).  Using these expressions, we introduced a
spectroscopic basis for these operators by defining a new set of
LECs $S_{1, \ldots , 13}$ in a close analogy to what is done in the
two-nucleon sector. Compared to the original parametrization of
the 3NF in Eq.~(\ref{DefEi}), the spectroscopic basis provides a
simple and transparent interpretation of the corresponding LECs, which
contribute only to specific partial waves in the $J^P = 1/2^\pm$,
$3/2^-$ and $5/2^-$ channels of elastic Nd scattering. More precisely,
leaving aside the LECs $S_{12,13}$, which only contribute to the
isospin $T=3/2$ channel, and the LEC $S_4$ that has vanishing matrix
elements in Nd states, the remaining LECs contribute to the 
$^2S_{1/2}\text{-,}$ $^2P$- and $^4P$-waves and the corresponding mixing angles
as summarized in Table \ref{tab:Spectro_summary}. As an independent
check of our results, we have calculated the $^3$H expectation values
of the linear combinations of the $E_i$-terms
given in Eq.~(\ref{SpectroLECs}) to verify that the $S_{5, \ldots
  , 13}$-terms indeed yield vanishing contributions. Our expressions for
the linear combinations of $E_i$'s contributing to specific Nd partial
waves are consistent with those given in Ref.~\cite{Filandri:2026ori}, see also
Ref.~\cite{Zuo:2025zgg} for a related discussion.
\item
We carried out a sensitivity analysis of vector and tensor analyzing powers
in Nd scattering to the spectroscopic LECs $s_i$. The
strongest impact is found from the LEC $s_{11}$, which governs the
$^4P_{5/2}$ partial wave, followed by the LECs contributing
to the $3/2^-$, $1/2^+$ and $1/2^-$ channels. Our results show that
the well-known disagreements for the nucleon and deuteron vector
analyzing power $A_y$ at low energies can, in principle, be removed via appropriately
chosen LECs $s_{10}$ and $s_{11}$ of natural size. On the other hand,
we found that the considered 3NFs have small effect
on the cross section in the SST breakup configuration, which
constitutes another well-known low-energy puzzle in the 3N continuum. 
\item
Owing to a strong reduction of the dimensionality of the parameter space in the
spectroscopic basis, we introduced a simple emulator for Nd scattering
observables using the RBF interpolation technique. This emulator was
then employed to perform exploratory fits of the LECs $c_D$, $c_E$ and
$s_i$ to the experimental data for the cross
section and analyzing powers at $10$, $70$ and $135$~MeV, in order to
assess the feasibility of fixing the short-range part of the 3NF in
elastic Nd scattering. We found that all LECs contributing to
isospin-$1/2$ states, apart from $s_2$ and $s_4$, can be reliably
determined from the Nd elastic scattering data. The LEC
$s_4$ contributes to elastic scattering observables only through
intermediate three-nucleon scattering states, see Table
\ref{tab:Spectro_summary}, making them largely insensitive to $s_4$, see also
sec.~\ref{sec:Sensitivity}. On the other hand, the LEC $s_2$ directly
contributes to the doublet $S$-wave, but its effect is, to a large extent, 
absorbable into a redefinition of the LECs $c_E$ and $s_3$ based on
Eq.~(\ref{SpectroscopicTransitions}). These expectations are in line
with our fit results. For the remaining LECs, 
one observes small to moderate (anti-)correlations that do not exceed $\sim 0.7$. 
\item
Our most complete fit of the LECs $c_D$, $c_E$, $s_1$, $s_3$ and $s_{5,
\ldots ,11}$ is shown to result in a significantly improved description
of elastic Nd scattering observables over the whole considered energy
range, as compared with the predictions
based on the SMS N$^4$LO$^+$ NN potential with or without the
parameter-free long-range 3NF at N$^2$LO. The resulting values of the
LECs are of natural size except for $s_6$, for which a somewhat
large value of $s_6 \approx 6$ is found. However, the values
of the LECs are expected to change upon the inclusion of the long- and
intermediate-range 3NF contributions beyond N$^2$LO, so that no
conclusive statements regarding their naturalness are possible at this
stage. The currently employed model of the nuclear Hamiltonian
partly resolves the $A_y$-puzzle by reducing the discrepancy between data and
theoretical predictions based on high-precision NN interactions by more than a
half. On the other hand, the description of some observables like,
e.g., $A_{xx}$ at $70$~MeV and $A_{yy}$ at $135$~MeV leaves room
for improvement.
\item
We have tested the resulting 3NF model by calculating selected
observables not used in the fitting procedure, including the cross
section and analyzing powers at $200$~MeV and spin correlation
coefficients at $135$ and $200$~MeV. For all considered cases, we
found a reasonable description of the data, which is visibly improved
compared to the calculations including only the long-range N$^2$LO
3NF. The remaining discrepancies are consistent with the expected size
of long- and intermediate-range 3NF contributions beyond N$^2$LO. 
\item
Another interesting outcome of our study concerns the LEC $c_D$, whose
value is found
to change from $c_D \approx 0.89$ to  $c_D \approx -2.51$ upon
including the subleading short-range 3NF contributions.  This negative
value is considerably closer to the one preferred by the tritium
half-life \cite{Jimenez:2026xgh}. We, however, again emphasize that
definitive conclusions regarding the discrepancy for the
tritium Gamow-Teller reduced matrix element will require a more
complete treatment of the 3NF and weak current operator beyond
N$^2$LO. 
\end{itemize}  
In summary, we found that $9$ out of $13$ LECs $s_i$ that contribute to the
subleading short-range 3NF can be fixed from elastic Nd scattering
data. The LECs $s_{2,4}$ can possibly be constrained by experimental data for Nd breakup
observables, while the determination of the isospin-$3/2$ LECs
$s_{12,13}$ will require going beyond the three-nucleon system.
The new measurement system developed at the RIKEN RI Beam Factory
\cite{Saito:2026dzr} is
expected to further improve the situation with elastic scattering data
at intermediate energies in the near future. 

In the
next step, a {\it complete}
N$^3$LO analysis of Nd scattering should be performed by taking into
account the leading loop contributions to the 3NF, calculated using the
symmetry-preserving gradient-flow regularization method \cite{Krebs:2023ljo,Krebs:2023gge}. Their
partial-wave decomposition and numerical implementation are in progress
by the LENPIC Collaboration. At this chiral order, one will have to
include the (formally enhanced) linear combinations of $S_i$  and
$F_i$ \cite{Huesmann:2026khj} 3NF terms to account for the 
choice made in Ref.~\cite{Reinert:2017usi} regarding the off-shell
behavior of the short-range NN
potential, see Refs.~\cite{Girlanda:2020pqn,Huesmann:2026khj} for details. The determination of $3$
linear combinations of $s_i$'s can be carried out using the methodology
presented in this paper, while fixing the  enhanced $c_D$-like interactions 
can be done using a more general emulation technique proposed in
Ref.~\cite{Heihoff:2026ycq}.  

\acknowledgments
We appreciate useful discussions with Hermann Krebs, Yuko Saito, Kimiko Sekiguchi,
and Patrick Walkowiak. We are also grateful to all members
of the LENPIC Collaboration  for sharing their insights into the
considered topics. 
This work has been supported by the European Research Council (ERC)
under the European Union's Horizon 2020 research and innovation
programme (grant agreement No.~885150), by the MKW NRW under the
funding code NW21-024-A, by JST ERATO (Grant No. JPMJER2304), by JSPS
KAKENHI (Grant No. JP20H05636) and by BMBF through the ErUM-Data project DEMOS.

\newpage
\onecolumngrid
\appendix
\renewcommand{\theequation}{\thesection.\arabic{equation}}
                
		\section{Partial wave decomposition of the subleading
                  contact 3NF}\label{Appendix:PWD}

                In the following, we provide explicit expressions for
                partial wave decomposition of the subleading contact
                3NFs defined in Eq.~(\ref{DefEi}). We choose the Faddeev components to be invariant with respect to the interchange of nucleons $2$ and $3$ and we use the partial-wave basis states
\beq
\ket{pq\alpha} \equiv \bigg| pq (ls)j \bigg( \lambda \frac{1}{2} \bigg) I (jI) J \bigg( t \frac{1}{2} \bigg) T \bigg\rangle,
\eeq
where $\boldsymbol{p}$ ($\boldsymbol{p}'$) and $\boldsymbol{q}$ ($\boldsymbol{q}'$) denote the initial (final) Jacobi momenta, $s$, $l$ and $j$ refer to the spin, orbital momentum and total angular momentum of the two-nucleon subsystem, $\lambda$ and $I$ are the orbital and total angular momenta of the spectator nucleon relative to the center-of-mass of the two-nucleon subsystem, while $J$ is the total angular momentum. Furthermore, $t$ is the isospin quantum number of the two-nucleon subsystem, while $T$ denotes the total isospin.  
Furthermore, for the sake of compactness, we use the notation with  $\hat{X} \equiv 2X+1$ and introduce
\beq
\{a,b,c\} \; \equiv \; \bigg\{ \begin{array}{l} 1 \quad \text{if} \quad |a-b|<c<a+b\,, \\[4pt]
                                 0 \quad \text{otherwise}.
\end{array}                                 
\eeq                
The result of each partial wave decomposition should be multiplied with the trivial factor of $\delta_{T T'} \delta_{m_T m_T'} \delta_{J J'} \delta_{m_J m_J'}$, which is not shown explicitly to keep the formulas more compact. 
\begin{itemize}
  \setlength{\itemsep}{0pt}
\item
  For the $E_1$-term
\begin{equation}
    V_{E_1} = \sum_{i\neq j \neq k} \boldsymbol{q}_i^2,
\end{equation}
we choose the Faddeev component
\begin{equation}
    V_{E_1}^{(1)} = 2 \boldsymbol{q}_1^2
\end{equation}
and obtain
\beqa
\mel{p' q' \alpha'}{V_{E_1}^{(1)}}{p q \alpha}  
&=&\frac{32}{3} \pi ^2   \delta _{l 0} \delta _{l' 0}  \delta _{j' j} \delta _{s j} \delta _{s' j} \delta _{t t'} \left\{\frac{1}{2},\frac{1}{2},j\right\} \left\{\frac{1}{2},\frac{1}{2},t'\right\} \left\{\frac{1}{2},t',T\right\}
\nn
&\times & \left(3 \left(q^2+q'^2\right) \delta _{I \frac{1}{2}} \delta _{I' \frac{1}{2}} \delta _{\lambda 0} \delta _{\lambda' 0} \left\{\frac{1}{2},j,J\right\}-2 q q'  \delta _{\lambda 1} \delta _{\lambda' 1} \delta _{I I'} \left\{\frac{1}{2},1,I'\right\} \left\{I',j,J\right\}\right) .
\eeqa
\item
For the $E_2$-term
\begin{equation}
    V_{E_2} = \sum_{i\neq j \neq k} \boldsymbol{q}_i^2 \boldsymbol{\tau}_i \cdot \boldsymbol{\tau}_j,
\end{equation}
we choose the Faddeev component
\begin{equation}
    V_{E_2}^{(1)} = \boldsymbol{q}_1^2 (\boldsymbol{\tau}_1 \cdot \boldsymbol{\tau}_2 + \boldsymbol{\tau}_1 \cdot \boldsymbol{\tau}_3)
\end{equation}
to obtain
\beqa
&&\mel{p' q' \alpha'}{V_{E_2}^{(1)}}{p q \alpha}  
\; =\;  32 \pi ^2 \hat{t} \hat{t'} (-1)^{T+\frac{1}{2}} \left((-1)^{t+t'}+ 1 \right) \delta _{l 0} \delta _{l' 0} \delta _{j' j} \delta _{s j} \delta _{s' j} \left\{\frac{1}{2},\frac{1}{2},j\right\} \begin{Bmatrix}1 & t & t' \\ \frac{1}{2} & \frac{1}{2} & \frac{1}{2}\end{Bmatrix} \nn
&& \hspace{0.7cm} {} \times  \begin{Bmatrix}t' & T & \frac{1}{2} \\ \frac{1}{2} & 1 & t\end{Bmatrix} \left(3 \left(q^2+q'^2\right) \delta _{I \frac{1}{2}} \delta _{I' \frac{1}{2}} \delta _{\lambda 0} \delta _{\lambda' 0} \left\{\frac{1}{2},j,J\right\}-2 q q'  \delta _{\lambda 1} \delta _{\lambda' 1} \delta _{I I'} \left\{\frac{1}{2},1,I'\right\} \left\{I',j,J\right\}\right) .
\eeqa
\item
For the $E_3$-term
\begin{equation}
    V_{E_3} = \sum_{i\neq j \neq k} \boldsymbol{q}_i^2 \boldsymbol{\sigma}_i \cdot \boldsymbol{\sigma}_j,
\end{equation}
we choose the Faddeev component
\begin{equation}
    V_{E_3}^{(1)} = \boldsymbol{q}_1^2 (\boldsymbol{\sigma}_1 \cdot \boldsymbol{\sigma}_2 + \boldsymbol{\sigma}_1 \cdot \boldsymbol{\sigma}_3)
\end{equation}
and obtain
\beqa
&& \hspace{-0.5cm} \mel{p' q' \alpha'}{V_{E_3}^{(1)}}{p q \alpha}  
\; =\; 32 \pi ^2 \hat{j} \hat{j'} (-1)^{J+\frac{1}{2}} \left((-1)^{j+j'} +1 \right) \delta _{l 0} \delta _{l' 0}  \delta _{s j} \delta _{s' j'} \delta _{t t'} \left\{\frac{1}{2},\frac{1}{2},t'\right\} \left\{\frac{1}{2},t',T\right\} \begin{Bmatrix}1 & j & j' \\ \frac{1}{2} & \frac{1}{2} & \frac{1}{2} \end{Bmatrix} \\
&& \hspace{2.25cm} {} \times \left(3 \left(q^2+q'^2\right)  \delta _{I \frac{1}{2}} \delta _{I' \frac{1}{2}} \delta _{\lambda 0} \delta _{\lambda' 0}   \begin{Bmatrix}j' & J & \frac{1}{2} \\ \frac{1}{2} & 1 & j\end{Bmatrix}  +2 q q' \hat{I} \hat{I'} \delta _{\lambda 1} \delta _{\lambda' 1}  \begin{Bmatrix}I & I' & 1 \\ \frac{1}{2} & \frac{1}{2} & 1\end{Bmatrix} \begin{Bmatrix}j' & I' & J \\ I & j & 1\end{Bmatrix}\right) .
\nonumber
\eeqa
\item
For the $E_4$-term
\begin{equation}
    V_{E_4} = \sum_{i\neq j \neq k} \boldsymbol{q}_i^2 \boldsymbol{\sigma}_i \cdot \boldsymbol{\sigma}_j \boldsymbol{\tau}_i \cdot \boldsymbol{\tau}_j\,,
\end{equation}
we choose the Faddeev component
\begin{equation}
    V_{E_4}^{(1)} = \boldsymbol{q}_1^2 (\boldsymbol{\sigma}_1 \cdot \boldsymbol{\sigma}_2 \boldsymbol{\tau}_1 \cdot \boldsymbol{\tau}_2 + \boldsymbol{\sigma}_1 \cdot \boldsymbol{\sigma}_3 \boldsymbol{\tau}_1 \cdot \boldsymbol{\tau}_3)
\end{equation}
to get 
\beqa
&&\mel{p' q' \alpha'}{V_{E_4}^{(1)}}{p q \alpha}  
\;=\; 192 \pi ^2 \hat{j} \hat{j'} \hat{t} \hat{t'} (-1)^{J+T+1} \left((-1)^{j+j'+t+t'}+1\right) \delta _{l 0} \delta _{l' 0} \delta _{s j} \delta _{s' j'}
\\
&& \hspace{0.7cm} {} \times
 \begin{Bmatrix}1 & j & j' \\ \frac{1}{2} & \frac{1}{2} & \frac{1}{2}\end{Bmatrix}  \begin{Bmatrix}1 & t & t' \\ \frac{1}{2} & \frac{1}{2} & \frac{1}{2}\end{Bmatrix} \begin{Bmatrix}t' & T & \frac{1}{2} \\ \frac{1}{2} & 1 & t\end{Bmatrix} \nn
&& \hspace{0.7cm} {} \times \left(3 \left(q^2+q'^2\right) \delta _{I \frac{1}{2}} \delta _{I' \frac{1}{2}} \delta _{\lambda 0} \delta _{\lambda' 0}  \begin{Bmatrix}j' & J & \frac{1}{2} \\ \frac{1}{2} & 1 & j\end{Bmatrix}+2 q q' \hat{I} \hat{I'} \delta _{\lambda 1} \delta _{\lambda' 1} \begin{Bmatrix}I & I' & 1 \\ \frac{1}{2} & \frac{1}{2} & 1\end{Bmatrix}  \begin{Bmatrix}j' & I' & J \\ I & j & 1\end{Bmatrix}\right) .
\nonumber
\eeqa
\item
For the $E_5$-term
\begin{equation}
    V_{E_5} = \sum_{i\neq j \neq k} 3 \boldsymbol{q}_i \cdot \boldsymbol{\sigma}_i \boldsymbol{q}_i \cdot \boldsymbol{\sigma}_j - \boldsymbol{q}_i^2 \boldsymbol{\sigma}_i \cdot \boldsymbol{\sigma}_j \,,
\end{equation}
we choose the Faddeev component
\begin{equation}
    V_{E_5}^{(1)} = 3 \boldsymbol{q}_1 \cdot \boldsymbol{\sigma}_1 (\boldsymbol{q}_1 \cdot \boldsymbol{\sigma}_2 + \boldsymbol{q}_1 \cdot \boldsymbol{\sigma}_3) - \boldsymbol{q}_1^2 (\boldsymbol{\sigma}_1 \cdot \boldsymbol{\sigma}_2 + \boldsymbol{\sigma}_1 \cdot \boldsymbol{\sigma}_3 )
\end{equation}
to get 
\beqa
 &&\mel{p' q' \alpha'}{V_{E_5}^{(1)}}{p q \alpha}  \;
     =\; 16 \pi ^2 \hat{j} \hat{j'} (-1)^{J}  \left((-1)^{j+j'}+1\right) \delta _{l 0} \delta _{l' 0}  \delta _{s j} \delta _{s' j'} \delta _{t t'} \left\{\frac{1}{2},\frac{1}{2},t'\right\}  \left\{\frac{1}{2},t',T\right\} \begin{Bmatrix}1 & j & j' \\ \frac{1}{2} & \frac{1}{2} & \frac{1}{2}\end{Bmatrix}
     \nn
&& \hspace{0.7cm} {} \times 
      \left(3 \delta _{I \frac{1}{2}} \left[-2 \delta _{\lambda 0} \left(3 q'^2 \hat{I'} \hat{\lambda'} (-1)^{I'+ \lambda'}  \begin{pmatrix}1 & 1 & \lambda' \\ 0 & 0 & 0\end{pmatrix} \begin{Bmatrix}1 & \frac{1}{2} & I' \\ \frac{1}{2} & \lambda' & 1\end{Bmatrix}  \begin{Bmatrix}j' & I' & J \\ \frac{1}{2} & j & 1\end{Bmatrix} \right. \right. \right.
     \nn
&& \hspace{0.7cm} {}  
     +  \left. \left. \left. i \left(q^2+q'^2\right) \delta _{I' \frac{1}{2}} \delta _{\lambda' 0}  \begin{Bmatrix}j' & J & \frac{1}{2} \\ \frac{1}{2} & 1 & j\end{Bmatrix}\right)-i \sqrt{2} q q' \hat{I'} \delta _{\lambda 1} \delta _{\lambda' 1}  \begin{Bmatrix}j & J & \frac{1}{2} \\ I' & 1 & j'\end{Bmatrix}\right] \right.
     \nn
&& \hspace{0.7cm} {} 
      + \left. q \hat{I} \left[-18 q \hat{\lambda} (-1)^{I}  \delta _{I' \frac{1}{2}} \delta _{\lambda' 0} \begin{pmatrix}1 & 1 & \lambda \\ 0 & 0 & 0\end{pmatrix} \begin{Bmatrix}1 & I & \frac{1}{2} \\ \frac{1}{2} & 1 & \lambda\end{Bmatrix}  \begin{Bmatrix}j' & J & \frac{1}{2} \\ I & 1 & j\end{Bmatrix} \right. \right.
     \nn
&& \hspace{0.7cm} {} 
    + \left. \left. q' \delta _{\lambda 1} \delta _{\lambda' 1} \left(-3 i \sqrt{2} \delta _{I' \frac{1}{2}}  \begin{Bmatrix}j' & J & \frac{1}{2} \\ I & 1 & j\end{Bmatrix}-4 i \hat{I'}  \begin{Bmatrix}I & I' & 1 \\ \frac{1}{2} & \frac{1}{2} & 1\end{Bmatrix}  \begin{Bmatrix}j' & I' & J \\ I & j & 1\end{Bmatrix}\right)\right]\right) .
\eeqa
\item
 For the $E_6$-term
\begin{equation}
    V_{E_6} = \sum_{i\neq j \neq k} (3 \boldsymbol{q}_i \cdot \boldsymbol{\sigma}_i \boldsymbol{q}_i \cdot \boldsymbol{\sigma}_j - \boldsymbol{q}_i^2 \boldsymbol{\sigma}_i \cdot \boldsymbol{\sigma}_j ) \boldsymbol{\tau}_i \cdot \boldsymbol{\tau}_j\,,
\end{equation}
we choose the Faddeev component
\begin{equation}
    V_{E_6}^{(1)} = (3 \boldsymbol{q}_1 \cdot \boldsymbol{\sigma}_1 \boldsymbol{q}_1 \cdot \boldsymbol{\sigma}_2  - \boldsymbol{q}_1^2 \boldsymbol{\sigma}_1 \cdot \boldsymbol{\sigma}_2 ) \boldsymbol{\tau}_1 \cdot \boldsymbol{\tau}_2 + (3 \boldsymbol{q}_1 \cdot \boldsymbol{\sigma}_1 \boldsymbol{q}_1 \cdot \boldsymbol{\sigma}_3  - \boldsymbol{q}_1^2 \boldsymbol{\sigma}_1 \cdot \boldsymbol{\sigma}_3 ) \boldsymbol{\tau}_1 \cdot \boldsymbol{\tau}_3
\end{equation}
and obtain
\beqa
&&\mel{p' q' \alpha'}{V_{E_6}^{(1)}}{p q \alpha}  \;
=\;  96 \pi ^2 \hat{j} \hat{j'} \hat{t} \hat{t'} (-1)^{J+T} \left((-1)^{j+j'+t+t'}+1\right) \delta _{l 0} \delta _{l' 0} \delta _{s j} \delta _{s' j'}    \begin{Bmatrix}1 & j & j' \\ \frac{1}{2} & \frac{1}{2} & \frac{1}{2}\end{Bmatrix}
\nn
&& \hspace{0.7cm} {} \times
\begin{Bmatrix}1 & t & t' \\ \frac{1}{2} & \frac{1}{2} & \frac{1}{2}\end{Bmatrix} \begin{Bmatrix}t' & T & \frac{1}{2} \\ \frac{1}{2} & 1 & t\end{Bmatrix}
    \left[3 \delta _{I \frac{1}{2}} \left(2 \delta _{\lambda 0} \left[\left(q^2+q'^2\right) \delta _{I' \frac{1}{2}} \delta _{\lambda' 0}  \begin{Bmatrix}j' & J & \frac{1}{2} \\ \frac{1}{2} & 1 & j\end{Bmatrix} \right. \right. \right.
    \nn
&& \hspace{0.7cm} {} 
- \left. \left. \left. 3 q'^2 \hat{I'} \hat{\lambda'} (-1)^{I'+ \lambda'+\frac{1}{2}}  \begin{pmatrix}1 & 1 & \lambda' \\ 0 & 0 & 0\end{pmatrix} \begin{Bmatrix}1 & \frac{1}{2} & I' \\ \frac{1}{2} & \lambda' & 1\end{Bmatrix}  \begin{Bmatrix}j' & I' & J \\ \frac{1}{2} & j & 1\end{Bmatrix}\right]
    +\sqrt{2} q q' \hat{I'} \delta _{\lambda 1} \delta _{\lambda' 1}  \begin{Bmatrix}j & J & \frac{1}{2} \\ I' & 1 & j'\end{Bmatrix}\right) \right.
    \nn
&& \hspace{0.7cm} {} 
    + \left. q \hat{I} \left(q' \delta _{\lambda 1} \delta _{\lambda' 1} \left(3 \sqrt{2} \delta _{I' \frac{1}{2}}  \begin{Bmatrix}j' & J & \frac{1}{2} \\ I & 1 & j\end{Bmatrix}+4 \hat{I'}  \begin{Bmatrix}I & I' & 1 \\ \frac{1}{2} & \frac{1}{2} & 1\end{Bmatrix}  \begin{Bmatrix}j' & I' & J \\ I & j & 1\end{Bmatrix}\right) \right. \right.
    \nn
&& \hspace{0.7cm} {} 
    - \left. \left. 18 q (-1)^{I+\frac{1}{2}} \hat{\lambda} \delta _{I' \frac{1}{2}} \delta _{\lambda' 0}  \begin{pmatrix}1 & 1 & \lambda \\ 0 & 0 & 0\end{pmatrix} \begin{Bmatrix}1 & I & \frac{1}{2} \\ \frac{1}{2} & 1 & \lambda\end{Bmatrix}  \begin{Bmatrix}j' & J & \frac{1}{2} \\ I & 1 & j\end{Bmatrix}\right)\right] .
\eeqa
\item
For the $E_7$-term
\begin{equation}
    V_{E_7} = i \sum_{i\neq j \neq k} \boldsymbol{q}_i \times (\boldsymbol{k}_i - \boldsymbol{k}_j) \cdot (\boldsymbol{\sigma}_i + \boldsymbol{\sigma}_j)\,,
\end{equation}
we choose the Faddeev component
\begin{equation}
    V_{E_7}^{(1)} = i ( \boldsymbol{q}_1 \times (\boldsymbol{k}_1 - \boldsymbol{k}_2) \cdot (\boldsymbol{\sigma}_1 + \boldsymbol{\sigma}_2) + \boldsymbol{q}_1 \times (\boldsymbol{k}_1 - \boldsymbol{k}_3) \cdot (\boldsymbol{\sigma}_1 + \boldsymbol{\sigma}_3) )
\end{equation}
and obtain
\beqa
&&\mel{p' q' \alpha'}{V_{E_7}^{(1)}}{p q \alpha}  \;
=\; -16 \pi ^2  \delta _{t t'} \left\{\frac{1}{2},\frac{1}{2},t'\right\} \left\{\frac{1}{2},t',T\right\} \left((-1)^{J+\frac{1}{2}} \hat{j} \hat{j'} \left[p \hat{s} \left((-1)^{j'+s}-1\right) \delta _{l 1} \delta _{l' 0} \delta _{s' j'}   \right. \right.
\nn
&& \hspace{0.7cm} {} \times
\begin{Bmatrix}s & j' & 1 \\ \frac{1}{2} & \frac{1}{2} & \frac{1}{2}\end{Bmatrix}
\begin{Bmatrix}j & j' & 1 \\ 1 & 1 & s\end{Bmatrix} 
\left. \left. \left(q' \hat{I'} \delta _{I \frac{1}{2}} \delta _{\lambda 0} \delta _{\lambda' 1}  \begin{Bmatrix}j' & I' & J \\ \frac{1}{2} & j & 1\end{Bmatrix}-q \hat{I} \delta _{I' \frac{1}{2}} \delta _{\lambda 1} \delta _{\lambda' 0}  \begin{Bmatrix}j' & J & \frac{1}{2} \\ I & 1 & j\end{Bmatrix}\right) \right. \right.
\nn
&& \hspace{0.7cm} {} 
 \left. \left.  +\delta _{l 0} \delta _{s j} \left(3 q q' \hat{I} \hat{I'} (-1)^{I+I'} \left((-1)^{j+j'}+1\right) \delta _{\lambda 1} \delta _{\lambda' 1} \delta _{l' 0} \delta _{s' j'}  \begin{Bmatrix}1 & I' & I \\ \frac{1}{2} & 1 & 1\end{Bmatrix} \begin{Bmatrix}1 & j & j' \\ \frac{1}{2} & \frac{1}{2} & \frac{1}{2}\end{Bmatrix} \begin{Bmatrix}J & I' & j' \\ 1 & j & I\end{Bmatrix} \right. \right. \right.
 \nn
 && \hspace{0.7cm} {}
 \left. \left. \left. -p' \hat{s'} \left((-1)^{j+s'}-1\right) \delta _{l' 1} \begin{Bmatrix}j & s' & 1 \\ \frac{1}{2} & \frac{1}{2} & \frac{1}{2}\end{Bmatrix} \begin{Bmatrix}j' & j & 1 \\ 1 & 1 & s'\end{Bmatrix} \left[q' \hat{I'} \delta _{I \frac{1}{2}} \delta _{\lambda 0} \delta _{\lambda' 1}  \begin{Bmatrix}j' & I' & J \\ \frac{1}{2} & j & 1\end{Bmatrix} \right.\right.\right.\right.
 \nn
&& \hspace{0.7cm} {} 
  \left.\left.\left.\left. -q \hat{I} \delta _{I' \frac{1}{2}} \delta _{\lambda 1} \delta _{\lambda' 0}  \begin{Bmatrix}j' & J & \frac{1}{2} \\ I & 1 & j\end{Bmatrix}\right]\right)\right] \right.
  \nn
  && \hspace{0.7cm} {} 
   \left. +6 q q' (-1)^{I'+\frac{3}{2}} \delta _{\lambda 1} \delta _{\lambda' 1} \delta _{l 0} \delta _{l' 0} \delta _{I I'} \delta _{j' j} \delta _{s j} \delta _{s' j} \left\{I',j,J\right\}  \left\{\frac{1}{2},\frac{1}{2},j\right\} \begin{Bmatrix}I' & 1 & \frac{1}{2} \\ 1 & \frac{1}{2} & 1\end{Bmatrix}  \right) .
\eeqa
\item
For the $E_8$-term
\begin{equation}
    V_{E_8} = i \sum_{i\neq j \neq k} \boldsymbol{q}_i \times (\boldsymbol{k}_i - \boldsymbol{k}_j) \cdot (\boldsymbol{\sigma}_i + \boldsymbol{\sigma}_j) \boldsymbol{\tau}_j \cdot \boldsymbol{\tau}_k\,,
\end{equation}
we choose the Faddeev component
\begin{equation}
    V_{E_8}^{(1)} = i ( \boldsymbol{q}_1 \times (\boldsymbol{k}_1 - \boldsymbol{k}_2) \cdot (\boldsymbol{\sigma}_1 + \boldsymbol{\sigma}_2) + \boldsymbol{q}_1 \times (\boldsymbol{k}_1 - \boldsymbol{k}_3) \cdot (\boldsymbol{\sigma}_1 + \boldsymbol{\sigma}_3) )  \boldsymbol{\tau}_2 \cdot \boldsymbol{\tau}_3
\end{equation}
to obtain
\beqa
&&\mel{p' q' \alpha'}{V_{E_8}^{(1)}}{p q \alpha}  
    \; = \; 96 \pi ^2 (-1)^{t'} \delta _{t t'}   \left\{\frac{1}{2},t',T\right\}\begin{Bmatrix} 1/2 & 1/2 & t' \\ 1/2 & 1/2 & 1\end{Bmatrix}  \left((-1)^{J+\frac{1}{2}} \hat{j} \hat{j'} \left[p \hat{s} \left((-1)^{j'+s}-1\right) \delta _{l 1} \delta _{l' 0} \delta _{s' j'} \phantom{\frac{0}{0}} \right. \right.
    \nn
    && \hspace{0.7cm} {} \times
    \begin{Bmatrix}s & j' & 1 \\ \frac{1}{2} & \frac{1}{2} & \frac{1}{2}\end{Bmatrix}  \begin{Bmatrix}j & j' & 1 \\ 1 & 1 & s\end{Bmatrix}
     \left. \left. \left(q' \hat{I'} \delta _{I \frac{1}{2}} \delta _{\lambda 0} \delta _{\lambda' 1}  \begin{Bmatrix}j' & I' & J \\ \frac{1}{2} & j & 1\end{Bmatrix}-q \hat{I} \delta _{I' \frac{1}{2}} \delta _{\lambda 1} \delta _{\lambda' 0}  \begin{Bmatrix}j' & J & \frac{1}{2} \\ I & 1 & j\end{Bmatrix}\right) \right. \right.
     \nn
  && \hspace{0.7cm} {} 
      \left.\left. +\delta _{l 0} \delta _{s j} \left(3 q q' \hat{I} \hat{I'} (-1)^{I+I'} \left((-1)^{j+j'}+1\right) \delta _{\lambda 1} \delta _{\lambda' 1} \delta _{l' 0} \delta _{s' j'} \begin{Bmatrix}1 & I' & I \\ \frac{1}{2} & 1 & 1\end{Bmatrix} \begin{Bmatrix}1 & j & j' \\ \frac{1}{2} & \frac{1}{2} & \frac{1}{2}\end{Bmatrix}  \begin{Bmatrix}J & I' & j' \\ 1 & j & I\end{Bmatrix} \right. \right. \right.
      \nn
  && \hspace{0.7cm} {} 
       \left. \left. \left.  -p' \hat{s'} \left((-1)^{j+s'}-1\right) \delta _{l' 1} \begin{Bmatrix}j & s' & 1 \\ \frac{1}{2} & \frac{1}{2} & \frac{1}{2}\end{Bmatrix} \begin{Bmatrix}j' & j & 1 \\ 1 & 1 & s'\end{Bmatrix} \left[q' \hat{I'} \delta _{I \frac{1}{2}} \delta _{\lambda 0} \delta _{\lambda' 1}  \begin{Bmatrix}j' & I' & J \\ \frac{1}{2} & j & 1\end{Bmatrix} \right.\right.\right.\right.
       \nn
  && \hspace{0.7cm} {} 
        \left.\left.\left.\left.   -q \hat{I} \delta _{I' \frac{1}{2}} \delta _{\lambda 1} \delta _{\lambda' 0} \left\{\frac{1}{2},j',J\right\} \begin{Bmatrix}j' & J & \frac{1}{2} \\ I & 1 & j\end{Bmatrix}\right]\right)\right] \right.
        \nn
  && \hspace{0.7cm} {} 
         \left. +6 q q' (-1)^{I'+\frac{3}{2}} \delta _{\lambda 1} \delta _{\lambda' 1}  \delta _{l 0} \delta _{l' 0} \delta _{I I'} \delta _{j' j} \delta _{s j} \delta _{s' j} \left\{I',j,J\right\}  \left\{\frac{1}{2},\frac{1}{2},j\right\} \begin{Bmatrix}I' & 1 & \frac{1}{2} \\ 1 & \frac{1}{2} & 1\end{Bmatrix}   \right) .
\eeqa
\item
For the $E_9$-term
\begin{equation}
    V_{E_9} = \sum_{i\neq j \neq k}  \boldsymbol{q}_i \cdot \boldsymbol{\sigma}_i \boldsymbol{q}_j \cdot \boldsymbol{\sigma}_j\,,
\end{equation}
we choose the Faddeev component
\begin{equation}
    V_{E_9}^{(1)} = \boldsymbol{q}_1 \cdot \boldsymbol{\sigma}_1 (\boldsymbol{q}_2 \cdot \boldsymbol{\sigma}_2 + \boldsymbol{q}_3 \cdot \boldsymbol{\sigma}_3)
\end{equation}
and obtain
\beqa
&&\mel{p' q' \alpha'}{V_{E_9}^{(1)}}{p q \alpha}  \;
=\;-8 \pi ^2 \delta _{t t'} \left\{\frac{1}{2},\frac{1}{2},t'\right\}\nn
 && \hspace{0.7cm} {} \times
\left\{\frac{1}{2},t',T\right\} \bigg(\delta _{I \frac{1}{2}} \bigg[-2 \sqrt{2} p \hat{s} \left((-1)^{j}-(-1)^{s}\right) \delta _{I' \frac{1}{2}} \delta _{l 1} \delta _{l' 0} \delta _{j' j} \delta _{s' j} \left(q \delta _{\lambda 1} \delta _{\lambda' 0}-q' \delta _{\lambda 0} \delta _{\lambda' 1}\right)  
     \nn
 && \hspace{0.7cm} {} \times
     \left. \left. \left\{\frac{1}{2},j,J\right\}   \begin{Bmatrix}s & j & 1 \\ \frac{1}{2} & \frac{1}{2} & \frac{1}{2}\end{Bmatrix} +\delta _{l 0} \delta _{s j} \left(2 \sqrt{2} p' \hat{s'} \left((-1)^{j}-(-1)^{s'}\right) \delta _{I' \frac{1}{2}} \delta _{l' 1} \delta _{j' j} \left(q \delta _{\lambda 1} \delta _{\lambda' 0}-q' \delta _{\lambda 0} \delta _{\lambda' 1}\right)   \left\{\frac{1}{2},j,J\right\} \right. \right.\right.
     \nn
     && \hspace{0.7cm} {}\times
      \left.\left.\left. \begin{Bmatrix}j & s' & 1 \\ \frac{1}{2} & \frac{1}{2} & \frac{1}{2}\end{Bmatrix} +q' \hat{I'} (-1)^{J} \hat{j} \hat{j'} \left((-1)^{j+j'}+1\right) \delta _{l' 0} \delta _{s' j'}  \begin{Bmatrix}1 & j & j' \\ \frac{1}{2} & \frac{1}{2} & \frac{1}{2}\end{Bmatrix} \left[-6 q' \hat{\lambda'} (-1)^{I'+ \lambda'} \delta _{\lambda 0}  \begin{pmatrix}1 & 1 & \lambda' \\ 0 & 0 & 0\end{pmatrix} \right. \right.\right. \right.
      \nn
      && \hspace{0.7cm} {}\times
      \left.\left.\left. \begin{Bmatrix}1 & \frac{1}{2} & I' \\ \frac{1}{2} & \lambda' & 1\end{Bmatrix}  \begin{Bmatrix}j' & I' & J \\ \frac{1}{2} & j & 1\end{Bmatrix}-i \sqrt{2} q \delta _{\lambda 1} \delta _{\lambda' 1}  \begin{Bmatrix}j & J & \frac{1}{2} \\ I' & 1 & j'\end{Bmatrix}\right]\right)\bigg]+q \hat{I} (-1)^{J} \hat{j} \hat{j'} \left((-1)^{j+j'}+1\right) \delta _{I' \frac{1}{2}} \delta _{l 0} \delta _{l' 0} \right.
      \nn
 && \hspace{0.7cm} {} \times
        \delta _{s j} \delta _{s' j'} \begin{Bmatrix}1 & j & j' \\ \frac{1}{2} & \frac{1}{2} & \frac{1}{2}\end{Bmatrix}  \begin{Bmatrix}j' & J & \frac{1}{2} \\ I & 1 & j\end{Bmatrix} \left(-6 q (-1)^{I} \hat{\lambda} \delta _{\lambda' 0} \begin{pmatrix}1 & 1 & \lambda \\ 0 & 0 & 0\end{pmatrix} \begin{Bmatrix}1 & I & \frac{1}{2} \\ \frac{1}{2} & 1 & \lambda\end{Bmatrix}-i \sqrt{2} q' \delta _{\lambda 1} \delta _{\lambda' 1}\right)\bigg) .
\eeqa
\item
For the $E_{10}$-term
\begin{equation}
    V_{E_{10}} = \sum_{i\neq j \neq k}  \boldsymbol{q}_i \cdot \boldsymbol{\sigma}_i \boldsymbol{q}_j \cdot \boldsymbol{\sigma}_j \boldsymbol{\tau}_i \cdot \boldsymbol{\tau}_j\,,
\end{equation}
we choose the Faddeev component
\begin{equation}
    V_{E_{10}}^{(1)} = \boldsymbol{q}_1 \cdot \boldsymbol{\sigma}_1 (\boldsymbol{q}_2 \cdot \boldsymbol{\sigma}_2 \boldsymbol{\tau}_1 \cdot \boldsymbol{\tau}_2 + \boldsymbol{q}_3 \cdot \boldsymbol{\sigma}_3 \boldsymbol{\tau}_1 \cdot \boldsymbol{\tau}_3)
\end{equation}
and obtain
\beqa
&& \mel{p' q' \alpha'}{V_{E_{10}}^{(1)}}{p q \alpha} 
\; =\;-48 \pi ^2 \hat{t} \hat{t'} (-1)^{T} \nn
&& \hspace{0.7cm} {} \times
 \begin{Bmatrix}1 & t & t' \\ \frac{1}{2} & \frac{1}{2} & \frac{1}{2}\end{Bmatrix}  \begin{Bmatrix}t' & T & \frac{1}{2} \\ \frac{1}{2} & 1 & t\end{Bmatrix} \left(q \hat{I} (-1)^{J} \hat{j} \hat{j'} \left((-1)^{j+j'+t+t'}+1\right) \delta _{I' \frac{1}{2}} \delta _{l 0} \delta _{l' 0} \delta _{s j} \delta _{s' j'}  \right.
   \nn
&& \hspace{0.7cm} {} \times   
   \left. \begin{Bmatrix}1 & j & j' \\ \frac{1}{2} & \frac{1}{2} & \frac{1}{2}\end{Bmatrix}   \begin{Bmatrix}j' & J & \frac{1}{2} \\ I & 1 & j\end{Bmatrix} \left(\sqrt{2} q' \delta _{\lambda 1} \delta _{\lambda' 1}-6 q (-1)^{I+\frac{1}{2}} \hat{\lambda} \delta _{\lambda' 0} \begin{pmatrix}1 & 1 & \lambda \\ 0 & 0 & 0\end{pmatrix} \begin{Bmatrix}1 & I & \frac{1}{2} \\ \frac{1}{2} & 1 & \lambda\end{Bmatrix}\right) \right.
   \nn
 && \hspace{0.7cm} {}   
   +  \left.\delta _{I \frac{1}{2}} \left[\delta _{l 0} \delta _{s j} \left(q' \hat{I'} (-1)^{J} \hat{j} \hat{j'} \delta _{l' 0} \delta _{s' j'} \left((-1)^{j+j'+t+t'}+1\right)  \begin{Bmatrix}1 & j & j' \\ \frac{1}{2} & \frac{1}{2} & \frac{1}{2}\end{Bmatrix}  \left[\sqrt{2} q \delta _{\lambda 1} \delta _{\lambda' 1}  \begin{Bmatrix}j & J & \frac{1}{2} \\ I' & 1 & j'\end{Bmatrix} \right.\right.\right.\right.
\nn
&& \hspace{0.7cm} {} 
- \left.\left.\left.\left. 6 q' \hat{\lambda'} (-1)^{I'+ \lambda'+\frac{1}{2}} \delta _{\lambda 0} \begin{pmatrix}1 & 1 & \lambda' \\ 0 & 0 & 0\end{pmatrix} \begin{Bmatrix}1 & \frac{1}{2} & I' \\ \frac{1}{2} & \lambda' & 1\end{Bmatrix}  \begin{Bmatrix}j' & I' & J \\ \frac{1}{2} & j & 1\end{Bmatrix}\right] \right.\right.\right.
\\
&& \hspace{0.7cm} {} 
      +2 i \sqrt{2} p' \hat{s'} \delta _{I' \frac{1}{2}} \delta _{l' 1}  \delta _{j' j} \left\{\frac{1}{2},j,J\right\} 
      \left.\left.\left.  \left((-1)^{j}-(-1)^{s'+t+t'}\right) \left(q \delta _{\lambda 1} \delta _{\lambda' 0}-q' \delta _{\lambda 0} \delta _{\lambda' 1}\right) \begin{Bmatrix}j & s' & 1 \\ \frac{1}{2} & \frac{1}{2} & \frac{1}{2}\end{Bmatrix} \right) \right.\right.
          \nn
  && \hspace{0.7cm} {}         
          -2 i \sqrt{2} p \hat{s} \delta _{I' \frac{1}{2}} \delta _{l 1} \delta _{l' 0} \delta _{j' j} \delta _{s' j} \left((-1)^{j}-(-1)^{s+t+t'}\right) 
          \left.\left.  \left(q \delta _{\lambda 1} \delta _{\lambda' 0}-q' \delta _{\lambda 0} \delta _{\lambda' 1}\right)   \left\{\frac{1}{2},j,J\right\} \begin{Bmatrix}s & j & 1 \\ \frac{1}{2} & \frac{1}{2} & \frac{1}{2}\end{Bmatrix} \right]\right) .
          \nonumber
\eeqa
\item
For the $E_{11}$-term
\begin{equation}
    V_{E_{11}} = \sum_{i\neq j \neq k}  \boldsymbol{q}_i \cdot \boldsymbol{\sigma}_j \boldsymbol{q}_j \cdot \boldsymbol{\sigma}_i\,,
\end{equation}
we choose the Faddeev component
\begin{equation}
    V_{E_{11}}^{(1)} = \boldsymbol{q}_1 \cdot \boldsymbol{\sigma}_2 \boldsymbol{q}_2 \cdot \boldsymbol{\sigma}_1 + \boldsymbol{q}_1 \cdot \boldsymbol{\sigma}_3 \boldsymbol{q}_3 \cdot \boldsymbol{\sigma}_1 
\end{equation}
and obtain
\beqa
&&\mel{p' q' \alpha'}{V_{E_{11}}^{(1)}}{p q \alpha}  \;
=\; -8 \pi ^2 \hat{j} \hat{j'} \delta _{t t'} \left\{\frac{1}{2},\frac{1}{2},t'\right\}  \left\{\frac{1}{2},t',T\right\}
\nn
 && \hspace{0.7cm} {}         \times
 \left(q' \hat{I'} \delta _{I \frac{1}{2}} \left[-4 p \hat{s} (-1)^{j}  \left((-1)^{j'+s}-1\right) \delta _{\lambda 0} \delta _{\lambda' 1} \delta _{l 1} \delta _{l' 0} \delta _{s' j'} \begin{Bmatrix}s & j' & 1 \\ \frac{1}{2} & \frac{1}{2} & \frac{1}{2}\end{Bmatrix} \right. \right.
  \begin{Bmatrix}s & J & \frac{1}{2} \\ \frac{1}{2} & 1 & j\end{Bmatrix} \begin{Bmatrix}s & J & \frac{1}{2} \\ I' & 1 & j'\end{Bmatrix}
\nn
 && \hspace{0.7cm} {}         
    \left.\left. +\delta _{l 0} \delta _{s j} \Bigg(4 p' \hat{s'} (-1)^{j'} \left((-1)^{j+s'}-1\right) \delta _{\lambda 0} \delta _{\lambda' 1} \delta _{l' 1}  \begin{Bmatrix}j & s' & 1 \\ \frac{1}{2} & \frac{1}{2} & \frac{1}{2}\end{Bmatrix} \begin{Bmatrix}1/2 & j & J \\ 1/2 & 1 & I' \\ 1 & s' & j'\end{Bmatrix}  \right. \right. 
\nn
 && \hspace{0.7cm} {}         
    +\left.\left.\left.(-1)^{J} \left((-1)^{j+j'}+1\right) \delta _{l' 0} \delta _{s' j'}\begin{Bmatrix}1 & j & j' \\ \frac{1}{2} & \frac{1}{2} & \frac{1}{2}\end{Bmatrix} \left[-6 q' \hat{\lambda'} (-1)^{I'+ \lambda'} \delta _{\lambda 0}  \begin{pmatrix}1 & 1 & \lambda' \\ 0 & 0 & 0\end{pmatrix} \begin{Bmatrix}1 & \frac{1}{2} & I' \\ \frac{1}{2} & \lambda' & 1\end{Bmatrix}  \right. \right.\right. \right.
\nn
 && \hspace{0.7cm} {}         \times
\begin{Bmatrix}j' & I' & J \\ \frac{1}{2} & j & 1\end{Bmatrix} 
 -\left.\left.\left.\left.i \sqrt{2} q \delta _{\lambda 1} \delta _{\lambda' 1}  \begin{Bmatrix}j & J & \frac{1}{2} \\ I' & 1 & j'\end{Bmatrix}\right]\right)\right.\Bigg]+q \hat{I} \delta _{I' \frac{1}{2}} \left[  4 p \hat{s} (-1)^{j'+s} \delta _{\lambda 1} \delta _{\lambda' 0} \delta _{l 1} \delta _{l' 0} \delta _{s' j'} \phantom{\frac{1}{2}} \right. \right.
\nn
&& \hspace{0.7cm} {}  \times
\left.\left. \left((-1)^{I+J}  +(-1)^{1+J+I+s+j'} \right)  \begin{Bmatrix}s & j' & 1 \\ \frac{1}{2} & \frac{1}{2} & \frac{1}{2}\end{Bmatrix} \begin{Bmatrix}1/2 & J & j' \\ 1/2 & I & 1 \\ 1 & j & s\end{Bmatrix} \right. \right.
\nn
&& \hspace{0.7cm} {}  
    +\delta _{l 0} \delta _{s j} \left[-4 p' \hat{s'} (-1)^{j'}  \left((-1)^{j+s'}-1\right) \delta _{\lambda 1} \delta _{\lambda' 0} \delta _{l' 1} \right. 
\begin{Bmatrix}j & s' & 1 \\ \frac{1}{2} & \frac{1}{2} & \frac{1}{2}\end{Bmatrix} \begin{Bmatrix}s' & \frac{1}{2} & J \\ \frac{1}{2} & j' & 1\end{Bmatrix} \begin{Bmatrix}s' & J & \frac{1}{2} \\ I & 1 & j\end{Bmatrix}
    \nn
 && \hspace{0.7cm} {}         
    \left.\left.\left.  +(-1)^{J} \left((-1)^{j+j'}+1\right) \delta _{l' 0} \delta _{s' j'}  \begin{Bmatrix}1 & j & j' \\ \frac{1}{2} & \frac{1}{2} & \frac{1}{2}\end{Bmatrix} \begin{Bmatrix}j' & J & \frac{1}{2} \\ I & 1 & j\end{Bmatrix} \right. \right. \right.
\nn
 && \hspace{0.7cm} {}  \times       
   \left.\left.\left. \left(-6 q \hat{\lambda} (-1)^{I}  \delta _{\lambda' 0} \begin{pmatrix}1 & 1 & \lambda \\ 0 & 0 & 0\end{pmatrix} \begin{Bmatrix}1 & I & \frac{1}{2} \\ \frac{1}{2} & 1 & \lambda\end{Bmatrix}-i \sqrt{2} q' \delta _{\lambda 1} \delta _{\lambda' 1}\right)\right]\right]\right) .
\eeqa
\item
For the $E_{12}$-term
\begin{equation}
    V_{E_{12}} = \sum_{i\neq j \neq k}  \boldsymbol{q}_i \cdot \boldsymbol{\sigma}_j \boldsymbol{q}_j \cdot \boldsymbol{\sigma}_i \boldsymbol{\tau}_i \cdot \boldsymbol{\tau}_j\,,
\end{equation}
we choose the Faddeev component
\begin{equation}
    V_{E_{12}}^{(1)} = \boldsymbol{q}_1 \cdot \boldsymbol{\sigma}_2 \boldsymbol{q}_2 \cdot \boldsymbol{\sigma}_1 \boldsymbol{\tau}_1 \cdot \boldsymbol{\tau}_2 + \boldsymbol{q}_1 \cdot \boldsymbol{\sigma}_3 \boldsymbol{q}_3 \cdot \boldsymbol{\sigma}_1 \boldsymbol{\tau}_1 \cdot \boldsymbol{\tau}_3
\end{equation}
and obtain
\beqa
&& \notag\mel{p' q' \alpha'}{V_{E_{12}}^{(1)}}{p q \alpha}  
\; =\; -48 \pi ^2 \hat{j} \hat{j'} \hat{t} \hat{t'} (-1)^{T}
\nn
 && \hspace{0.7cm} {} \times
 \begin{Bmatrix}1 & t & t' \\ \frac{1}{2} & \frac{1}{2} & \frac{1}{2}\end{Bmatrix} \begin{Bmatrix}t' & T & \frac{1}{2} \\ \frac{1}{2} & 1 & t\end{Bmatrix} \left(q' \hat{I'} \delta _{I \frac{1}{2}} \left[\delta _{l 0} \delta _{s j} \left(4 p' \hat{s'} (-1)^{j'+\frac{1}{2}}   \left((-1)^{j+s'+t+t'}-1\right) \delta _{\lambda 0} \delta _{\lambda' 1} \delta _{l' 1} \phantom{\frac{0}{0}} \right.\right.\right.
  \nn
  && \hspace{0.7cm} {} \times
  \left.\left.\left.  \begin{Bmatrix}j & s' & 1 \\ \frac{1}{2} & \frac{1}{2} & \frac{1}{2}\end{Bmatrix}  \begin{Bmatrix}1/2 & j & J \\ 1/2 & 1 & I' \\ 1 & s' & j'\end{Bmatrix}+(-1)^{J} \left((-1)^{j+j'+t+t'}+1\right) \delta _{l' 0}  \delta _{s' j'} \begin{Bmatrix}1 & j & j' \\ \frac{1}{2} & \frac{1}{2} & \frac{1}{2}\end{Bmatrix}
      \right.\right.\right.
  \nn
  && \hspace{0.7cm} {} \times
\left(\sqrt{2} q \delta _{\lambda 1} \delta _{\lambda' 1} \right.
  \left.\left.\left.\left. \begin{Bmatrix}j & J & \frac{1}{2} \\ I' & 1 & j'\end{Bmatrix}-6 q' \hat{\lambda'} (-1)^{I'+ \lambda'+\frac{1}{2}} \delta _{\lambda 0} \begin{pmatrix}1 & 1 & \lambda' \\ 0 & 0 & 0\end{pmatrix} \begin{Bmatrix}1 & \frac{1}{2} & I' \\ \frac{1}{2} & \lambda' & 1\end{Bmatrix}  \begin{Bmatrix}j' & I' & J \\ \frac{1}{2} & j & 1\end{Bmatrix}\right)\right) \right. \right.
\nn
 && \hspace{0.7cm} {}  
  -\left.\left.4 p \hat{s} (-1)^{j+\frac{1}{2}}  \left((-1)^{j'+s+t+t'}-1\right) \delta _{\lambda 0} \delta _{\lambda' 1} \delta _{l 1} \delta _{l' 0} \delta _{s' j'}  \begin{Bmatrix}s & j' & 1 \\ \frac{1}{2} & \frac{1}{2} & \frac{1}{2}\end{Bmatrix} \begin{Bmatrix}s & J & \frac{1}{2} \\ \frac{1}{2} & 1 & j\end{Bmatrix} \begin{Bmatrix}s & J & \frac{1}{2} \\ I' & 1 & j'\end{Bmatrix}\right] \right.
\nn
 && \hspace{0.7cm} {}  
  +\left. q \hat{I} \delta _{I' \frac{1}{2}} \left[4 p \hat{s} (-1)^{j'+s+\frac{1}{2}} \left((-1)^{j+t+t'} + (-1)^{1+j+j'+s}   \right) \delta _{\lambda 1} \delta _{\lambda' 0} \delta _{l 1} \delta _{l' 0} \delta _{s' j'} \phantom{\frac{0}{0}} \right.\right.
\nn
 && \hspace{0.7cm} {}  
 \left.\left.
\times \ \begin{Bmatrix}s & j' & 1 \\ \frac{1}{2} & \frac{1}{2} & \frac{1}{2}\end{Bmatrix}
       \begin{Bmatrix}1/2 & j' & J \\ 1/2 & 1 & I \\ 1 & s & j\end{Bmatrix} +\delta _{l 0} \delta _{s j} \left((-1)^{J} \left((-1)^{j+j'+t+t'}+1\right) \delta _{l' 0} \delta _{s' j'}  \begin{Bmatrix}1 & j & j' \\ \frac{1}{2} & \frac{1}{2} & \frac{1}{2}\end{Bmatrix}  \right.\right.\right.
\nn
&& \hspace{0.7cm} {}\times
\begin{Bmatrix}j' & J & \frac{1}{2} \\ I & 1 & j\end{Bmatrix} \left(\sqrt{2} q' \delta _{\lambda 1} \delta _{\lambda' 1} \right.
-\left.\left.\left.\left.6 q \hat{\lambda} (-1)^{I+\frac{1}{2}} \delta _{\lambda' 0} \begin{pmatrix}1 & 1 & \lambda \\ 0 & 0 & 0\end{pmatrix} \begin{Bmatrix}1 & I & \frac{1}{2} \\ \frac{1}{2} & 1 & \lambda\end{Bmatrix}\right)-4 p' \hat{s'} (-1)^{j'+\frac{1}{2}}  \delta _{\lambda 1} \delta _{\lambda' 0} \delta _{l' 1}
 \right.\right.\right.
\nn
&& \hspace{0.7cm} {}  \times
    \left((-1)^{j+s'+t+t'}-1\right)  \begin{Bmatrix}j & s' & 1 \\ \frac{1}{2} & \frac{1}{2} & \frac{1}{2}\end{Bmatrix}
  \left.\left.\left. \begin{Bmatrix}s' & \frac{1}{2} & J \\ \frac{1}{2} & j' & 1\end{Bmatrix} \begin{Bmatrix}s' & J & \frac{1}{2} \\ I & 1 & j\end{Bmatrix}\right)\right]\right) .
\eeqa
\item
Finally, for the $E_{13}$-term
\begin{equation}
    V_{E_{13}} = \sum_{i\neq j \neq k}  \boldsymbol{q}_i \cdot \boldsymbol{\sigma}_j \boldsymbol{q}_j \cdot \boldsymbol{\sigma}_i \boldsymbol{\tau}_i \cdot \boldsymbol{\tau}_k\,,
\end{equation}
we choose the Faddeev component
\begin{equation}
    V_{E_{13}}^{(1)} = \boldsymbol{q}_1 \cdot \boldsymbol{\sigma}_2 \boldsymbol{q}_2 \cdot \boldsymbol{\sigma}_1 \boldsymbol{\tau}_1 \cdot \boldsymbol{\tau}_3 + \boldsymbol{q}_1 \cdot \boldsymbol{\sigma}_3 \boldsymbol{q}_3 \cdot \boldsymbol{\sigma}_1 \boldsymbol{\tau}_1 \cdot \boldsymbol{\tau}_2
\end{equation}
and obtain
\beqa
&&\mel{p' q' \alpha'}{V_{E_{13}}^{(1)}}{p q \alpha}  \;
=\; -48 \pi ^2 \hat{j} \hat{j'} \hat{t} \hat{t'} (-1)^{T}
\nn
&& \hspace{0.7cm} {} \times
 \begin{Bmatrix}1 & t & t' \\ \frac{1}{2} & \frac{1}{2} & \frac{1}{2}\end{Bmatrix}  \begin{Bmatrix}t' & T & \frac{1}{2} \\ \frac{1}{2} & 1 & t\end{Bmatrix} \left(q' \hat{I'} \delta _{I \frac{1}{2}} \left[\delta _{l 0} \delta _{s j} \left(4 p' \hat{s'} (-1)^{j'+\frac{1}{2}} \left((-1)^{j+s'}-(-1)^{t+t'}\right)  \delta _{\lambda 0} \delta _{\lambda' 1}   \delta _{l' 1} \phantom{\frac{0}{0}} \right.\right.\right.
\nn
&& \hspace{0.7cm} {} \times
    \left.\left.\left. \begin{Bmatrix}j & s' & 1 \\ \frac{1}{2} & \frac{1}{2} & \frac{1}{2}\end{Bmatrix}  \begin{Bmatrix}1/2 & j & J \\ 1/2 & 1 & I' \\ 1 & s' & j'\end{Bmatrix}+(-1)^{J} \left((-1)^{j+j'}+(-1)^{t+t'}\right) \delta _{l' 0} \delta _{s' j'}\begin{Bmatrix}1 & j & j' \\ \frac{1}{2} & \frac{1}{2} & \frac{1}{2}\end{Bmatrix}   \left[\sqrt{2} q \delta _{\lambda 1} \delta _{\lambda' 1} \phantom{\frac{0}{0}} \right.\right.\right.\right.
\nn
&& \hspace{0.7cm} {} \times
    \left.\left.\left.\left.  \begin{Bmatrix}j & J & \frac{1}{2} \\ I' & 1 & j'\end{Bmatrix}-6 q' \hat{\lambda'} (-1)^{I'+ \lambda'+\frac{1}{2}} \delta _{\lambda 0} \begin{pmatrix}1 & 1 & \lambda' \\ 0 & 0 & 0\end{pmatrix} \begin{Bmatrix}1 & \frac{1}{2} & I' \\ \frac{1}{2} & \lambda' & 1\end{Bmatrix}  \begin{Bmatrix}j' & I' & J \\ \frac{1}{2} & j & 1\end{Bmatrix}\right]\right) \right.\right.
\nn
&& \hspace{0.7cm} {}
    - \left.\left. 4 p \hat{s} (-1)^{j+\frac{1}{2}} \left((-1)^{j'+s}-(-1)^{t+t'}\right) \delta _{\lambda 0} \delta _{\lambda' 1} \delta _{l 1} \delta _{l' 0} \delta _{s' j'}  \begin{Bmatrix}s & j' & 1 \\ \frac{1}{2} & \frac{1}{2} & \frac{1}{2}\end{Bmatrix} \begin{Bmatrix}s & J & \frac{1}{2} \\ \frac{1}{2} & 1 & j\end{Bmatrix} \begin{Bmatrix}s & J & \frac{1}{2} \\ I' & 1 & j'\end{Bmatrix}\right] \right.
\nn
&& \hspace{0.7cm} {}
    +\left.q \hat{I} \delta _{I' \frac{1}{2}} \left[4 p \hat{s} (-1)^{j'+s+\frac{1}{2}} \delta _{\lambda 1} \delta _{\lambda' 0} \delta _{l 1} \delta _{l' 0} \delta _{s' j'} \left( (-1)^{1+j+j'+s+t+t'}  + (-1)^{j}  \right) \right.\right.
\nn
&& \hspace{0.7cm} {}
    \times \left.\left. \begin{Bmatrix}s & j' & 1 \\ \frac{1}{2} & \frac{1}{2} & \frac{1}{2}\end{Bmatrix} \begin{Bmatrix}1/2 & j' & J \\ 1/2 & 1 & I \\ 1 & s & j\end{Bmatrix} +\delta _{l 0} \delta _{s j} \left((-1)^{J} \left((-1)^{j+j'}+(-1)^{t+t'}\right) \delta _{l' 0} \delta _{s' j'}  \begin{Bmatrix}1 & j & j' \\ \frac{1}{2} & \frac{1}{2} & \frac{1}{2}\end{Bmatrix} \right.\right.\right.
\nn
&& \hspace{0.7cm} {} \times
\begin{Bmatrix}j' & J & \frac{1}{2} \\ I & 1 & j\end{Bmatrix} 
    \left.\left.\left. \left(\sqrt{2} q' \delta _{\lambda 1} \delta _{\lambda' 1}-6 q \hat{\lambda} (-1)^{I+\frac{1}{2}}  \delta _{\lambda' 0} \begin{pmatrix}1 & 1 & \lambda \\ 0 & 0 & 0\end{pmatrix} \begin{Bmatrix}1 & I & \frac{1}{2} \\ \frac{1}{2} & 1 & \lambda\end{Bmatrix}\right)-4 p' \hat{s'}(-1)^{j'+\frac{1}{2}}  \delta _{\lambda 1} \delta _{\lambda' 0} \delta _{l' 1} \right.\right.\right.
\nn
&& \hspace{0.7cm} {} \times
 \left((-1)^{j+s'}-(-1)^{t+t'}\right) 
    \left.\left.\left.  \begin{Bmatrix}j & s' & 1 \\ \frac{1}{2} & \frac{1}{2} & \frac{1}{2}\end{Bmatrix}  \begin{Bmatrix}s' & \frac{1}{2} & J \\ \frac{1}{2} & j' & 1\end{Bmatrix} \begin{Bmatrix}s' & J & \frac{1}{2} \\ I & 1 & j\end{Bmatrix}\right)\right]\right) .
\eeqa
\end{itemize}

\section{S-matrix for elastic Nd scattering}\label{Appendix:Smatrix}

To calculate the S-matrix for elastic Nd scattering, we solve the
momentum-space Faddeev equation in the partial-wave basis as described
in Ref.~\cite{Gloeckle:1995jg} in order to obtain the transition operator $U$. In the
channel-spin representation, the S-matrix is related to the transfer
operator via \cite{Seyler:1969sii,Gloeckle:1995jg}
\beq
\label{UMatrix}
S^J_{\lambda' \Sigma', \, \lambda \Sigma} = \delta_{\lambda ' \lambda}
\delta_{\Sigma ' \Sigma} - \frac{4 \pi}{3} i q_0 m i^{\lambda' -
  \lambda}  U^J_{\lambda ' \Sigma', \lambda \Sigma}\,,
\eeq
where $q_0$ and $m$ denote the initial momentum and mass of the
nucleon, respectively.  
For $J \geq 3/2$ states with the parity $P= (-1)^{J \pm 1/2}$, the  S-matrix $S^J_{\lambda ' \Sigma', \lambda \Sigma}$
has the form
\beq
S  = \left( \begin{array}{ccc} S^J_{J\mp \frac{3}{2}\,  \frac{3}{2}, \, J\mp
            \frac{3}{2}\,  \frac{3}{2}} & S^J_{J\mp \frac{3}{2}\,  \frac{3}{2}, \, J\pm
                                          \frac{1}{2}\,  \frac{1}{2}}
             & S^J_{J\mp \frac{3}{2}\,  \frac{3}{2}, \, J\pm
                                          \frac{1}{2}\,  \frac{3}{2}}  \\
 S^J_{J\pm \frac{1}{2}\,  \frac{1}{2}, \, J\mp
            \frac{3}{2}\,  \frac{3}{2}} & S^J_{J\pm \frac{1}{2}\,  \frac{1}{2}, \, J\pm
                                          \frac{1}{2}\,  \frac{1}{2}}
             & S^J_{J\pm \frac{1}{2}\,  \frac{1}{2}, \, J\pm
                                          \frac{1}{2}\,  \frac{3}{2}}  \\
S^J_{J\pm \frac{1}{2}\,  \frac{3}{2}, \, J\mp
            \frac{3}{2}\,  \frac{3}{2}} & S^J_{J\pm \frac{1}{2}\,  \frac{3}{2}, \, J\pm
                                          \frac{1}{2}\,  \frac{1}{2}}
             & S^J_{J\pm \frac{1}{2}\,  \frac{3}{2}, \, J\pm
                                          \frac{1}{2}\,  \frac{3}{2}}  
           \end{array}
          \right),
\eeq
and can be diagonalized via a suitable unitary transformation $U = v \, w \,
x$,
\beq
S \; =\;  U^T \, \left( \begin{array}{ccc} S^J_{J \mp \frac{3}{2}\, 
      \frac{3}{2}} &0 &0 \\ 0
      & 
S^J_{J \pm \frac{1}{2} \, \frac{1}{2}} & 0 \\
0 & 0 & S^J_{J \pm \frac{1}{2} \, \frac{3}{2}} \end{array} \right) \, U\,,
\eeq
where the matrices $v$,
$u$ and $x$ are parametrized in terms of the mixing
angles $\epsilon$, $\xi$ and $\eta$ as follows:   
\beq
v = \left( \begin{array}{ccc} 1 & 0 & 0 \\
             0 & \cos \epsilon & \sin \epsilon \\
             0, & - \sin \epsilon & \cos \epsilon  \end{array}
         \right), \qquad
         w = \left( \begin{array}{ccc}  \cos \xi & 0 & \sin \xi \\
           0 & 1 & 0 \\
           - \sin \xi & 0 & \cos \xi  \end{array}
       \right), \qquad
             x = \left( \begin{array}{ccc} \cos \eta & \sin \eta & 0
                          \\ - \sin \eta & \cos \eta & 0 \\ 0 & 0 & 1  \end{array}
       \right).
\eeq
Similarly, in the $J^P = 1/2^+$ channel, we have
\beq
S = \left( \begin{array}{cc} S^{\frac{1}{2}}_{2\,  \frac{3}{2}, \, 2\,  \frac{3}{2}} & S^{\frac{1}{2}}_{2\,  \frac{3}{2}, \, 0\,  \frac{1}{2}}  \\
S^{\frac{1}{2}}_{0\,  \frac{1}{2}, \, 2\,  \frac{3}{2}} & S^{\frac{1}{2}}_{0\,  \frac{1}{2}, \, 0\,  \frac{1}{2}} 
          \end{array}
        \right), \qquad
S = x^T \, \left( \begin{array}{cc}  e^{2 i \delta_{^4D_{1/2}}} & 0 \\
                    0 &
                    e^{2 i \delta_{^2S_{1/2}}} \end{array} \right) \,
                x, \quad \text{with} \quad
 x = \left( \begin{array}{cc} \cos \eta & \sin \eta 
                          \\ - \sin \eta & \cos \eta   \end{array}
                      \right)\,,               
\eeq
while in the  $J^P = 1/2^-$ channel, the parametrization acquires the
form 
\beq
S = \left( \begin{array}{cc} S^{\frac{1}{2}}_{1\,  \frac{1}{2}, \, 1\,  \frac{1}{2}} & S^{\frac{1}{2}}_{1\,  \frac{1}{2}, \, 1\,  \frac{3}{2}}  \\
S^{\frac{1}{2}}_{1\,  \frac{3}{2}, \, 1\,  \frac{1}{2}} & S^{\frac{1}{2}}_{1\,  \frac{3}{2}, \, 1\,  \frac{3}{2}} 
          \end{array}
        \right), \qquad
S = v^T \, \left( \begin{array}{cc}  e^{2 i \delta_{^2P_{1/2}}} & 0 \\
                    0 &
                    e^{2 i \delta_{^4P_{1/2}}} \end{array} \right) \,
                v, \quad \text{with} \quad
 v = \left( \begin{array}{cc} \cos \epsilon & \sin \epsilon 
                          \\ - \sin \epsilon & \cos \epsilon   \end{array}
                      \right)\,.               
\eeq
Thus, the S-matrix in the $J\geq 3/2$ ($J=1/2$) channels is
parametrized in terms of $3$ ($2$) phase shifts and $3$ ($1$) mixing
angles, which become complex-valued above the deuteron breakup energy.

\bigskip 

 \twocolumngrid

 \bigskip
 
\section{Inverse basis transformation}\label{Appendix:InverseSpectroBasis}

Given the LECs $S_i$ in the spectroscopic basis, the ones in the
original operator basis defined in Eq.~(\ref{DefEi}) can be obtained via 
\beqa
E_1 &=& \frac{1}{240}   (100 S_1 - 45 S_2 + 40\sqrt{2} S_3 - 40 S_5
-50  S_6 \nn
&-&  20\sqrt{2} S_7 -80 S_{8} - 
64 S_{9} + 8\sqrt{5} S_{10} - 126 S_{11}
),\nn
E_2 &=& \frac{1}{120}   (-15 S_2 + 150\sqrt{2} S_3 - 135 S_4 + 100
S_5 - 5 S_6\nn 
&-&  10\sqrt{2} S_7 - 40 S_8 - 4 S_9 -  20\sqrt{5} S_{10} +9 S_{11} - 
40 S_{12} \nn
&-& 20 S_{13}
),\nn
E_3 &=& \frac{1}{120}   (20 S_1 -30 S_2 - 150\sqrt{2} S_3 + 135 S_4
-60 S_5 \nn
&-& 15 S_6 +  10\sqrt{2} S_7 +120 S_8 - 36 S_9 +  20\sqrt{5} S_{10}
\nn
&-&69 S_{11} +  40 S_{12} 
+ 20 S_{13}
),\nn
E_4 &=& \frac{1}{360}   (40 S_1 - 75 S_2 - 110\sqrt{2} S_3 + 135 S_4 - 140
S_5 \nn
&-& 15 S_6 -  10\sqrt{2} S_7 - 40 S_8 + 28 \sqrt{5} S_{10} -45 S_{11}
\nn
&+& 
40 S_{12} + 20 S_{13}
),\nn
E_5 &=& \frac{1}{240}   (140 S_1 - 105 S_2 - 260\sqrt{2} S_3 + 270 S_4 - 160
S_5 \nn
&+& 10 S_6 +  160 S_8 - 64 S_9 + 48 \sqrt{5} S_{10} -66 S_{11} + 80 S_{12} 
\nn
&+& 
40 S_{13}
),\nn
E_6 &=& \frac{1}{144}   (28 S_1 - 21 S_2 - 68\sqrt{2} S_3 + 54 S_4 - 32
S_5 -6 S_6 \nn
&-&   16 \sqrt{2} S_7 + 32 S_8 + 16 \sqrt{5} S_{10} -18 S_{11} + 16 S_{12} 
\nn
&+& 
8 S_{13}
),\nn
E_7 &=& \frac{1}{8}   (-4 S_1 +3  S_2 +20\sqrt{2} S_3 -18 S_4 + 16
S_5 - 16 S_8 ),\nn
E_8 &=& \frac{1}{120}   (-20 S_1 +15 S_2 +100\sqrt{2} S_3 -90 S_4 +80
S_5 \nn
&-& 10 S_6 -  80 S_8 -8 S_9 +18 S_{11}
), \nn
E_9 &=& \frac{1}{80}   (60 S_1 - 45 S_2 - 100\sqrt{2} S_3 + 90 S_4 - 120
S_5  \nn
&-& 5 S_6+   120 S_8 -16 S_9 -39 S_{11} + 40 S_{12} + 20 S_{13} 
), \nn
E_{10} &=& \frac{1}{240}   (80 S_1 -60 S_2 - 160\sqrt{2} S_3 + 180 S_4
-120 S_5 \nn
&-& 25 S_6 -  40\sqrt{2} S_7 +120 S_8 - 8 S_9 -32 \sqrt{5} S_{10}
\nn
&-&27 S_{11} +  40 S_{12} 
+ 20 S_{13}
),\nn
E_{11} &=& \frac{1}{80}   (20 S_1 -15 S_2 + 100\sqrt{2} S_3 -90 S_4
+120 S_5 \nn
&-& 15 S_6  -120 S_8 - 24 S_9 -  21 S_{11} +  60 S_{13}
),\nn
E_{12} &=& \frac{1}{80}   (80\sqrt{2} S_3 -60 S_4
+40 S_5 -5 S_6  -40 S_8 \nn
&+& 16 \sqrt{5} S_{10 }- 15 S_{11} +  20 S_{13}
),\nn
E_{13} &=& \frac{1}{120}   (-60 S_1 +45 S_2 + 300\sqrt{2} S_3 -270 S_4
+160 S_5 \nn
&-& 10 S_6  -20 \sqrt{2} S_7 - 160 S_8 - 8 S_9 -  40 \sqrt{5} S_{10}
\nn
&+&
18 S_{11} -  40 S_{12} +  40 S_{13} 
).
\eeqa

\section{Quality assurance of the emulator}\label{Appendix:AccuracyEmulator}

To verify the accuracy of the emulator described in
sec.~\ref{sec:Emulator}, $300$ points in the space of variable LECs
were randomly selected to provide a testing dataset (while maintaining
the triton binding energy constraint using the appropriate values of $c_E$).
Then, Nd elastic scattering observables were emulated for
the test points and compared to the exact results. Given the vastly
different scale of the differential cross section and polarization
observables, this comparison was performed separately and then repeated
for all three energies. 
The relative deviation for the differential
cross section is defined via 
\beqa
    \text{Error}\left[\left.\frac{d\sigma}{d\theta}\right\vert_{\theta=\theta_i}\right]
    &\equiv& \delta \sigma_i^{\rm emul} \;
    = \; \frac{|\sigma_i^\text{emul}-\sigma_i^\text{exact}|}{\sigma_i^\text{exact}}.
    \label{eq:error_interpolator_ds}
    \nonumber
\eeqa
For polarization observables, collectively denoted with $P^\alpha$,
we use the absolute difference as the accuracy measure by defining
\beqa
    \text{Error}\left[P^\alpha(\theta_i)\right]&\equiv&
    \delta P^{\alpha, \, \rm emul}_i \\
    &=& |(P^\alpha_i)^\text{emul} - (P^\alpha_i)^\text{exact}|.
    \nonumber
    \label{eq:error_interpolator_pol}
\eeqa

However, considering only these error measures might be
misleading. Indeed, a small
relative difference to the exact result does not necessarily confirm
a high accuracy of the emulator, since Nd scattering observables
are strongly dominated by the parameter-independent part of the
Hamiltonian.  
It is, therefore, necessary to assess the accuracy of the emulator
in relationship with the 3NF contributions we are interested in here. 
To this aim, we compare the deviations between the emulated and exact results
with the size of the 3NF contributions that depend on variable LECs.
Let $\sigma^0_i$ and $(P^\alpha)^0_i$ denote the
parameter-independent predictions for the cross section and
polarization observables, respectively,  at some angle $\theta_i$,
which are obtained using $c_D = s_i = 0$
(and the value of $c_E$ fixed from the $^3$H binding energy). 
Then, we measure the size of the parameter-dependent 3NF in a close
analogy with Eqs.~(\ref{eq:error_interpolator_ds})
and (\ref{eq:error_interpolator_pol}) via 
\begin{equation}
  \delta \sigma_i^{\rm 3NF}  = \frac{|\sigma_i^\text{exact}-\sigma^0_i|}{\sigma^0_i},
    \label{eq:size_3nf_ds}
\end{equation}
and
\begin{equation}
  \delta P^{\alpha , \, \rm 3NF}_i   = |(P^\alpha_i)^\text{exact} - (P^\alpha_i)^\text{0}|.
    \label{eq:size_3nf_pol}
\end{equation}
Using the above measures, we build histograms by calculating the
emulation error and the 3NF size for the differential cross section
and the analyzing powers for different scattering angles
at the energies of $E_N = 10$, $70$ and $135$~MeV. If the emulator is accurate, then the size of
the error in the description of observables should be much smaller
than the size of the 3NF contributions, and one expects a
small overlap between the histograms. Conversely, if the histograms
overlap significantly, then the interpolation errors are comparable to
the 3NF contributions, meaning that the emulator is not capable of
accurately capturing 3NF effects we are interested in here.

\begin{figure}[t]
	\centering
	\includegraphics[width=0.99\linewidth]{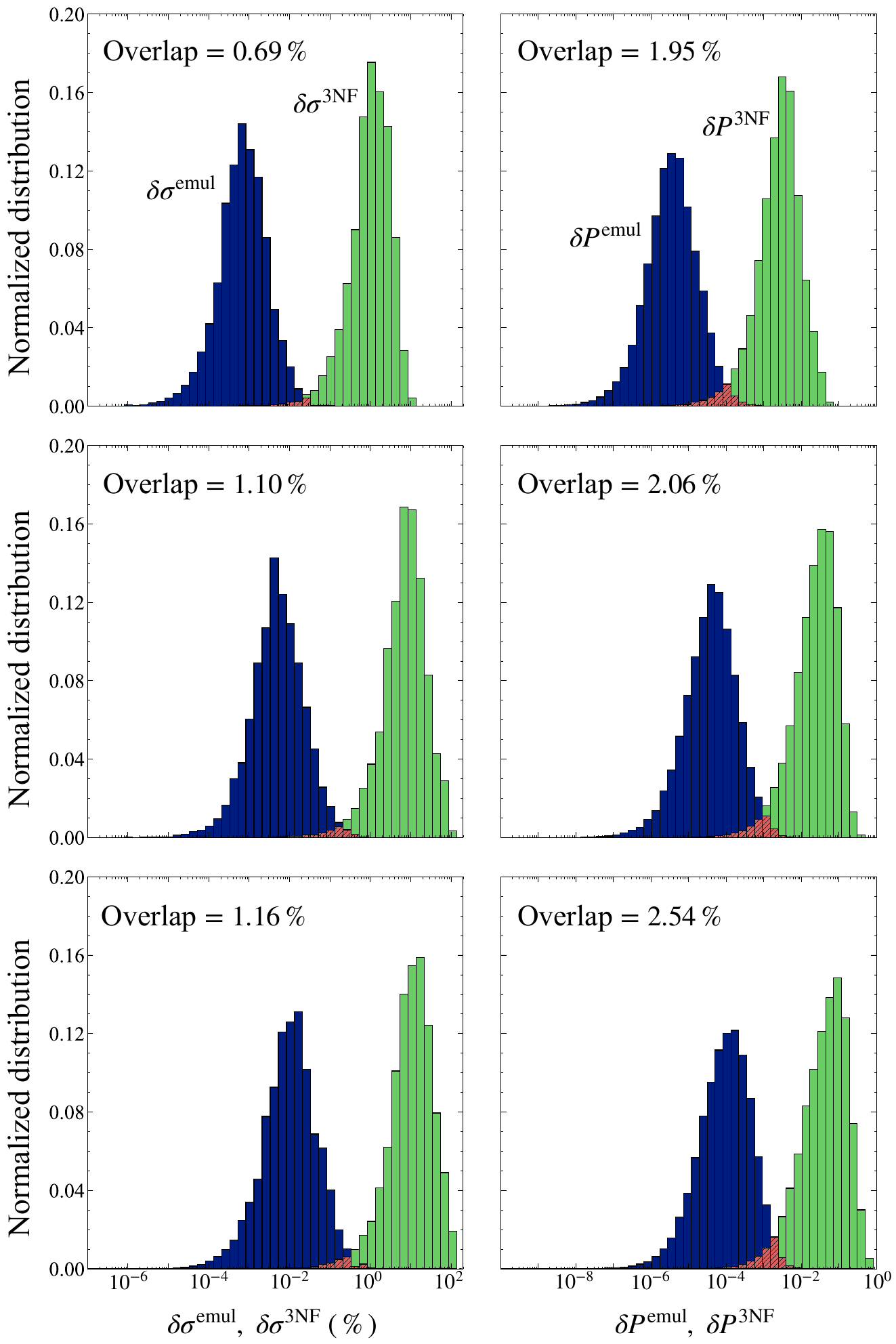}
        \caption{Relative errors for the emulation of the differential
      cross section (left plots) and absolute errors for the emulation
      of the analyzing powers (right
      plots) as defined in the text. The results at $E_N = 10$, $70$ and $135$~MeV are shown
      in the first, second and third row, respectively.  The
      histograms are based on a sample of $300$ randomly chosen values
      of the variable LECs $c_D$ and $s_i$ within the interpolation
      domain. Dark- and light-shaded histograms show
      the emulation errors and the size of the 3NF contributions,
      respectively. The hatched areas mark the overlap regions between the
      histograms, with the corresponding values given in each plot.  
      }
	\label{fig:interpolation}
\end{figure}

In Fig.~\ref{fig:interpolation}, we show the accuracy tests performed on the emulator. 
One observes that the average error in all cases is $3$ to $4$ orders of
magnitude smaller than the size of the 3NF, yielding a very
small overlap between histograms. This precision is comparable to the
systematic error introduced through the numerical solution of the
Faddeev equation. We, therefore, conclude that the constructed emulator
provides a sufficiently accurate reproduction of Nd elastic scattering observables for all values of the
LECs within the interpolation domain.

\section{Angular resolution in the center-of-mass
  frame}\label{Appendix:angular_error}

As already mentioned in the main text, we take into
account angular uncertainties of the  experimental data in our fits. Except for
the data of Ref.~\cite{Sperisen:1984bjw}, all other references listed in Table
\ref{tab:exp_database_sources} provide an estimation of the angular
resolution. These angular uncertainties are given in the
laboratory (lab)
frame, so they have to be transformed into the center-of-mass (CoM)
system by means of relativistic kinematical relations. The transformation depends
on whether the detected particle is a proton or a deuteron. For most
of the data, the outgoing deuteron and proton are detected in coincidence, while
in some cases, like for the data gathered by the SMART magnetic
spectrograph of Refs.~\cite{Sekiguchi:2002sf,Sekiguchi:2005vq}, each
data point corresponds to a detection of either a deuteron or a proton. Thus,
the transformations, whose form is described in Ref.~\cite{dick2009singularity},
need to be applied slightly differently for those two cases. Keeping
the convention of the authors of Ref.~\cite{dick2009singularity}, the relationships between the lab 
and the CoM angles read: 
\beqa
\label{eq:cm_to_lab}
    \cos{\theta_{jl}} &=& \frac{\gamma_{cl}(\cos{\theta_{jc}} + \alpha_{jc})}{\sqrt{\gamma_{cl}^2(\cos{\theta_{jc}}+\alpha_{jc})^2+\sin^2{\theta_{jc}}}},\\[10pt]
\label{eq:lab_to_cm}    
    \cos{\theta_{jc}} &=& \frac{-\alpha_{jc}\gamma_{cl}^2(1-\cos^2{\theta_{jl}})\pm\cos^2{\theta_{jl}}\sqrt{D}}{\gamma_{cl}^2(1-\beta_{cl}^2\cos^2{\theta_{jl}})}, \phantom{XX}
\eeqa
where
\begin{equation}
    D\equiv1+\gamma_{cl}^2(1-\alpha_{jc}^2)\frac{1-\cos^2{\theta_{jl}}}{\cos^2{\theta_{jl}}}.
\end{equation}
Here, for a reaction $1+2\rightarrow3+4$ with particle $1$ being the projectile and particle $2$ the target, $j=3,4$ refers to any of the
outgoing particles, while $\theta_{jl}$ and $\theta_{jc}$ are their
respective lab and CoM scattering angles. Further, $\gamma_{cl} = (1 -
\beta_{cl}^2 )^{-1/2}$ where 
$\beta_{cl} = v_{cl}/c$ with $\fett v_{cl}$ being the velocity of the
CoM system relative to the lab frame and $c$ denoting the speed of light.  Finally,
$\alpha_{jc}$ is defined as  
\begin{equation}
    \alpha_{jc}=\frac{\beta_{cl}}{\beta_{jc}}.
  \end{equation}
  
Notice that the transformation to the CoM system features a singularity when the lab
scattering angle approaches its maximum possible value. 
In this kinematical region, a small change in $\theta_{jl}$ leads to
large shifts in $\theta_{jc}$, see Ref.~\cite{dick2009singularity} for
more details.

Regarding the experiments carried out with a coincidence detection,
either the lab angles of both outgoing particles are provided, see  
e.g.~Ref.~\cite{SUDA2007745}, or the angular ranges for the detection of the 
scattered and recoil particles in the lab frame are specified. 
In some cases only the CoM angle is given, see e.g.~Ref.~\cite{Sekiguchi:2002sf}, but
the detection of protons and deuterons can still be identified from the provided information. 

Thus, when the scattering angle is given in the CoM, we transformed it
to the lab frame using Eq.~(\ref{eq:cm_to_lab}) for the detected
particle. Then, having the scattering angle in the lab frame $\theta_\text{LAB}$
and an angular resolution $\Delta\theta_\text{LAB}$, we determine
$\Delta\theta_\text{CoM}$ via 
\begin{equation}
    \Delta\theta_\text{CoM}=\frac{\theta_\text{CoM}\big|_{\theta_\text{LAB}+\Delta\theta_\text{LAB}}-\theta_\text{CoM}
      \big|_{\theta_\text{LAB}-\Delta\theta_\text{LAB}}}{2},
\end{equation}
where $\theta_\text{CoM}\big|_{\theta_\text{LAB}}$ is obtained from Eq.~(\ref{eq:lab_to_cm}).



\end{document}